\pdfoutput=1
\documentclass[12pt]{article}
\usepackage{graphicx}
\usepackage{amsmath}
\usepackage{amsfonts}
\usepackage{amsthm}
\usepackage{enumerate,setspace}
\usepackage{bbm,comment}
\usepackage{authblk}
\usepackage{dcolumn,booktabs}
\usepackage{siunitx}
\usepackage{bm}

\usepackage{natbib}
\usepackage{xcolor}
\usepackage{hyperref}
\usepackage{enumitem}
\definecolor{LightGray}{gray}{0.9}
\usepackage{multirow}
\usepackage{placeins}
\usepackage{float}
\usepackage[english]{babel}
\usepackage{longtable}
\usepackage{microtype}
\usepackage{titlesec}
\titlespacing*{\section}{0pt}{1.82ex}{1ex}
\titlespacing*{\subsection}{0pt}{1.45ex}{0.8ex}
\titlespacing*{\subsubsection}{0pt}{1ex}{0.6ex}

\newcommand{\blind}{0}

\usepackage{silence}
\definecolor{comcolor}{rgb}{0,0.5,0}
\definecolor{funcolor}{rgb}{0,0.4,1}
\definecolor{concolor}{rgb}{0,0,1}

\definecolor{codegreen}{rgb}{0,0.6,0}
\definecolor{codegray}{rgb}{0.5,0.5,0.5}
\definecolor{codepurple}{rgb}{0.58,0,0.82}
\definecolor{backcolour}{rgb}{0.93,0.93,0.93}


\begin{document}

\def\spacingset#1{\renewcommand{\baselinestretch}%
{#1}\small\normalsize} \spacingset{1}

\if0\blind
{
  \title{\bf Bayesian precision factor models for high-dimensional macroeconomic time series}
  \author[1]{Peter Dunson}
  \affil[1]{Department of Mathematics \& Statistics, Kenyon College}
  \maketitle
} \fi

\if1\blind
{
  \bigskip
  \bigskip
  \bigskip
  \begin{center}
    {\LARGE\bf Bayesian precision factor models for high-dimensional macroeconomic time series}
\end{center}
  \medskip
} \fi

\bigskip

\begin{abstract}
\noindent Series in a macroeconomic panel are almost all correlated with one another, which makes correlation a poor guide to which pairs are directly linked. For example, Treasury rates at different maturities are all highly correlated, yet a short and a long rate may be related only through the rates at maturities between them; similarly, sectors of the economy such as output, employment, prices, and interest rates may be related directly or only through other series. Both are questions about the inverse covariance matrix. Recovering it from quarterly data is difficult: contemporaneous dependence is entangled with temporal dynamics, the innovations remain densely dependent once those dynamics are removed, and the number of series is comparable to the number of quarters observed. We propose VAPOR (Vector Autoregressive Precision factOR), a Bayesian model that estimates vector autoregressive dynamics jointly with a factor-structured innovation precision matrix, inferring the partial correlations directly from the fitted precision rather than imposing a zero pattern. Applied to 100 quarterly FRED-QD series, VAPOR learns from the data that each Treasury rate is directly related only to the rates at adjacent maturities, places all 26 selected dependencies within a single FRED-QD group, and attains the lowest held-out prediction error among the estimators compared.
\end{abstract}

\noindent%
\vfill

\newpage
\spacingset{1.8}

\section{Introduction}\label{sec:intro}

Many macroeconomic series move together only because they respond to the same shocks, not because they are directly linked. \citet{acemoglu2012network} show that the pattern of linkages across sectors determines whether idiosyncratic shocks average out or aggregate into fluctuations. A small number of common factors accounts for most variation in large macroeconomic panels \citep{stock2002forecasting, forni2000generalized}, so pairwise correlation is a poor guide to direct dependence. The partial correlation between two series given all the others measures that dependence: its magnitude gives the strength, and the partial correlations distinguishable from zero define an estimated conditional independence graph. In contrast, most network methods for macroeconomic and financial panels rely on forecast error variance decompositions \citep{diebold2014network, billio2012econometric}, which describe how shocks propagate over time rather than which pairs remain dependent when the remaining series are held fixed.

We study 100 quarterly U.S.\ macroeconomic series from the FRED-QD database \citep{mccracken2020fredqd}, a public panel maintained by the Federal Reserve Bank of St.\ Louis and assembled from the statistical releases of the Bureau of Economic Analysis, the Bureau of Labor Statistics, the Federal Reserve Board, and the Census Bureau. The series cover output, employment, consumption, prices, interest rates, money and credit, and financial markets, and are transformed to stationarity using the transformation codes distributed with the database. Observations accumulate one per quarter and cannot be increased by collecting more data, so over a window running from 1996 to 2025 the panel is short relative to its width. Section~\ref{sec:motivation} describes the data in detail.

Our analysis addresses two questions about this panel. The first is whether the maturities of the Treasury curve remain dependent on one another once the intervening maturities are conditioned on, given that a few factors explain most of the curve's variation \citep{litterman1991common}. The second is whether sectors of the economy, such as output, employment, prices, and interest rates, are related directly or only through other series. Marginal correlations cannot answer either question. Among the five policy and Treasury rate series, sample correlations run from $0.28$ to $0.95$ with a median of $0.68$, which would place every maturity in dependence with every other.

\begin{figure}[h!]
\centering
\includegraphics[width=0.9\textwidth,alt={Two panels. Left, a six-by-six matrix of marginal correlations among the federal funds rate, the three-month, one-, five- and ten-year Treasury rates, and the thirty-year mortgage rate, in which every pair is substantially correlated. Right, the conditional dependence graph estimated by VAPOR for the same six series: a single chain joining adjacent maturities, with the mortgage rate attached to the five-year yield and no non-adjacent maturity pair selected.}]{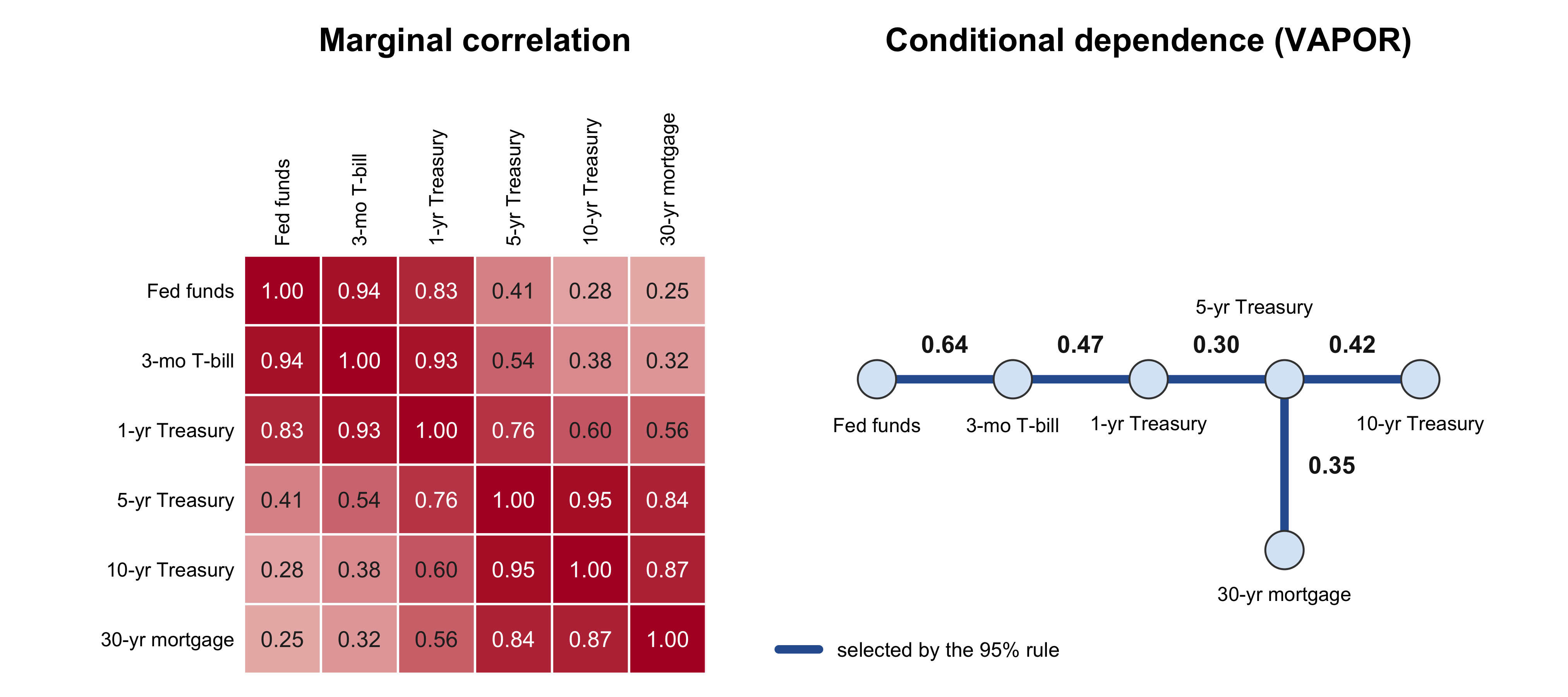}
\caption{Marginal and conditional dependence among six interest-rate series in the $p = 100$ panel of economic indicators: the federal funds rate, the three-month Treasury bill rate, the one-, five- and ten-year Treasury rates, and the thirty-year fixed mortgage rate. These six series form one connected component of the graph reported by VAPOR. Left: sample correlations of the transformed series over the training window. Right: the edges VAPOR reports among them, those whose central 95\% posterior credible interval excludes zero, labeled with posterior mean partial correlations. Only adjacent maturities are connected, and the mortgage rate is connected only to the five-year Treasury rate.}
\label{fig:termstructure_p100}
\end{figure}

Figure~\ref{fig:termstructure_p100} contrasts marginal and conditional dependence among these series. VAPOR, the model we develop below, estimates partial correlations of $0.30$ to $0.64$ between adjacent maturities and shrinks all six non-adjacent pairs to $0.12$ or less in absolute value, so the curve reduces to a nearest-neighbor chain. Across the full panel, all 26 pairs whose posterior credible intervals exclude zero join series within a single FRED-QD group, so the groups of this economy are close to conditionally independent of one another.

Competing estimators recover this structure only in part. The graphical lasso applied to vector-autoregressive residuals recovers the chain but sets both negative dependencies to exactly zero, while a factor stochastic volatility model of the innovation covariance ranks the negative dependencies 15th and 33rd but ranks three of the four adjacent-maturity pairs 754th, 760th and 1{,}753rd of 4{,}950. VAPOR attains the lowest held-out prediction error, with a mean squared error of $0.563$, against $0.592$ for the graphical lasso and $218.8$ for the covariance factor model.

This panel poses three problems for standard methods. First, the contemporaneous dependence we want to estimate is mixed with temporal dynamics, and removing those dynamics imperfectly distorts the estimated precision matrix. Second, the series move together strongly even after the dynamics are removed, so the remaining dependence is dense and low rank rather than sparse. Third, the number of series is close to the number of quarters observed, so the sample covariance is poorly conditioned, its inverse is dense and hard to interpret, and a full vector autoregression has more coefficients than the data can support.

\subsection{Related Work}

For a mean-zero Gaussian vector with covariance $\Sigma$, the precision matrix $\Omega = \Sigma^{-1}$ encodes conditional independence: a zero off-diagonal entry $\Omega_{ij} = 0$ implies that variables $i$ and $j$ are conditionally independent given all others, an absent edge in the Gaussian graphical model \citep{lauritzen1996graphical}. Inferring which entries are zero is the problem of Gaussian graphical model selection, studied since \citet{dempster1972covariance} considered estimation of the precision matrix under zero constraints, and it remains important in applications across genomics, neuroimaging, and finance. Penalized approaches impose exact zeros on $\Omega$ directly: neighborhood selection regresses each variable on the remainder \citep{meinshausen2006high}; the graphical lasso maximizes an $\ell_1$-penalized likelihood in the concentration matrix \citep{yuan2007model, friedman2008sparse}; and constrained $\ell_1$ minimization targets the same support \citep{cai2011constrained}. Each is designed for independent observations.

Bayesian treatments follow two routes. The canonical one places a prior on the graph $G$ and a conditionally conjugate prior on $\Omega$ given $G$, such as the hyper-inverse Wishart or G-Wishart \citep{roverato2002hyper, atay2005monte}, and explores the space of graphs by stochastic search \citep{jones2005experiments}; see \citet{mohammadi2015bayesian} for a review and the \texttt{BDgraph} package for software. This delivers posterior probabilities over graphs, but the search is expensive in high dimensions and harder still for non-decomposable graphs, where the normalizing constant is unavailable in closed form. At $p = 100$ there are $2^{4950}$ possible graphs, so posterior mass spreads thinly enough that no single graph carries appreciable probability, and reporting a single estimated $\hat{G}$ invites over-interpretation. The alternative replaces the point mass at zero with a continuous shrinkage prior on the entries of $\Omega$, as in the Bayesian graphical lasso \citep{wang2012bayesian} and the graphical horseshoe \citep{li2019graphical}, trading exact sparsity for tractable computation while maintaining positive definiteness. Our approach builds on Precision Factor Analysis (PrecFact) \citep{chandra2026precfact}, which regularizes $\Omega$ through a low-rank plus diagonal (LRD) factorization with a Dirichlet-Laplace prior on the loadings, instead of shrinking its elements toward zero. This captures dependence shared across many pairs that scattered zeros cannot express.

Precision estimators designed for independent and identically distributed (iid) data, including PrecFact, ignore temporal dependence. When applied directly to $\bm{y}_t$, without first removing the dynamics, they conflate the temporal dynamics linking $\bm{y}_t$ to its past with the contemporaneous structure that remains once those dynamics are removed, which distorts the estimated precision matrix, as our simulations show.

Alternatively, there is a rich literature on modeling multivariate time series, with vector autoregressions (VARs) a particularly popular class since \citet{sims1980macroeconomics}. Bayesian shrinkage has been widely used to improve VARs since \citet{litterman1986forecasting}. \citet{banbura2010large} show that a Bayesian VAR scales to large panels when shrinkage increases with the cross-sectional dimension, and \citet{giannone2015prior} learn the degree of shrinkage adaptively through a hierarchical model. \citet{basu2015regularized} establish nonasymptotic guarantees for $\ell_1$-penalized estimation of high-dimensional VARs.

For high-dimensional time series, \citet{kastner2020sparse} combine a VAR with factor stochastic volatility on the residuals \citep{kastner2019sparse}, using Dirichlet-Laplace shrinkage priors on the VAR coefficients. Their factor model targets the residual covariance matrix. Although the covariance can be inverted to obtain the precision matrix, we find that covariance factor models can perform poorly in inferring nearly sparse precision matrices. The same limitation applies to factor-based covariance estimators for independent observations, which pair a factor structure with a sparse residual covariance \citep{fan2011high, fan2013large}. Sparsity in the residual covariance does not transfer to the precision matrix: by the Woodbury identity \citep{woodbury1950inverting} the inverse of a factor-plus-sparse covariance is dense whenever the loadings are, so the zero pattern that identifies the graph is not preserved under inversion.

Low-rank structure in the precision matrix also arises in latent-variable graphical models. \citet{chandrasekaran2012latent} show that marginalizing a Gaussian graphical model over latent variables yields an observed-variable precision matrix expressible as a sparse matrix minus a low-rank one, and give conditions under which the two components are separately identifiable. There the low-rank term is a nuisance induced by unobserved confounders rather than the target of estimation.

The work closest to ours estimates contemporaneous networks from the innovations of a VAR. \citet{barigozzi2019nets} estimate a VAR in which both the coefficient matrices and the innovation precision matrix are sparse, recovering a directed Granger network and an undirected contemporaneous network by a joint penalized algorithm. \citet{ahelegbey2016bayesian} give a Bayesian graphical treatment of structural VARs that samples the contemporaneous and temporal causal structures jointly with the model parameters, and \citet{vahap2025bayesian} selects conditional independence structure in a time-varying VAR by Bayesian graphical variable selection, applied to ten series. All three follow the traditional graphical modeling approach of inferring the locations of exact zeros in the innovation precision matrix, which faces the computational and inferential hurdles described above.

\subsection{The Proposed Approach}

We propose VAPOR (Vector Autoregressive Precision factOR), a joint Bayesian model for the conditional independence structure of high-dimensional time series. A VAR(1) models the temporal dynamics while a low-rank-plus-diagonal factorization models the precision matrix of the innovations, and both components are estimated jointly in a single Bayesian model. Estimating the VAR coefficients first and then applying a precision estimator to the residuals treats the fitted residuals as data, ignoring estimation uncertainty in the VAR step and biasing recovery of $\Omega$ when the dynamics are strong or hard to estimate. VAPOR instead updates $B$ and $\Omega$ jointly, so that uncertainty in each informs the other. Sparsity also enters differently than in the estimators above: rather than imposing exact zeros, VAPOR shrinks continuously through the LRD parameterization, concentrating the off-diagonal mass into a small number of directions, and obtains the reported graph by posterior selection afterwards.

A reduced-rank form of the coefficient matrix extends the model to settings in which the number of series approaches or exceeds the number of observations. In simulations, VAPOR's advantage over competing methods grows with dimension and reverses in several settings, most clearly when the innovation precision has no factor structure. An open-source R implementation accompanies the paper.

\section{Data}\label{sec:motivation}

The FRED-QD panel is released monthly and each release is a distinct vintage; we use the May 2026 vintage throughout, in which observations run from Q1 1996 to Q4 2025. Because source agencies revise historical values, estimates from a different vintage will differ from those reported here. Each series carries a transformation code assigned by \citet{mccracken2020fredqd} to induce stationarity, leaving a few series untransformed and otherwise taking first differences of levels, first differences of logarithms of most real quantities, and second differences of logarithms of price indices; we apply these codes as distributed rather than selecting transformations ourselves. The FRED-QD appendix also assigns each series to a group, such as NIPA (National Income and Product Accounts), Prices, or Interest Rates; we use these groups, rather than a classification of our own, wherever we refer to sectors. 

To assess how methods behave as the dimension grows relative to the sample size, we consider two variable sets: a curated set of $p = 29$ series ($T/p \approx 3.4$) and an expanded set of $p = 100$ series ($T/p \approx 1.0$). The second regime is harder: as $p$ approaches $T$, the sample covariance becomes poorly conditioned. The $p = 29$ panel comprises headline indicators spanning output, employment, consumption, prices, interest rates, and financial markets. The $p = 100$ panel extends it by first retaining every FRED-QD series observed without gaps over the full window, then selecting from these by hand the 100 most widely reported indicators, keeping coverage across the same sectors. Requiring complete observations from 1996 excludes series introduced or discontinued during the window, so the panel is weighted toward long-established indicators. Both panels were fixed before any model was fit.

For both dimensions the training set runs from Q1 1996 to Q4 2020 ($T_{\mathrm{train}} = 100$ quarters) and the test set from Q1 2021 to Q4 2025 ($T_{\mathrm{test}} = 20$ quarters), so the two panels differ only in the number of series. The training period covers the dot-com bust, the 2008 financial crisis, and the COVID shock; the test period covers the post-COVID recovery. Each series is standardized by its training mean and standard deviation, which are also applied to the test quarters.

\begin{figure}[h!]
\centering
\includegraphics[width=\textwidth,alt={Two panels. Left, a heat map of the 100 standardized series by quarter from 1996 to 2025, grouped by FRED-QD group, with recessions marked above and the training and test periods bracketed; entire columns turn blue together in 2008 and blue then red at the COVID shock, and some rows keep one color for years. Right, the sorted eigenvalues of the correlation matrix on a log scale for 29 and 100 series, each with a dashed line at the largest eigenvalue expected for independent series; a few eigenvalues lie above the lines and the rest decay steeply, much further at 100 series.}]{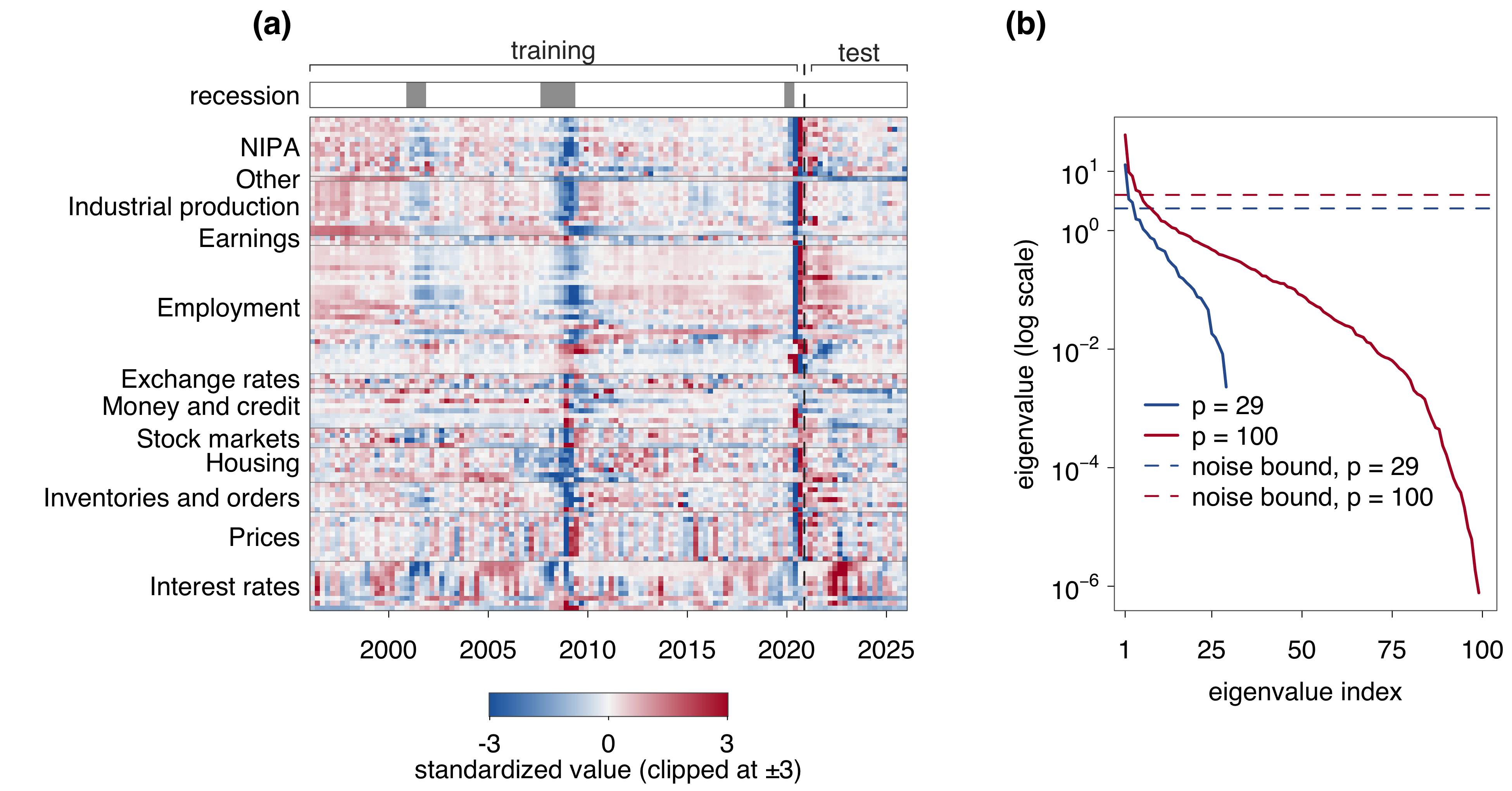}
\caption{The FRED-QD panel. (a) The $p = 100$ standardized series by quarter, grouped by FRED-QD group; red marks values above a series' training mean and blue values below it, in standard deviations, clipped at $\pm 3$, and recessions dated by the National Bureau of Economic Research shaded gray above. (b) Eigenvalues of the training-window correlation matrix on a log scale; dashed lines mark the largest eigenvalue expected for independent series, $(1 + \sqrt{p/T})^2$. At $p = 100$ the 100th eigenvalue, zero, is not shown.}
\label{fig:data_overview}
\end{figure}

Figure~\ref{fig:data_overview} shows the panel. In panel (a), whole columns change color together at the 2001 and 2008 recessions and at the COVID shock, as most series respond to the same shocks; the median absolute pairwise correlation over the training window is $0.35$ at $p = 29$ and $0.30$ at $p = 100$. Other rows change slowly and keep one sign for years, and lag-1 autocorrelation ranges from $-0.47$ to $0.92$ across the $p = 29$ series. In the COVID quarters most series lie beyond $\pm 3$, and the median excess kurtosis is $7.7$ and $10.2$.

Panel (b) shows that the dependence is concentrated in a few directions: the first principal component explains $44\%$ and $41\%$ of the total variance, and the leading eigenvalues lie well above the largest expected for independent series. The remaining eigenvalues decay toward zero. At $p = 100$ the smallest is exactly zero, so the sample covariance has rank 99 and cannot be inverted, and at $p = 29$ its condition number is $5.6 \times 10^{3}$. These features shape the model of Section~\ref{sec:model}: a VAR for the persistence, a low-rank-plus-diagonal precision matrix for dependence concentrated in a few directions, and regularization of both components because $T$ is close to $p$. Table~\ref{tab:data_summary} in the Supplementary Material gives further summaries of both panels.

Although almost every pair of series is correlated, the conditional dependence that VAPOR recovers at $p = 100$ is sparse (Section~\ref{subsec:p100}). Only 26 of the $4{,}950$ pairs have partial correlations distinguishable from zero, every one of them joins two series in the same FRED-QD group, and the interest rates form a chain in which each maturity is linked only to its neighbors (Figure~\ref{fig:termstructure_p100}).

\FloatBarrier

\section{Model}\label{sec:model}

\subsection{Notation}

Let $p$ denote the number of variables, $T$ the number of time points, and $k$ the number of latent precision factors. We observe a $p$-dimensional time series $\bm{y}_1, \dots, \bm{y}_T \in \mathbb{R}^p$. Let $K_x = p \cdot \mathrm{lags}$ denote the number of predictors per equation, $B \in \mathbb{R}^{p \times K_x}$ the VAR coefficient matrix, $\Lambda \in \mathbb{R}^{p \times k}$ the factor loadings matrix, $\Delta = \mathrm{diag}(\delta_1^2, \dots, \delta_p^2)$ a diagonal matrix of idiosyncratic precisions, and $\Omega \in \mathbb{R}^{p \times p}$ the precision matrix of the VAR residuals.

\subsection{Likelihood}

We model the data as a VAR(1) process:
\begin{equation}
    \bm{y}_t = B\bm{y}_{t-1} + \bm{\epsilon}_t, \quad \bm{\epsilon}_t \overset{\mathrm{iid}}{\sim} N_p(\bm{0}, \Omega^{-1}), \quad t = 2, \dots, T.
\end{equation}
The precision matrix $\Omega$ encodes the contemporaneous conditional dependence structure of the residuals after accounting for temporal dynamics. We model $\Omega$ via the LRD decomposition of \citet{chandra2026precfact}, with the priors on its components given in Section~\ref{subsec:priors}:
\begin{equation}
    \Omega = \Lambda\Lambda^\top + \Delta
\end{equation}
with $\Lambda$ and $\Delta$ as defined above. As in any factor model, $\Lambda$ is identified only up to right-orthogonal rotation, since $\Lambda Q$ with $QQ^\top = I_k$ yields the same $\Omega$. This is immaterial here: every quantity we report is a functional of $\Omega = \Lambda\Lambda^\top + \Delta$, which is invariant to the rotation, and we do not interpret individual loadings or factors. The diagonal entries of $\Omega$ are not directly interpretable as conditional dependencies; the off-diagonal structure of $\Omega$ defines the Gaussian graphical model.

This specification treats $\Omega$ as static. A body of work on dynamic Gaussian graphical models instead lets the dependence structure evolve over time: some allow the edge set itself to change, at estimated changepoints \citep{gibberd2017regularized} or through latent regime states \citep{liu2022dynamic}, while others hold the graph fixed and let only the magnitudes of the dependencies evolve \citep{carvalho2007dynamic}. Our aim is a single interpretable graph summarizing contemporaneous dependence once temporal dynamics are removed, so we estimate a static $\Omega$; we return to the time-varying extension in Section~\ref{sec:discussion}.

The LRD form also does not impose exact zeros in $\Omega$: for generic loadings every off-diagonal entry is nonzero, so $\Lambda\Lambda^\top + \Delta$ does not by itself define a sparse graph. Instead of fixing a zero pattern in advance, the LRD parameterization concentrates the off-diagonal mass into a low-dimensional set of directions, shrinking most partial correlations toward zero while allowing a small number to be large. Sparsity in the reported graph is then obtained by a separate selection step. In the application we report an edge when the central 95\% posterior credible interval of its partial correlation excludes zero; in the simulations, where sensitivity and specificity are computed against a known graph, we use the posterior false-discovery-rate procedure of \citet{chandra2026precfact}, which controls the positive false discovery rate (pFDR) of \citet{storey2003positive}.

\subsection{Priors}\label{subsec:priors}

\paragraph{VAR coefficients.}\label{par:var-coefficients}

We place a Dirichlet-Laplace (DL) prior \citep{bhattacharya2015dirichlet} on each row of $B$, following the equation-by-equation scheme of \citet{kastner2020sparse}. The DL prior is well suited to the VAR coefficients because it combines a spike of mass near zero with heavy tails: the many coefficients that reflect no genuine dynamic relationship are shrunk aggressively, while the few large ones are left essentially unpenalized. This behavior is important in the high-dimensional VAR, where the number of coefficients grows with $p^2$ while the available sample grows only with $T$. For the $j$-th row $\bm{b}_j \in \mathbb{R}^p$, the DL prior takes the form:
\begin{equation}
    b_{jh} \mid \zeta_{jh}, \eta_{jh} \sim N(0, \zeta_{jh}\eta_{jh}^2), \quad
    \zeta_{jh} \sim \mathrm{Exp}(1/2), \quad
    \eta_{jh} \sim \mathrm{Ga}(a_B, \xi_j), \quad \xi_j = 1/2,
\end{equation}
where $\xi_j = 1/2$ is a fixed row-level hyperparameter and $a_B = 1/(p K_x)$ with $K_x = p \cdot \mathrm{lags}$ the number of predictors per equation, so that $a_B$ is the inverse of the total number of VAR coefficients, following \citet{bhattacharya2015dirichlet}, with the sampling steps as corrected by \citet{gruber2025note}. For the VAR(1) model presented here, $K_x = p$, so $a_B = 1/p^2$.

\paragraph{Factor loadings.} We also shrink the high-dimensional loadings $\Lambda$ using a DL prior. Each entry has the hierarchical form
\begin{equation}
\begin{gathered}
    \lambda_{jl} \mid \psi_{jl}, \phi_{jl}, \tau \sim N(0, \psi_{jl}\phi_{jl}^2\tau^2), \quad
    \psi_{jl} \sim \mathrm{Exp}(1/2), \\
    \phi \sim \mathrm{Dir}(a_L, \dots, a_L), \quad
    \tau \sim \mathrm{Ga}(pk a_L, 1/2),
\end{gathered}
\end{equation}
where $\phi$ is the vector of local scales with shape $a_L = 0.5$ by default and $\tau$ is a global scale. Equivalently, the products $T_{jl} = \phi_{jl}\tau$ are independent $\mathrm{Ga}(a_L, 1/2)$ variables \citep{bhattacharya2015dirichlet}, and the sampler updates each $T_{jl}$ from its generalized inverse Gaussian (GIG) full conditional, in the order established by \citet{gruber2025note}.

\paragraph{Idiosyncratic precisions.} We place a Dirichlet process (DP) prior \citep{ferguson1973bayesian, sethuraman1994constructive} on the diagonal entries $\delta_1^2, \dots, \delta_p^2$. The DP encourages the idiosyncratic precisions to cluster into a small number of distinct values rather than being estimated freely, which stabilizes the diagonal in high dimensions by borrowing strength across variables with similar residual scales. The DP concentration parameter $\alpha$ governs the number of clusters and is assigned a prior updated within the sampler.

\paragraph{Hyperparameters.} The DL shape $a_B = 1/(pK_x)$, the inverse of the number of VAR coefficients \citep{bhattacharya2015dirichlet}, scales the prior's concentration with the dimension of the coefficient matrix, so that shrinkage strengthens automatically as $p$ grows and the coefficients become harder to identify. The DP base measure, the prior on each idiosyncratic precision $\delta_j^2$, is $\mathrm{Ga}(1, 1)$, which has mean one on the scale of the standardized data. The PrecFact default is $\mathrm{Ga}(0.1, 0.1)$, which places a third of its mass below $10^{-4}$; under that choice the sampled precisions of some series stayed near zero for long stretches of the chain in our application, and the partial correlations involving those series mixed slowly. The concentration parameter $\alpha$ receives a $\mathrm{Ga}(0.1, 0.1)$ prior under the \citet{escobar1995bayesian} update, imposing little prior information on the number of residual-precision clusters.

\subsection{Reduced-Rank Parameterization of the VAR Coefficients}\label{subsec:reduced-rank}

The parameterization described in this subsection is designed for the setting where $T \approx p$ or $T < p$. The DL prior of the previous section places a shrinkage prior on each entry of $B$ directly, but the number of free parameters in $B$ is $p K_x$, which for the VAR(1) model considered here is $p^2$. When $T$ is not comfortably larger than $p$, this parameterization becomes difficult to estimate: the generalized least squares (GLS) system in the $B$-update becomes ill-conditioned. Shrinking each coefficient toward zero is only one way to reduce effective dimensionality, and it may be insufficiently flexible when the dynamics are governed by a small number of shared directions rather than by sparsity in the individual entries of $B$. In that case a different notion of dimension reduction is more appealing: restricting $B$ to a low-rank structure, so that the dynamics are driven by a few latent factors.

We address this by replacing the free $p \times K_x$ matrix $B$ with a reduced-rank factorization,
\begin{equation}
    B = UV^\top, \qquad U \in \mathbb{R}^{p \times r}, \quad V \in \mathbb{R}^{K_x \times r},
\end{equation}
for a rank $r \le \min(p, K_x)$. This reduces the number of free parameters governing the VAR dynamics from $pK_x$ to $r(p + K_x)$, which is much smaller when $r$ is small relative to $p$. The columns of $U$ and $V$ act as latent factors driving the coefficient matrix: each series' dynamics are expressed through a shared $r$-dimensional subspace rather than estimated freely, so information is pooled across equations. Reduced-rank structure of this kind has a long history in vector autoregression \citep{velu1986reducedrank, carriero2011forecasting}. Only the product $B = UV^\top$ is identified: for any invertible $r \times r$ matrix $R$, $(UR)(VR^{-\top})^\top = UV^\top$, so $U$ and $V$ individually carry a rotational ambiguity. This is immaterial here, since $U$ and $V$ enter the likelihood and every downstream quantity (the innovations $\bm\epsilon_t = \bm y_t - B\bm y_{t-1}$, and hence $\Omega$) only through their product.

We place a ridge prior on $U$, $\mathrm{vec}(U) \sim N(\bm 0, \sigma_U^2 I)$, and consider two priors on $V$.

\paragraph{Ridge.} The simplest choice places a constant-variance Gaussian prior on every entry of $V$, $\mathrm{vec}(V) \sim N(\bm 0, \sigma_V^2 I)$. This requires $r$ to be fixed in advance and treats every retained dimension identically; it is fast and stable but offers no mechanism for determining whether $r$ is well chosen.

\paragraph{Dirichlet-Laplace.} Alternatively, we place the same DL prior described in the previous section on $V$ in place of $B$, row-by-row over the $K_x$ rows of $V$, with the DL shape parameter set to $a_V = 1/(K_x r)$ following the same convention as $a_B$. This gives entry-level local and global shrinkage within a fixed rank $r$, but, like ridge, does not itself inform the choice of $r$.

\paragraph{Estimating the rank.} We fix $r$ rather than estimate it. Thresholding the singular values of a ridge estimate of $B$ recovers the true rank only once $T$ is several times larger than $p$, because at $T \approx p$ estimation noise spreads across every singular direction. A multiplicative gamma process shrinkage prior \citep{bhattacharya2011sparse}, which infers the rank adaptively, was also unreliable at $T \approx p$ and added substantial computational cost. We use $r = 3$ in the application and examine $r \in \{2, 3, 4\}$ in sensitivity analysis.

\paragraph{When to use this.} We use the reduced-rank parameterization in both applications and in Simulations E and F, and the full-rank parameterization in Simulations A through D, where the data-generating coefficient matrix is sparse rather than low rank. The reason for preferring reduced rank on the FRED-QD panels is posterior geometry rather than accuracy. Under the full-rank parameterization at $p = 29$, most posterior draws of $B$ have spectral radius above one, on data transformed to stationarity, and the unconstrained $B$ absorbs contemporaneous co-movement that belongs in $\Omega$. Reduced rank removes both problems. Supplement~\ref{sec:app-supplement} gives the diagnostics.

\subsection{Joint Gibbs Sampler}

Estimation proceeds via a joint Gibbs sampler that updates $B$ and $\Omega$ within a single chain, so that uncertainty propagates between the two components rather than being fixed in a first stage. The sampler cycles through three blocks:

\begin{enumerate}
    \item \textbf{Sample $B \mid \bm{y}_{1:T}, \Omega$}: Given current $\Omega$, form GLS-transformed data and sample each row $\bm{b}_j$ under the DL prior using the Gaussian sampler of \citet{bhattacharya2016fast}, whose cost is linear in the number of predictors per equation, followed by GIG updates for the local scales $\eta_{jh}$ and auxiliary variables $\zeta_{jh}$.

    \item \textbf{Sample $\Omega \mid \bm{y}_{1:T}, B$}: Given current $B$, compute residuals $\bm{\epsilon}_t = \bm{y}_t - B\bm{y}_{t-1}$ and update $\Lambda$ and $\Delta$ using the latent variable representation of PrecFact, including GIG updates for the loading shrinkage parameters.

    \item \textbf{Update $\alpha$}: Update the DP concentration parameter by the auxiliary-variable scheme of \citet{escobar1995bayesian}, and the cluster allocations by the conjugate Gibbs scheme reviewed in \citet{neal2000markov}.
\end{enumerate}

When the reduced-rank parameterization of Section~\ref{subsec:reduced-rank} is used, Block 1 is replaced by two conditional draws: sampling $U \mid V$ row-by-row, then $V \mid U$ jointly, with the residuals $\bm\epsilon_t = \bm y_t - UV^\top \bm y_{t-1}$ passed to Blocks 2 and 3 unchanged.

\section{Conditional Dependence Among U.S.\ Macroeconomic Series}\label{sec:application}

We analyze the panel at two dimensions, each of independent economic interest. The $p = 29$ set of headline indicators ($T/p = 3.45$) summarizes the major sectors, with a well-conditioned sample covariance, a baseline in which unstructured and sequential estimators can be expected to perform well. The $p = 100$ set ($T/p = 1.0$), the regime VAPOR is designed for, resolves the same economy at finer granularity, distinguishing durable from nondurable production and short from long maturities, at the cost of a covariance matrix that is no longer invertible. In each we report the estimated dependence structure, its out-of-sample predictive content, and the adequacy of the fitted model. We present the $p = 29$ results first, in Section~\ref{subsec:p29}, and the higher-dimensional case in Section~\ref{subsec:p100}.

\subsection{Curated Panel: $p = 29$}\label{subsec:p29}

\subsubsection{Predictive evaluation}\label{subsec:pred}

We use the reduced-rank parameterization with $r = 3$ and $k = 10$ precision factors; Supplement~\ref{sec:app-supplement} gives the diagnostics behind that choice. We first ask whether the fitted precision matrix can reconstruct series it has not seen. At each test quarter, 8 of the 29 variables are held out at random and predicted from the rest by their conditional mean under the estimated precision matrix,
\begin{equation}
    \hat{\bm{y}}_h \mid \bm{y}_o = -\Omega_{hh}^{-1}\Omega_{ho}\bm{y}_o .
\end{equation}

Performance is measured by the mean squared prediction error (MSPE) over the 20 test quarters, averaged over 100 independent draws of the held-out sets because the result depends on which variables are hidden. Each stochastic sampler (PrecFact, VAR + PrecFact, Kastner, VAPOR) runs four chains of 100{,}000 iterations, with a burn-in of 20{,}000 and thinning of 80, and we report the mean and standard deviation of the per-chain MSPE. One of Kastner's chains (seed 103) stopped with a Cholesky decomposition error and was replaced by a chain with seed 105. We compare against PrecFact applied to the raw series; VAR + Glasso, with its penalty chosen by the Bayesian information criterion, and VAR + PrecFact, each applied to the residuals of a first-stage Bayesian VAR(1), fit with the \texttt{bayesianVARs} package \citep{gruber2026bayesianvars} under its default priors and summarized by its posterior mean coefficient matrix; Kastner's VAR with factor stochastic volatility (FSV); the sample covariance; and a random walk. Truncating the first-stage coefficient matrix to rank three, matching VAPOR, gives MSPEs of $0.560$ for VAR + Glasso and $0.634$ for VAR + PrecFact, and leaves each comparison with VAPOR unchanged (Supplement~\ref{sec:app-supplement}). For the dependence comparisons below, the partial correlations of each Bayesian method are the posterior means of the per-draw partial correlations, pooled over the same four chains.

\begin{table}[h!]
\centering
\caption{FRED-QD held-out predictive evaluation, $p = 29$, $T_{\mathrm{train}} = 100$ ($T/p = 3.45$). MSPE over 20 test quarters, Q1 2021--Q4 2025, averaged over 100 draws of the held-out sets; stochastic samplers additionally averaged over four chains, with the standard deviation across chains in parentheses. VAR + Glasso and VAR + PrecFact use a full-rank first stage. The lowest MSPE is shown in bold.}
\label{tab:fredqd_mspe}
\vspace{4pt}
\begin{tabular}{lc}
\toprule
Method & MSPE \\
\midrule
      PrecFact (no VAR) & 0.660 (0.026) \\
      VAR + Glasso      & \textbf{0.560} \\
      VAR + PrecFact    & 0.700 (0.054) \\
      Kastner (VAR+FSV) & 0.568 (0.005) \\
      Sample covariance & 0.611 \\
      Random walk       & 0.891 \\
      VAPOR (proposed)  & 0.573 (0.007) \\
\bottomrule
\end{tabular}
\end{table}

VAR + Glasso attains the lowest MSPE at this dimension ($0.560$), followed by Kastner ($0.568$) and VAPOR ($0.573$); neither difference from VAPOR is statistically significant ($p = 0.62$ and $0.83$). VAPOR is significantly more accurate than the two estimators that use the same precision factor model without estimating the dynamics jointly, by $0.127$ against VAR + PrecFact ($p < 0.001$) and $0.087$ against PrecFact applied to the raw series ($p = 0.005$). With condition number $5.6 \times 10^{3}$, the sample covariance leaves little for structured shrinkage to exploit at this dimension. The full paired comparison is in Supplement~\ref{sec:app-supplement}.

\FloatBarrier

\subsubsection{The estimated dependence graph}\label{subsubsec:graph}

\begin{figure}[h!]
\centering
\includegraphics[width=\textwidth,alt={Four side-by-side 29 by 29 partial correlation matrices, for VAPOR, Kastner, VAR plus graphical lasso, and the sample covariance. VAPOR and Kastner concentrate their mass in a few small blocks and leave the rest of the matrix near zero; the graphical lasso is denser, and the sample covariance is nearly saturated. Boxes on the VAPOR panel mark the 13 reported pairs, each numbered as in the caption.}]{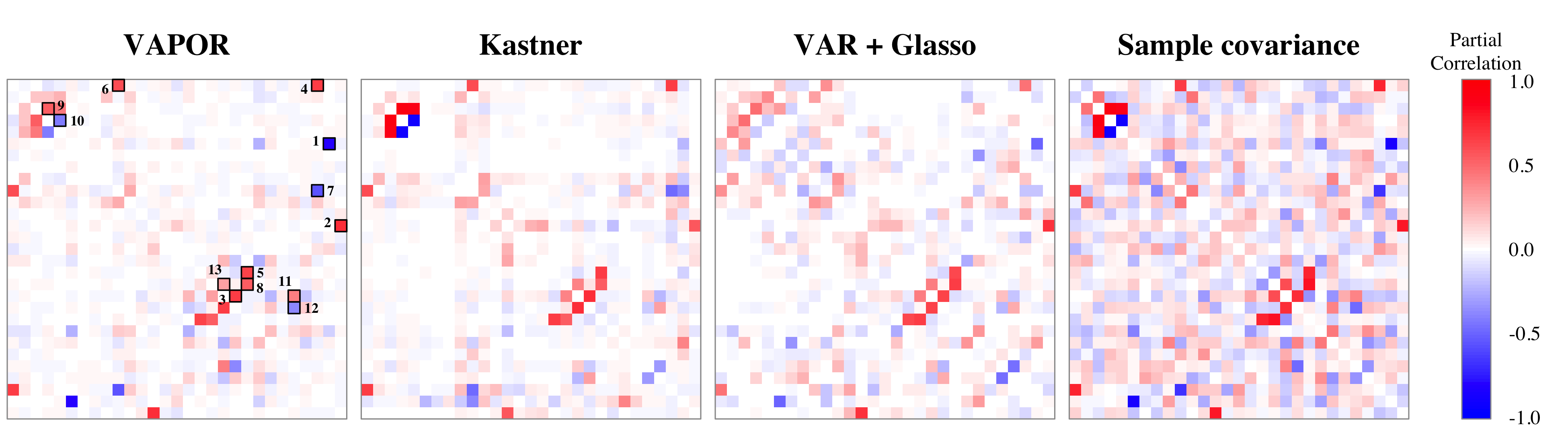}
\caption{Estimated partial correlation matrices at $p = 29$ (posterior mean for VAPOR and Kastner; no thresholding applied). Boxes on the VAPOR panel mark the 13 pairs whose central 95\% posterior credible interval excludes zero, numbered by decreasing magnitude: 1 payroll employment and unemployment rate ($-0.79$), 2 consumer price index and producer price index for finished goods ($0.72$), 3 ten-year Treasury and Baa corporate yields ($0.65$), 4 real GDP and real private investment ($0.65$), 5 federal funds and three-month Treasury rates ($0.63$), 6 real GDP and real consumption ($0.58$), 7 real consumption and real private investment ($-0.55$), 8 one-year and three-month Treasury rates ($0.53$), 9 industrial production of final products and of consumer goods ($0.53$), 10 industrial production of consumer goods and of business equipment ($-0.41$), 11 ten-year Treasury yield and S\&P 500 ($0.41$), 12 Baa yield and S\&P 500 ($-0.38$), 13 one-year and ten-year Treasury rates ($0.30$).}
\label{fig:heatmaps_p29}
\end{figure}

Figure~\ref{fig:heatmaps_p29} shows the estimated partial correlation matrices for VAPOR, Kastner, VAR + Glasso, and the sample covariance at $p = 29$, with no thresholding applied. VAPOR and Kastner, fit with the same number of factors, both concentrate their mass in a small number of blocks and leave the rest of the matrix near zero. VAR + Glasso is denser, and the sample covariance is essentially saturated, with a non-negligible partial correlation between nearly every pair of variables.

Under the credible-interval rule, 13 of the 406 pairs are reported. The strongest joins payroll employment and the unemployment rate ($-0.786$), followed by consumer prices and finished-goods producer prices ($0.721$). Real GDP is linked to real private investment ($0.649$) and real consumption ($0.585$), and consumption and investment share a negative edge ($-0.551$): because both are components of output, holding output fixed forces them to move in opposite directions. Two edges join industrial production indices, final products with consumer goods ($0.532$) and consumer goods with business equipment ($-0.412$), and the ten-year Treasury yield is linked to the Baa corporate bond yield ($0.653$). The ten-year and Baa yields are each also linked to the S\&P 500 ($0.409$ and $-0.378$), the only reported pairs at either dimension that join series in different FRED-QD groups. The remaining three edges join the federal funds rate to the three-month Treasury rate ($0.627$), the three-month rate to the one-year rate ($0.534$), and the one-year rate to the ten-year rate ($0.300$). Each joins adjacent maturities in this panel, which has no five-year yield, and the three non-adjacent pairs have posterior means of $0.09$ or less in absolute value, so the rates form a nearest-neighbor chain, as at $p = 100$. The competing estimators rank these 13 edges similarly at this dimension (Table~\ref{tab:edge_recovery_p29} in the Supplementary Material); at $p = 100$ they diverge.

\subsubsection{Model checking}\label{subsubsec:post_pred}\label{subsubsec:sensitivity}\label{subsec:checks}

We conducted posterior predictive checks of model fit, a sensitivity analysis to hyperparameter choice, and Markov chain Monte Carlo (MCMC) diagnostics, with full results in Supplement~\ref{sec:app-supplement}. Panels replicated one step ahead from the fitted model reproduce the scale of the series and the second eigenvalue of the correlation matrix, and slightly understate the first ($12.9$ observed, against a predictive interval of $9.3$ to $12.0$), which no Bayesian competitor reproduces either. The features not reproduced are heavier tails than Gaussian innovations allow and residual lag-one autocorrelation, both discussed in Section~\ref{subsubsec:post_pred_high}. The reported graph is stable: across factor counts, coefficient ranks, loading-prior shapes, and six random seeds, every configuration other than $k = 5$ is within $0.03$ of the reported MSPE, and 12 of the 13 edges are selected under every seed. All 13 edges have $\hat{R} < 1.05$ across the four chains \citep{gelman1992inference}, each of which takes about one minute on a laptop (Apple M2, 8\,GB).

\subsection{Expanded Panel: $p = 100$}\label{subsec:p100}

\subsubsection{Predictive evaluation}\label{subsec:pred_high}

We now analyze the 100-series panel. Retaining $p = 100$ FRED-QD series over the window of Section~\ref{sec:motivation} gives $T_{\mathrm{train}} = 100$ and $T/p = 1.0$: the number of series equals the number of training observations, the sample covariance is singular, and the $p^2$ free VAR coefficients cannot be estimated directly. We therefore use the reduced-rank parameterization of Section~\ref{subsec:reduced-rank} with $r = 3$ and $k = 25$ precision factors. Evaluation is identical to the $p = 29$ case, with 30 of the 100 series held out at each test quarter. Every stochastic sampler is run as at $p = 29$ except Kastner's: all eight 100{,}000-iteration chains we attempted (seeds 101--108) stopped with Cholesky decomposition errors, so we report four chains of 20{,}000 iterations with a burn-in of 10{,}000 and thinning of 10. Truncating the first-stage coefficient matrix to three singular directions gives MSPEs of $0.584$ for VAR + Glasso and $0.701$ for VAR + PrecFact, and again leaves each comparison with VAPOR unchanged.

\begin{table}[h!]
\centering
\caption{FRED-QD held-out predictive evaluation, $p = 100$, $T_{\mathrm{train}} = 100$ ($T/p = 1.0$). MSPE over 20 test quarters, Q1 2021--Q4 2025, averaged over 100 draws of the held-out sets; stochastic samplers additionally averaged over four chains, with the standard deviation across chains in parentheses; Kastner's chains run 20{,}000 iterations, the others 100{,}000. VAR + Glasso and VAR + PrecFact use a full-rank first stage. The lowest MSPE is shown in bold.}
\label{tab:fredqd_mspe_p100}
\vspace{4pt}
\begin{tabular}{lc}
\toprule
Method & MSPE \\
\midrule
      PrecFact (no VAR) & 0.666 (0.015) \\
      VAR + Glasso      & 0.592 \\
      VAR + PrecFact    & 0.677 (0.035) \\
      Kastner (VAR+FSV) & 218.790 (101.664) \\
      Sample covariance & 1.887 \\
      Random walk       & 1.113 \\
      VAPOR (proposed)  & \textbf{0.563 (0.004)} \\
\bottomrule
\end{tabular}
\end{table}

VAPOR attains the lowest MSPE, $0.563$. VAR + Glasso is next ($0.592$): VAPOR predicts more accurately in 13 of the 20 test quarters and in 81 of the 100 held-out sets, but the paired quarter-by-quarter difference of $0.029$ (standard error $0.018$) is not statistically significant ($p = 0.13$). VAPOR is ahead of VAR + PrecFact ($0.677$; paired advantage $0.114$, $p = 0.001$, 18 of 20 quarters) and of PrecFact on the raw series ($0.666$; paired advantage $0.103$, $p = 0.001$, 20 of 20 quarters), so within the precision factor family joint estimation improves on sequential estimation at both dimensions. The covariance-based approaches deteriorate sharply once $T = p$: the sample covariance is singular, and even with a ridge of $10^{-8}$ added so that it can be inverted, its MSPE is $1.887$; the random walk has a higher MSPE than every precision-based method ($1.113$), and Kastner's factor stochastic volatility model performs far worse ($218.8$, standard deviation $101.7$ across chains; paired advantage $218.2$, $p < 0.001$, 20 of 20 quarters). VAPOR is also the most stable across chains, with standard deviation $0.004$ against $0.015$ for PrecFact and $0.035$ for VAR + PrecFact. The full paired comparison is in Supplement~\ref{sec:app-supplement-p100}.

Kastner's method predicts as accurately as VAPOR at $p = 29$ but far worse than every other method at $p = 100$, in all four chains (fitting details in Supplement~\ref{sec:app-supplement-p100}). The method estimates the innovation covariance, while held-out prediction requires its inverse. At $T = p$ the inverted matrix has a condition number between $6.8 \times 10^{6}$ and $3.9 \times 10^{7}$ across the four chains, against $1.6 \times 10^{3}$ for the graphical lasso and $7.2 \times 10^{2}$ for VAPOR, so solving for the held-out series amplifies estimation error, a step that estimators of the precision matrix avoid.

\subsubsection{The estimated dependence graph}\label{subsubsec:graph_high}

\begin{figure}[h!]
\centering
\includegraphics[width=\textwidth,alt={Four side-by-side 100 by 100 partial correlation matrices, for VAPOR, Kastner, VAR plus graphical lasso, and the sample covariance. VAPOR shows compact blocks along the diagonal against a near-zero background; Kastner shows a few strong blocks against an almost empty background; the graphical lasso shows a block pattern similar to VAPOR's but more diffuse; and the sample covariance shows large values nearly everywhere with no discernible block structure. Boxes on the VAPOR panel mark the 26 reported pairs.}]{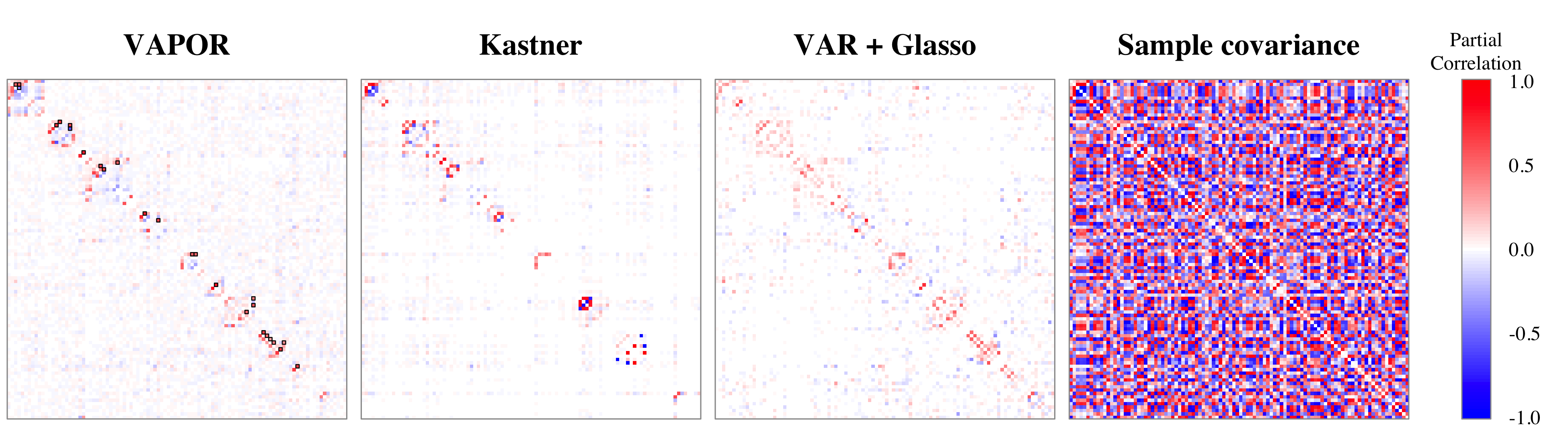}
\caption{Estimated partial correlation matrices at $p = 100$ (posterior mean for VAPOR and Kastner; no thresholding applied). Boxes on the VAPOR panel mark the 26 pairs whose central 95\% posterior credible interval excludes zero; they are listed in Table~\ref{tab:app-edges-p100} and shown as a graph in Figure~\ref{fig:network_p100}. VAPOR and Kastner are fit with the same number of factors. Color scale as in Figure~\ref{fig:heatmaps_p29}; the diagonal is omitted.}
\label{fig:heatmaps_p100}
\end{figure}

Figure~\ref{fig:heatmaps_p100} shows the estimated partial correlation matrices for VAPOR, Kastner, VAR + Glasso, and the sample covariance, again with no thresholding applied. VAPOR concentrates its mass into a set of blocks along the diagonal and leaves most of the matrix near zero. Kastner's estimate has a few strong blocks against an almost empty background, VAR + Glasso recovers a block structure similar to VAPOR's but more diffusely, and the sample covariance is entirely saturated, with large partial correlations between nearly every pair of variables and no graphical interpretation.

Table~\ref{tab:edge_counts_p100} quantifies this. Kastner's estimate has the fewest partial correlations above $0.05$, 228 of 4{,}950 pairs, against 409 for VAPOR, 476 for VAR + Glasso, and 4{,}712 for the sample covariance. The two have the same count above $0.30$, and above $0.40$ the order reverses: Kastner retains 30 partial correlations above $0.50$ against VAPOR's 19. The graphical lasso has more weak entries than VAPOR and fewer strong ones: it has fewer than VAPOR at the three largest thresholds, with 10 above $0.50$. Kastner's sparsity does not carry over to the other comparisons: it has the highest MSPE of the model-based estimators (Table~\ref{tab:fredqd_mspe_p100}), and it ranks below 750th three of the four adjacent-maturity edges that VAPOR and VAR + Glasso both place in their top 40 (Table~\ref{tab:edge_recovery_p100}).

\begin{table}[h!]
\centering
\caption{Number of off-diagonal partial correlations exceeding each magnitude, $p = 100$ (out of 4{,}950 possible pairs). VAPOR and Kastner are fit with the same number of factors.}
\label{tab:edge_counts_p100}
\begin{tabular}{lcccc}
\toprule
Threshold $|\rho_{ij}|$ & VAPOR & Kastner & VAR + Glasso & Sample covariance \\
\midrule
        $> 0.05$ & 409 & 228 & 476 & 4712 \\
        $> 0.10$ & 148 &  92 & 196 & 4474 \\
        $> 0.20$ &  64 &  55 &  65 & 4001 \\
        $> 0.30$ &  41 &  41 &  31 & 3490 \\
        $> 0.40$ &  27 &  35 &  20 & 2938 \\
        $> 0.50$ &  19 &  30 &  10 & 2400 \\
\bottomrule
\end{tabular}
\end{table}

\begin{figure}[h!]
\centering
\includegraphics[width=0.85\textwidth,alt={The 26 reported dependencies drawn as a graph of 40 labelled nodes shaded by FRED-QD group. Every component lies within a single group. The interest rates form a five-node chain, with the thirty-year mortgage rate attached to the five-year Treasury yield. Each edge is labelled with its posterior mean partial correlation; blue edges are positive and red edges negative.}]{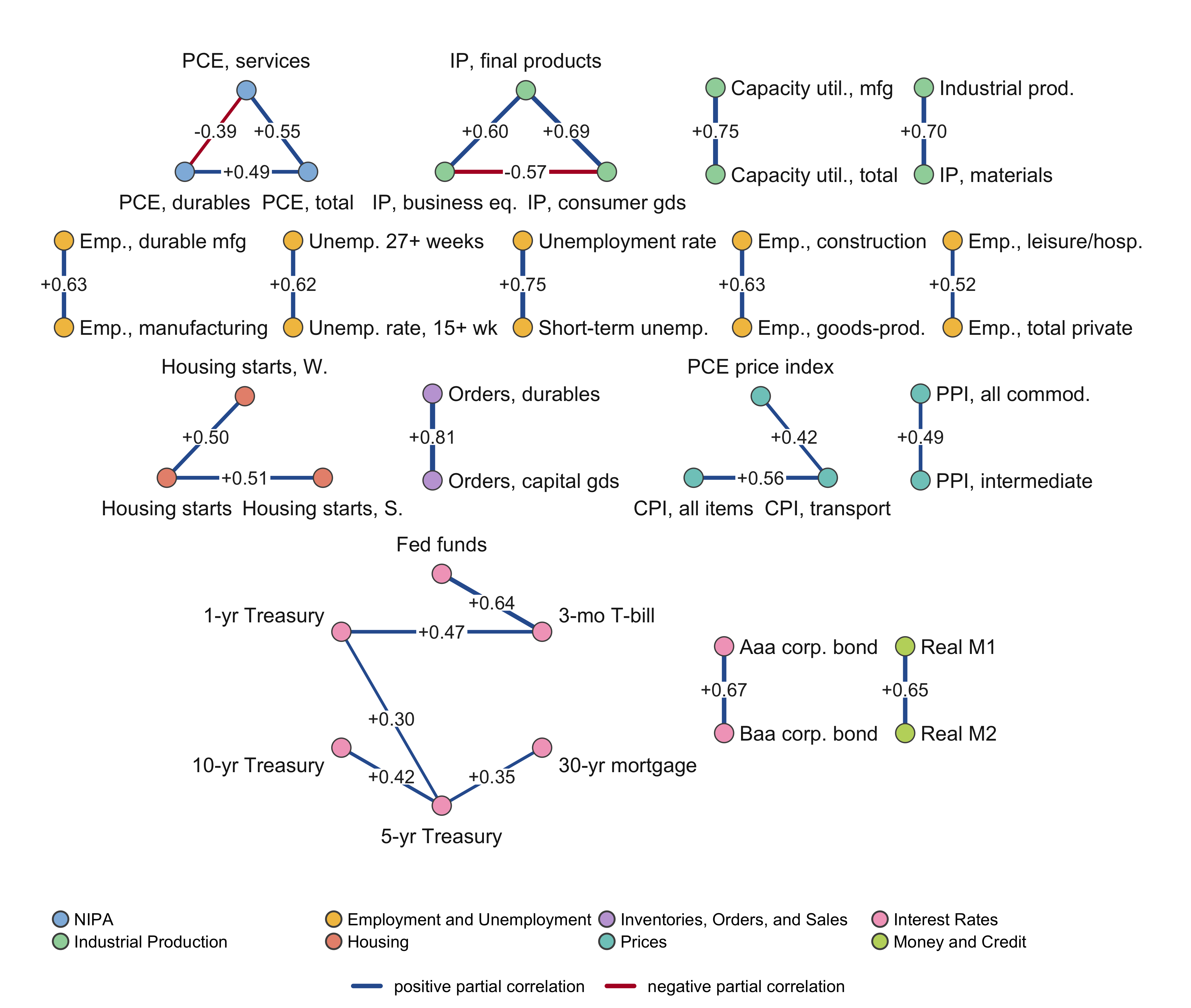}
\caption{The reported conditional dependence graph at $p = 100$, with 26 edges. Nodes are shaded by FRED-QD group as given in the legend; IP denotes industrial production, PCE personal consumption expenditures, CPI the consumer price index, and PPI the producer price index. Each edge is labeled with its posterior mean partial correlation; blue lines denote positive partial correlations, red lines negative.}
\label{fig:network_p100}
\end{figure}

The same credible-interval rule yields 26 edges, listed in Table~\ref{tab:app-edges-p100} and shown in Figure~\ref{fig:network_p100} with nodes shaded by FRED-QD group. The real-activity edges connect new orders for durable goods and for capital goods ($0.807$), total and manufacturing capacity utilization ($0.750$), industrial production and its materials index ($0.699$), production of final products with its consumer-goods and business-equipment components ($0.691$ and $0.605$), and consumer goods with business equipment ($-0.573$). The labor edges connect the unemployment rate with the short-term unemployment rate ($0.751$), two measures of long-term unemployment ($0.616$), and three employment aggregates with one of their components each: manufacturing with durable-goods manufacturing ($0.632$), goods-producing industries with construction ($0.630$), and total private employment with leisure and hospitality ($0.521$).

The price edges connect the consumer price index with its transportation component ($0.558$), the consumption price index with the same component ($0.419$), and the producer price index with intermediate materials prices ($0.486$). The consumption edges connect total real consumption with its services and durables components ($0.550$ and $0.492$), and durables with services ($-0.388$). Two housing edges connect total housing starts with starts in the South and in the West ($0.512$ and $0.504$), and a money edge connects the real M1 and M2 money stocks ($0.648$). The interest-rate edges connect adjacent maturities from the federal funds rate through the three-month, one-year, five-year and ten-year Treasury rates ($0.639$, $0.467$, $0.304$, and $0.422$), the Aaa and Baa corporate bond yields ($0.667$), and the five-year Treasury rate with the thirty-year mortgage rate ($0.355$). Each survives conditioning on the remaining 98 series.

The recovered graph answers both questions posed in Section~\ref{sec:intro}. On the term structure, marginal correlation among the five interest-rate series decays with the distance between maturities but remains substantial throughout, ranging from $0.28$ to $0.95$ with a median of $0.68$; Figure~\ref{fig:termstructure_p100} contrasts this with the conditional graph. Conditionally, the graph is a chain: all four adjacent maturity pairs are selected, and none of the six non-adjacent pairs is, the largest having a posterior mean partial correlation of $0.115$. The federal funds rate and the one-year Treasury rate correlate at $0.83$ and the one- and ten-year rates at $0.60$, yet neither pair survives conditioning on the intervening maturities. Each maturity is conditionally independent of the rest of the curve given its immediate neighbors, the nearest-neighbor structure implied by a smooth yield curve. VAR + Glasso ranks the four adjacent pairs among its 33 largest partial correlations. On connections between sectors, all 26 selected edges join two series in the same FRED-QD group, including the link from the five-year Treasury yield to the thirty-year mortgage rate, which lies within the interest-rate group. Conditional on the dynamics and on within-group dependence, the groups are close to conditionally independent of one another, so co-movement between them runs through intermediate series rather than directly. Both negative edges join components that compete within a group: consumer goods against business equipment in industrial production ($-0.573$), and durables against services in consumption ($-0.388$).

When all $4{,}950$ pairs are ranked by the magnitude of each method's estimated partial correlation, VAR + Glasso agrees most closely with VAPOR's 26 edges, with a median rank of 21.5, against 39.5 for Kastner and 54 for VAR + PrecFact (Table~\ref{tab:app-edge-recovery-p100} in the Supplementary Material). The disagreements are concentrated in the seven edges of Table~\ref{tab:edge_recovery_p100}. VAR + Glasso ranks the term-structure edges near the top, the three-month to one-year link 5th and the five- to ten-year link 8th, but sets both negative edges to exactly zero, so that they fall below all 1{,}369 of its nonzero entries. Kastner recovers the negative edges, ranking consumer goods with business equipment 33rd and durables with services 15th, but not the term structure, ranking the federal funds to three-month link 754th, the three-month to one-year link 760th, and the five- to ten-year link 1{,}753rd. VAR + PrecFact ranks the five-year Treasury to mortgage-rate link 2{,}599th, and Kastner ranks it 1{,}619th. Each competitor misses a different part of this structure: VAR + Glasso the negative edges, Kastner the term structure, and VAR + PrecFact the mortgage link. VAPOR is the only method that ranks all seven within its top 40. Across all 26 edges, the sample covariance does worse than every other method, with a median rank of 890.5 and only 9 of the 26 in its top 250.

\begin{table}[h!]
\centering
\footnotesize
\setlength{\tabcolsep}{4pt}
\caption{Ranks of seven edges reported by VAPOR at $p = 100$: the four links between adjacent interest-rate maturities, the five-year Treasury to thirty-year mortgage link, and the two negative edges, between consumer-goods and business-equipment industrial production and between real consumption of durables and of services. All $4{,}950$ pairs are ranked by the absolute value of each method's estimated partial correlation (rank 1 is the strongest). A dash marks a pair the graphical lasso sets to exactly zero.}
\label{tab:edge_recovery_p100}
\begin{tabular}{lrrrrrr}
\toprule
 & VAPOR & \multicolumn{5}{c}{Rank among $4{,}950$ pairs} \\
\cmidrule(l){3-7}
Edge & \shortstack{partial\\cor.} & VAPOR & \shortstack{VAR +\\Glasso} & \shortstack{VAR +\\PrecFact} & Kastner & \shortstack{Sample\\cov.} \\
\midrule
Fed funds -- 3-month bill & $0.64$ & 8 & 10 & 56 & 754 & 132 \\
3-month bill -- 1-year & $0.47$ & 23 & 5 & 62 & 760 & 3{,}660 \\
1-year -- 5-year & $0.30$ & 39 & 33 & 112 & 97 & 1{,}293 \\
5-year -- 10-year & $0.42$ & 25 & 8 & 7 & 1{,}753 & 18 \\
5-year -- 30-year mortgage & $0.35$ & 32 & 62 & 2{,}599 & 1{,}619 & 2{,}275 \\
\midrule
Consumer goods -- business equip. & $-0.57$ & 13 & -- & 6 & 33 & 867 \\
Durables -- services & $-0.39$ & 28 & -- & 128 & 15 & 8 \\
\bottomrule
\end{tabular}
\end{table}

Of VAPOR's 100 largest partial correlations in magnitude, 93 join two series in the same FRED-QD group, against 82 for PrecFact and for VAR + PrecFact, 77 for Kastner, 73 for VAR + Glasso, and 31 for the sample covariance. The selected graphs differ more. All 26 of VAPOR's edges lie within a group, while the graphical lasso's own penalized estimate has 1{,}369 nonzero off-diagonal entries, 294 (21\%) of them within a group. Of the 25 largest VAR + Glasso partial correlations that VAPOR does not select, two miss the selection rule by less than $0.02$: employment in trade, transportation and utilities with retail trade employment, with a posterior mean of $0.569$ and a 95\% interval of $(-0.012, 0.855)$, and real GDP with real private investment, with $0.491$ and $(-0.015, 0.868)$.

Because a conditional independence graph is defined relative to the variables conditioned on, the $p = 29$ and $p = 100$ panels need not report the same edges. Twenty-seven of the 29 series also appear in the larger panel, and the other two are prices that enter no reported edge, so all 13 edges reported at $p = 29$ are candidates at $p = 100$. Four survive: the policy to short-rate link (8th of $4{,}950$ pairs), the three-month to one-year link (23rd), and the two industrial production links (5th and 13th). The output to investment link ranks 21st but is not selected, for the reason given in Section~\ref{subsubsec:post_pred_high}. Six fall to ranks between 113 and 412, remaining in the upper tenth without reaching the reporting threshold. For four of them, the larger panel supplies series that lie between the two endpoints. The link from the ten-year Treasury yield to the Baa corporate yield is displaced by the direct link between the Aaa and Baa yields, which is reported at $p = 100$ and unavailable at $p = 29$. The consumer to producer price link is displaced by the transportation and intermediate-materials price indices, each of which has a reported edge at $p = 100$. The two consumption links are displaced by the durables and services components of consumption itself, and the one- to ten-year link falls to 1{,}444th once the five-year yield lies between them.

The payroll to unemployment link changes most. It is the strongest dependence at $p = 29$ at $-0.786$, and at $p = 100$ it falls to 2{,}623rd with a partial correlation of $-0.012$. The larger panel contains 22 additional series from the Employment and Unemployment group, four of them unemployment-rate or duration measures, and the labor dependencies reported there run between close substitutes rather than across the two aggregates: the unemployment rate to its short-duration counterpart, and manufacturing employment to its durable-goods component. Conditional on those, payroll employment and the unemployment rate are close to independent: the aggregate relationship in the curated panel runs through the finer series.

\subsubsection{Model checking}\label{subsubsec:post_pred_high}\label{subsubsec:sensitivity_high}\label{subsec:checks_high}

The same checks at $p = 100$ (Supplement~\ref{sec:app-supplement-p100}) show a closer fit. Five of the nine posterior predictive summaries fall inside their central 95\% intervals, including the two largest eigenvalues of the correlation matrix, so the model reproduces the dominant factor structure, and the mean and ninetieth percentile of absolute correlations miss by less than $0.01$. Applied to the competing models, the same check places both eigenvalues inside the interval only for VAPOR among the Bayesian methods. Two features are not reproduced at either dimension. The series are heavier-tailed than Gaussian innovations allow (kurtosis $19.3$, against $13.8$ to $15.4$); the gap remains when the 2008--09 and COVID quarters are removed, and Kastner's model, whose innovation variances change over time, is the only method that reproduces it. The residuals also retain lag-one autocorrelation ($0.208$), because a rank-three coefficient matrix cannot give each series its own autoregressive coefficient. At $p = 29$, a full-rank coefficient matrix lowers the mean residual autocorrelation, computed at the posterior mean coefficients, from $0.18$ to $0.08$, but most of its posterior draws are then non-stationary (Section~\ref{subsec:reduced-rank}).

The reported graph is stable across tuning. Across factor counts, coefficient ranks, loading-prior shapes, and six random seeds, the held-out MSPE ranges from $0.560$ to $0.607$, all but two configurations remain below VAR + Glasso ($0.592$), and 23 of the 26 edges are selected under every seed. Every reported edge has $\hat{R} \le 1.06$ across the four chains. Partial correlations between real GDP and its expenditure components mix slowly, because GDP is close to an exact linear combination of them; none of these pairs is reported, because each credible interval reaches zero. Each chain takes 14 minutes on a laptop (Apple M2, 8\,GB).

\section{Simulations}\label{sec:simulations}

We evaluate VAPOR in six simulation studies, reported in full in Supplement~\ref{sec:supplement_sims}. The competitors are those of Section~\ref{sec:application}, with the graphical lasso \citep{friedman2008sparse} applied to the raw series in place of the sample covariance and random walk. The graphical lasso penalty is tuned by minimizing Frobenius distance to the true precision matrix, which favors it. We report edge-selection sensitivity and specificity and the Frobenius distance between the estimated and true correlation-scaled precision matrices over 25 replicates per scenario, with paired per-replicate differences against VAPOR.

When data are generated from the proposed model (Simulation~A), VAPOR has the lowest Frobenius distance in every scenario. Its margin over methods that ignore the dynamics grows with the spectral radius of $B$, reaching 2.898 against PrecFact on the raw series at $\rho(B) = 0.99$. Its margin over sequential VAR + PrecFact is smaller, between 0.023 and 0.105, and not significant at $T = 50$. When the dynamics are approximately low rank and $T/p$ is between 1 and 2 (Simulation~E), reduced-rank VAPOR is closer to the truth than Kastner's method and the graphical lasso in every replicate, and at $p = T = 150$ its Frobenius error is roughly half of theirs. When $T < p$ (Simulation~F), VAPOR is more accurate for $p \geq 100$, and at $p = 50$ it is indistinguishable from the graphical lasso. In both simulations the graphical lasso returns an exactly diagonal estimate in 11 to 21 of the 25 replicates at each setting, and its error is within 0.10 of that of the identity matrix.

VAPOR does not perform best in three settings. When the signal strength varies (Simulation~D), sequential VAR + PrecFact is more accurate at the strongest signal levels, and VAR + Glasso at the weakest. When the data come from a factor stochastic volatility model (Simulation~C), which has no single true $\Omega$ and is compared on held-out MSPE, Kastner's correctly specified method predicts best unless the coefficient matrix is dense, where the sample covariance does. When $\Omega$ has no factor structure (Simulation~B), VAR + Glasso recovers it more accurately than VAPOR at every sample size, as does Kastner except at $T = 50$, where the two cannot be distinguished. VAPOR's gains come from the low-rank-plus-diagonal factorization of the precision matrix, and its performance declines when the true precision is not close to this form.

\section{Discussion}\label{sec:discussion}

VAPOR estimates vector autoregressive dynamics jointly with a factor-structured innovation precision matrix, so the conditional independence graph is inferred directly from the fitted precision rather than by inverting an estimated covariance. On the FRED-QD panel at $p = 100$, where the number of series equals the number of training quarters, this yields a sparse graph organized by FRED-QD group and the lowest held-out prediction error among the estimators compared. At $p = 29$, where the sample covariance is well conditioned, VAPOR's prediction error cannot be distinguished from that of VAR + Glasso or Kastner's model.

All 26 dependencies reported at $p = 100$ fall within a single FRED-QD group, so once temporal dynamics are removed the innovation precision is close to block diagonal and the disturbances to different sectors are close to independent. At $p = 29$, where each group is represented by fewer series, two of the 13 reported edges cross groups, joining the ten-year Treasury yield and the Baa corporate bond yield to the S\&P 500. Within the interest-rate block, each maturity is conditionally independent of the rest of the curve given its immediate neighbors at both dimensions, although marginally every maturity is correlated with every other.

One interesting future extension of VAPOR would be to let $\Omega_t = \Lambda V_t \Lambda^\top + \Delta$, with $V_t$ a diagonal matrix of factor variances that evolve through stochastic volatility, so that the precision changes over time through only the $k$ factor variances. Positive factor variances rescale the terms of $\Lambda V_t \Lambda^\top$ but do not change its zero pattern, so the support of $\Omega_t$ does not depend on the volatility path. Covariance factor stochastic volatility lacks this property: its implied precision graph changes with the volatility state. The precision-side factors are not independent of the idiosyncratic component \citep{chandra2026precfact}, so the equation-by-equation sampling that makes covariance factor stochastic volatility cheap does not carry over.

There are a variety of other potentially useful extensions. First, it would be interesting to obtain identifiable and interpretable precision factors that could be related to economic aggregates. Second, the reduced-rank factors $U$ and $V$ could be given a hierarchical or time-varying structure, so that the dominant modes of the VAR can shift across regimes. Third, a heavier-tailed or copula-based likelihood, combined with the LRD precision, would accommodate the heavy tails and skewness common in macroeconomic and financial data.

\section*{Reproducibility}

\if0\blind
VAPOR is implemented in the \texttt{R} package \texttt{VAPOR}, available at \url{https://github.com/peterdunson/vapor}, which provides the joint Gibbs sampler in \texttt{C++} through \texttt{RcppArmadillo}. Code reproducing every table and figure in this paper and its supplements is available in the same repository.
\fi
\if1\blind
VAPOR is implemented in an \texttt{R} package that provides the joint Gibbs sampler in \texttt{C++} through \texttt{RcppArmadillo}. Code reproducing every table and figure in this paper and its supplements is available in the same repository, which will be made public on acceptance.
\fi 
The application uses the FRED-QD database \citep{mccracken2020fredqd}, a public monthly-updated panel; we use the May 2026 vintage, and results may differ slightly under later vintages because historical values are revised.

\if0\blind
\section*{Acknowledgments}

I thank David Dunson for helpful comments on an earlier version of this paper, and Ciprian Crainiceanu for his mentorship and support. All errors are my own. The author used Claude (Anthropic; Claude Opus 4 and 5 and Claude Sonnet 4 and 5 model families) to assist with code development and debugging, checking computations, and editing the text for clarity. The author reviewed and verified all code, results, and text, and takes full responsibility for the content.
\fi

\section*{Supplementary Materials}
\noindent\textbf{Supplement S1} provides details for the $p=29$ panel: MCMC convergence diagnostics and trace plots, partial correlation magnitudes and edge recovery by the competing methods, posterior predictive checks, paired predictive comparisons, sensitivity analyses, and the choice of coefficient rank. \textbf{Supplement S2} provides the same for $p=100$, adding the complete list of reported edges with per-edge convergence diagnostics and a summary of both panels. \textbf{Supplement S3} reports Simulations~A through~F in full: data from the proposed model, a sparse precision matrix with no factor structure, a competing model, varying signal strength, approximately low-rank dynamics under limited sample size, and the $T<p$ regime.

\newpage

\begin{singlespace}

\bibliographystyle{apalike}
\bibliography{refs}

\end{singlespace}

\newpage
\appendix

\renewcommand{\thetable}{S\arabic{table}}
\renewcommand{\thefigure}{S\arabic{figure}}
\setcounter{table}{0}
\setcounter{figure}{0}

\renewcommand{\thesection}{S\arabic{section}}
\setcounter{section}{0}

\section{Application: Supplementary Diagnostics}\label{sec:app-supplement}

This appendix collects the full diagnostic output for the $p = 29$ application discussed in Section~\ref{subsec:p29}: MCMC convergence, partial correlation magnitudes and edge recovery, posterior predictive checks, paired predictive comparisons, sensitivity analyses, and the diagnostics behind the choice of coefficient rank.

\subsection{MCMC Convergence}

\begin{table}[h!]
\centering
\caption{MCMC convergence diagnostics, $p = 29$, from four independent chains of 100{,}000 iterations (burn-in 20{,}000, thinning 80). ESS is the bulk effective sample size and $\hat R$ the rank-normalized statistic. $\rho(B)$ is the spectral radius of the coefficient matrix. CPI and PPI denote the consumer and producer price indices, PCE personal consumption expenditures, and IP industrial production.}
\label{tab:app-convergence}
\begin{tabular}{lcc}
\toprule
Quantity & ESS & $\hat R$ \\
\midrule
Log-likelihood                            & 1061 & 1.008 \\
Payroll employment -- Unemployment rate   &  858 & 1.004 \\
CPI, all items -- PPI, finished goods     &  132 & 1.031 \\
10-yr Treasury -- Baa corp.\ bond         & 2897 & 1.000 \\
Real GDP -- Private investment            & 1089 & 1.003 \\
Fed funds -- 3-mo T-bill                  & 1416 & 1.002 \\
Real GDP -- PCE, total                    &  881 & 1.005 \\
PCE, total -- Private investment          & 1039 & 1.005 \\
1-yr Treasury -- 3-mo T-bill              & 1030 & 1.003 \\
IP, final products -- IP, consumer goods  & 1436 & 1.006 \\
IP, consumer goods -- IP, business eq.    & 1821 & 1.005 \\
10-yr Treasury -- S\&P 500                & 2523 & 1.002 \\
Baa corp.\ bond -- S\&P 500               & 2377 & 1.001 \\
1-yr Treasury -- 10-yr Treasury           & 2601 & 1.000 \\
$\rho(B)$                                 & 3672 & 0.999 \\
\bottomrule
\end{tabular}
\end{table}

\begin{figure}[h!]
\centering
\includegraphics[width=\textwidth,height=0.8\textheight,keepaspectratio,alt={Eight trace plots from four chains at p equals 29: the log-likelihood, the spectral radius of the coefficient matrix, and six reported partial correlations. The four chains overlap in every panel; for the consumer price to producer price pair, one chain lies below the others for stretches of the run. Each panel heading gives its R-hat and effective sample size.}]{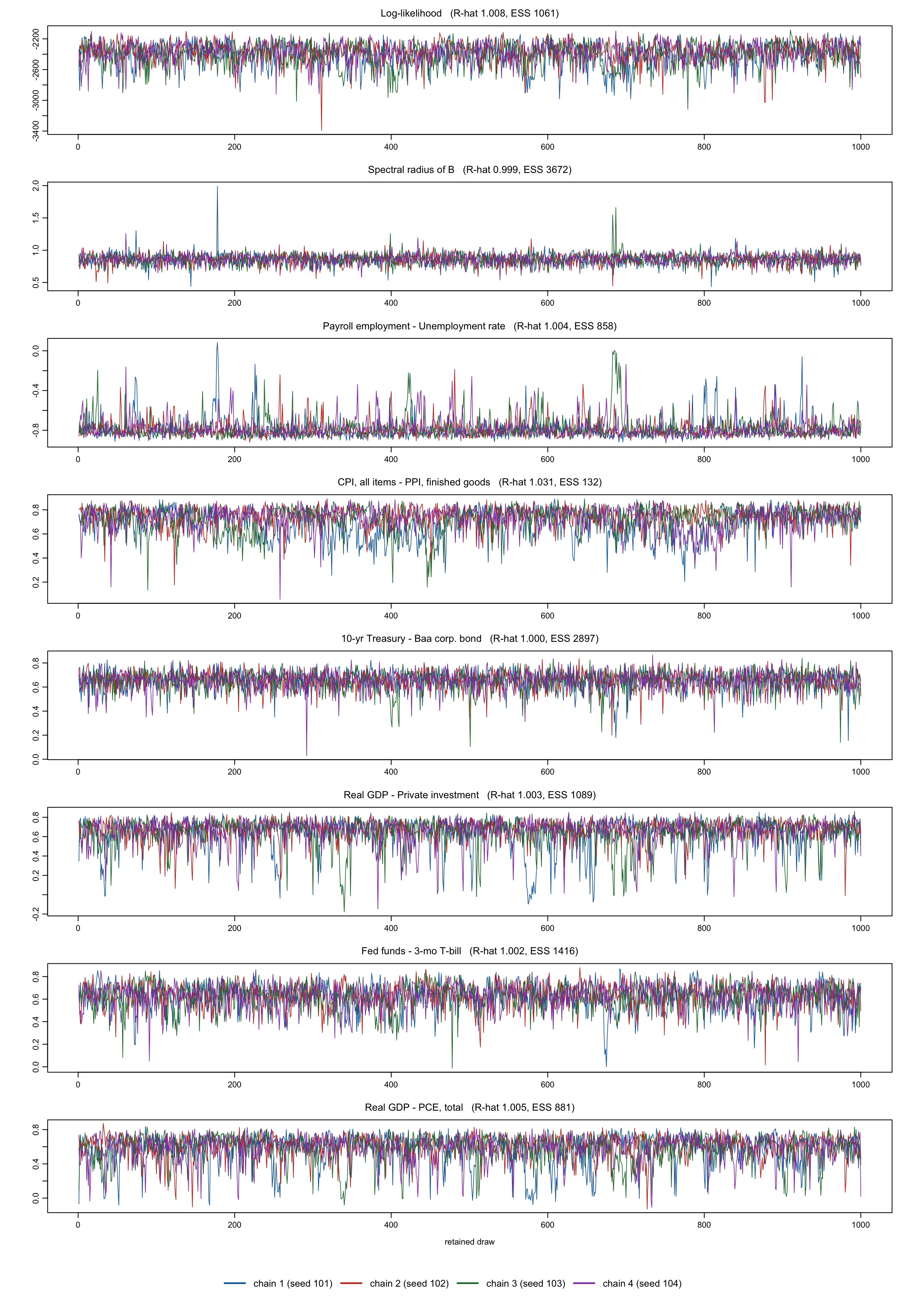}
\caption{Trace plots across four chains at $p = 29$: the log-likelihood, the spectral radius of the coefficient matrix, and six reported partial correlations, including the slowest-mixing edge, between the consumer price index (CPI) and the producer price index (PPI) for finished goods. Each panel gives its $\hat{R}$ and effective sample size.}
\label{fig:app-trace}
\end{figure}

Table~\ref{tab:app-convergence} reports bulk effective sample sizes and rank-normalized $\hat{R}$ statistics from four independent chains of 100{,}000 iterations, for the log-likelihood, each of the 13 reported partial correlations, and the spectral radius of the coefficient matrix. Figure~\ref{fig:app-trace} shows trace plots for the log-likelihood, the spectral radius, and six reported edges. All 406 partial correlations have $\hat{R} \le 1.05$, with median $\hat{R}$ of $1.0005$ and median effective sample size $3{,}684$, and 12 of the 13 reported edges are selected by each of the four chains when the credible-interval rule is applied to the chains separately. All 13 edges satisfy $\hat{R} < 1.05$, and 12 have $\hat{R} \le 1.006$. The slowest is the edge between the consumer price index and the producer price index for finished goods ($\hat{R} = 1.031$, effective sample size 132), for which one chain spends stretches of the run at lower values than the other three. The spectral radius is the best-mixing quantity monitored, and its posterior places probability $0.027$ above one. The sampler bounds each idiosyncratic precision below at $10^{-8}$ as a numerical safeguard for the Cholesky factorization in the update of $\Lambda$; no precision reaches this bound in any chain.

\FloatBarrier

\subsection{Partial Correlation Magnitudes and Edge Recovery}

Table~\ref{tab:edge_counts} counts the partial correlations exceeding a range of magnitudes at $p = 29$. Above 0.1, VAPOR has 42 of 406, against 49 for Kastner, 75 for VAR + Glasso, and 200 for the sample covariance, and VAPOR has the lowest count at every threshold up to 0.2. Table~\ref{tab:edge_recovery_p29} ranks all 406 pairs by the magnitude of each method's estimated partial correlation and records where VAPOR's 13 edges fall. Each competing estimator places the 13 edges at a median rank of 7 or 8, and VAR + Glasso, Kastner and the sample covariance place all 13 in their top 52. VAR + PrecFact ranks the ten-year Treasury to Baa link and the two S\&P 500 edges between 231st and 360th. VAPOR's own ordering places its 13 edges at ranks 1 to 14; the pair at rank 10, industrial production of final products and of business equipment, has a credible interval that includes zero.

\begin{table}[h!]
\centering
\caption{Number of off-diagonal partial correlations exceeding each magnitude, $p = 29$ (out of 406 possible pairs). VAPOR and Kastner are fit with the same number of factors.}
\label{tab:edge_counts}
\begin{tabular}{lcccc}
\toprule
Threshold $|\rho_{ij}|$ & VAPOR & Kastner & VAR + Glasso & Sample covariance \\
\midrule
        $> 0.05$ & 78 &  89 & 134 & 297 \\
        $> 0.10$ & 42 &  49 &  75 & 200 \\
        $> 0.20$ & 20 &  23 &  33 &  64 \\
        $> 0.30$ & 14 &  14 &  17 &  34 \\
        $> 0.40$ & 12 &  10 &   9 &  17 \\
        $> 0.50$ &  9 &   9 &   5 &  12 \\
\bottomrule
\end{tabular}
\end{table}

\begin{table}[h!]
\centering
\caption{Where VAPOR's 13 reported edges fall when all 406 variable pairs are ranked by the magnitude of each method's estimated partial correlation, $p = 29$.}
\label{tab:edge_recovery_p29}
\begin{tabular}{lcccc}
\toprule
Method & Median rank & In top 13 & In top 52 & In top 130 \\
\midrule
    VAPOR             & 7 & 12 & 13 & 13 \\
    VAR + Glasso      & 8 &  9 & 13 & 13 \\
    VAR + PrecFact    & 7 & 10 & 10 & 10 \\
    Kastner (VAR+FSV) & 8 & 10 & 13 & 13 \\
    Sample covariance & 8 & 12 & 13 & 13 \\
\bottomrule
\end{tabular}
\end{table}

\FloatBarrier

\subsection{Posterior Predictive Checks}

\begin{figure}[h!]
\centering
\includegraphics[width=0.85\textwidth,alt={Nine histograms, one per summary, each showing the posterior predictive distribution with a vertical line at the observed value. The observed value falls within the bulk of the histogram for the second eigenvalue and the mean standard deviation. It lies at or beyond the upper edge for the other seven, and far beyond it for kurtosis and residual lag-one autocorrelation.}]{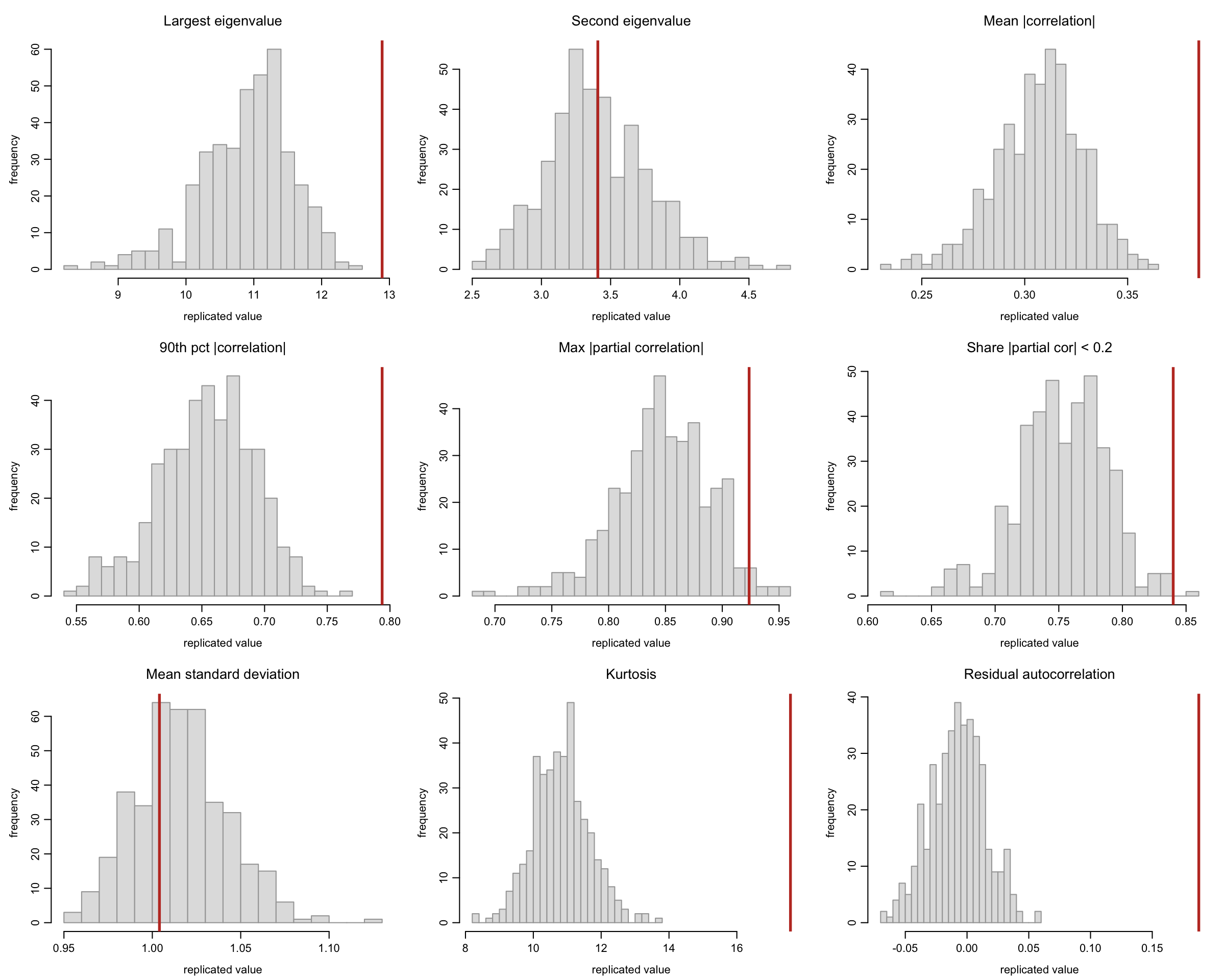}
\caption{Posterior predictive checks, $p = 29$. Histograms show the posterior predictive distribution of each of the nine summaries under the fitted model, from 400 replicated panels; vertical lines mark the observed values.}
\label{fig:app-ppc}
\end{figure}

Table~\ref{tab:app-ppc} and Figure~\ref{fig:app-ppc} compare nine summaries of the observed panel to their posterior predictive distributions under the fitted model. Replicates are generated one step ahead, drawing $\bm{y}_t^{\mathrm{rep}}$ from $N(B_s\bm{y}_{t-1}, \Omega_s^{-1})$ at each posterior draw $s$ and conditioning on the observed $\bm{y}_{t-1}$, so the check targets the innovation structure given the observed history. The summaries cover the factor structure (the two largest eigenvalues of the correlation matrix), the bulk and the tail of the dependence (mean and ninetieth percentile of absolute correlations, largest absolute partial correlation, proportion of absolute partial correlations below $0.2$), the marginals (mean standard deviation, kurtosis), and the dynamics (mean lag-one autocorrelation of the residuals $\bm{y}_t - B_s\bm{y}_{t-1}$, which the model implies should be zero).

The fitted model reproduces the second eigenvalue ($3.41$ against $2.74$ to $4.17$) and the mean standard deviation. It understates the largest eigenvalue ($12.89$ against $9.32$ to $12.05$) and the mean and ninetieth percentile of absolute correlations ($0.385$ against $0.260$ to $0.348$, and $0.794$ against $0.568$ to $0.722$), so the dominant common factor is stronger in the data than in the fitted model. The largest absolute partial correlation ($0.924$ against $0.752$ to $0.923$) and the share of absolute partial correlations below $0.2$ ($0.840$ against $0.672$ to $0.823$) lie just above their intervals. Kurtosis is understated by more ($17.6$ against $9.3$ to $12.6$), which reflects the Gaussian likelihood: the transformed series have heavier tails than Gaussian innovations admit. The residuals also retain lag-one autocorrelation of $0.188$ against a predictive range of $-0.051$ to $0.035$, because a rank-three coefficient matrix cannot give each series its own autoregressive coefficient (Section~\ref{subsubsec:post_pred_high}). The two COVID quarters 2020Q2 and 2020Q3 account for $59\%$ of the total squared standardized residual norm at $p = 29$ and $75\%$ at $p = 100$, but removing them, the 2008--09 quarters, or both from the check does not bring the summaries inside their intervals.

\begin{table}[h!]
\centering
\caption{Posterior predictive checks, $p = 29$. Observed values, posterior predictive means, and central 95\% predictive intervals for nine summaries.}
\label{tab:app-ppc}
\begin{tabular}{lcccc}
\toprule
Summary & Observed & Replicate mean & \multicolumn{2}{c}{95\% interval} \\
\midrule
Largest eigenvalue        & 12.892 & 10.902 &  9.320 & 12.046 \\
Second eigenvalue         &  3.408 &  3.407 &  2.742 &  4.166 \\
Mean $|$correlation$|$    &  0.385 &  0.307 &  0.260 &  0.348 \\
90th pct.\ $|$correlation$|$ & 0.794 & 0.654 & 0.568 & 0.722 \\
Largest $|$partial cor.$|$ & 0.924 &  0.847 &  0.752 &  0.923 \\
Prop.\ $|$partial cor.$| < 0.2$ & 0.840 & 0.753 & 0.672 & 0.823 \\
Mean standard deviation   &  1.004 &  1.016 &  0.967 &  1.070 \\
Kurtosis                  & 17.581 & 10.825 &  9.284 & 12.572 \\
Residual lag-1 autocor.   &  0.188 & $-0.008$ & $-0.051$ & 0.035 \\
\bottomrule
\end{tabular}
\end{table}

\FloatBarrier

\subsection{Paired Predictive Comparison}

Because every method predicts the same held-out variables in the same quarters, predictive losses are paired and the difference in accuracy can be assessed directly rather than by comparing means alone. Per-quarter losses are averaged across the 100 draws of the held-out sets before the comparison. Table~\ref{tab:dm-p29} reports the mean per-quarter difference in squared prediction error between each competitor and VAPOR, its standard error, and a Diebold--Mariano statistic \citep{diebold1995comparing} with the small-sample correction of \citet{harvey1997testing}. A positive difference favors VAPOR. VAPOR is significantly more accurate than the two PrecFact-based estimators, with either first stage, and the random walk. It cannot be distinguished from VAR + Glasso with either first stage, from Kastner's model, or from the sample covariance; VAR + Glasso and Kastner have slightly lower MSPE than VAPOR at this dimension, and VAPOR predicts more accurately than each of them in 9 of the 20 test quarters and in 46 and 42 of the 100 held-out sets. Rows marked ``rank 3'' truncate the first-stage coefficient matrix to its three leading singular directions.

\begin{table}[h!]
\centering
\caption{Paired predictive comparison against VAPOR, $p = 29$, over 20 test quarters. Difference is the competitor's mean squared prediction error minus VAPOR's, so positive values favor VAPOR. VAPOR's mean MSPE is $0.573$.}
\label{tab:dm-p29}
\begin{tabular}{lccccc c}
\toprule
Method & MSPE & Difference & SE & DM & $p$ & Quarters won \\
\midrule
  VAR + Glasso             & 0.560 & $-0.013$ & 0.024 & $-0.51$ & 0.615    &  9 / 20 \\
  VAR + Glasso (rank 3)    & 0.560 & $-0.013$ & 0.034 & $-0.36$ & 0.722    & 10 / 20 \\
  Kastner (VAR+FSV)        & 0.568 & $-0.005$ & 0.024 & $-0.22$ & 0.832    &  9 / 20 \\
  Sample covariance        & 0.611 &  0.038 & 0.041 &  0.90 & 0.377    & 13 / 20 \\
  VAR + PrecFact           & 0.700 &  0.127 & 0.029 &  4.23 & $<0.001$ & 18 / 20 \\
  VAR + PrecFact (rank 3)  & 0.634 &  0.061 & 0.019 &  3.17 & 0.005    & 14 / 20 \\
  PrecFact (no VAR)        & 0.660 &  0.087 & 0.027 &  3.14 & 0.005    & 17 / 20 \\
  Random walk              & 0.891 &  0.318 & 0.079 &  3.93 & 0.001    & 16 / 20 \\
\bottomrule
\end{tabular}
\end{table}

\FloatBarrier

\subsection{Sensitivity Analysis}

Table~\ref{tab:app-sensitivity} reports held-out MSPE, the number of reported edges, and the posterior probability that the coefficient matrix is non-stationary, across factor counts, coefficient ranks, loading-prior shapes, and random seeds, each from a single chain of 100{,}000 iterations; Table~\ref{tab:app-length} reports the effect of chain length. Across six random seeds at the reported configuration the MSPE ranges from $0.563$ to $0.579$ and the edge count from 12 to 13. The $k = 5$ setting is the outlier on both MSPE ($0.629$) and edge count (6), and every other configuration falls within $0.03$ of the reported value of $0.573$. The posterior probability of a non-stationary coefficient matrix is at most $0.052$ throughout. Over the eight distinct configurations, four of the 13 reported edges are selected in all eight, five more in seven, and the edge between industrial production of final products and of consumer goods, the least stable, in four. Across chain lengths from 7{,}500 to 100{,}000 iterations, the pooled graph has 12 or 13 edges and the MSPE ranges from $0.573$ to $0.582$, while the largest $\hat{R}$ among pairs selected by any single chain falls from $1.094$ to $1.031$.

\begin{table}[h!]
\centering
\caption{Sensitivity of the $p = 29$ results to tuning, each from a single chain of 100{,}000 iterations (seed 101 unless stated); the $k=10$, $r=3$ and $a_L=0.5$ rows are the same fit. Edges are those whose central 95\% posterior credible interval excludes zero. $\Pr(\rho(B) > 1)$ is the posterior probability that the coefficient matrix is non-stationary. The Dirichlet-Laplace shape on the VAR coefficients is fixed by dimension and not varied.}
\label{tab:app-sensitivity}
\begin{tabular}{lccc}
\toprule
Setting & MSPE & Edges & $\Pr(\rho(B) > 1)$ \\
\midrule
$k=5$         & 0.629 &  6 & 0.030 \\
$k=8$         & 0.576 & 13 & 0.028 \\
$k=10$        & 0.563 & 12 & 0.020 \\
$k=15$        & 0.546 & 11 & 0.045 \\
$r=2$         & 0.578 & 13 & 0.043 \\
$r=3$         & 0.563 & 12 & 0.020 \\
$r=4$         & 0.580 & 12 & 0.052 \\
$a_L=0.3$     & 0.564 & 10 & 0.038 \\
$a_L=0.5$     & 0.563 & 12 & 0.020 \\
$a_L=1.0$     & 0.578 & 15 & 0.025 \\
seed 102      & 0.579 & 13 & 0.030 \\
seed 103      & 0.575 & 12 & 0.038 \\
seed 104      & 0.575 & 13 & 0.022 \\
seed 105      & 0.573 & 12 & 0.025 \\
seed 106      & 0.577 & 13 & 0.050 \\
\bottomrule
\end{tabular}
\end{table}

\begin{table}[h!]
\centering
\caption{Effect of chain length, $p = 29$, four chains at each length. Edges are those selected by the pooled 95\% credible-interval rule; $\hat R$ and ESS are computed across the four chains for every pair selected by at least one chain; MSPE is the mean across chains with the standard deviation in parentheses.}
\label{tab:app-length}
\begin{tabular}{lcccccc}
\toprule
Iterations & Burn-in / thin & Edges & Max $\hat R$ & Median $\hat R$ & Min ESS & MSPE \\
\midrule
7{,}500   & 2{,}500 / 5   & 13 & 1.094 & 1.031 &  28 & 0.575 (0.012) \\
20{,}000  & 10{,}000 / 10 & 12 & 1.084 & 1.010 &  61 & 0.582 (0.001) \\
50{,}000  & 10{,}000 / 40 & 13 & 1.033 & 1.003 & 105 & 0.575 (0.005) \\
100{,}000 & 20{,}000 / 80 & 13 & 1.031 & 1.003 & 132 & 0.573 (0.007) \\
\bottomrule
\end{tabular}
\end{table}

\FloatBarrier

\subsection{Choice of Coefficient Rank}

Section~\ref{subsec:reduced-rank} states that the reduced-rank parameterization is used on both panels because of posterior geometry rather than accuracy. Table~\ref{tab:app-rank} gives the evidence at $p = 29$, averaged over three seeds at each setting.

Under the full-rank parameterization, between $68\%$ and $74\%$ of posterior draws of $B$ have spectral radius above one at every factor count, with median spectral radius between $1.07$ and $1.09$. The persistence is not a property of the data: ordinary least squares gives $0.911$ and the posterior mean of the default \texttt{bayesianVARs} specification $0.910$, both stationary. It is an accumulation effect in the posterior. A draw of $B$ carries estimation noise across all $p^2$ entries, and for a $p \times p$ matrix whose entries have standard deviation $\sigma$ the spectral radius grows like $\sigma\sqrt{p}$, so shrinking entries individually does not control the largest eigenvalue. Restricting $B$ to rank $r$ leaves only $r$ nonzero eigenvalues, and the posterior probability of a non-stationary draw falls below $0.05$ at every setting.

\begin{table}[h!]
\centering
\small
\caption{Coefficient rank and factor count at $p = 29$, averaged over three seeds. $\Pr(\rho(B) > 1)$ is the posterior probability of a non-stationary coefficient matrix; ``max eig.'' is the posterior predictive mean of the largest correlation eigenvalue, against $12.89$ in the data; ``Edges'' counts pairs whose 95\% credible interval excludes zero. MSPE is computed as in Table~\ref{tab:fredqd_mspe}, averaged over 100 draws of the held-out sets.}
\label{tab:app-rank}
\begin{tabular}{llccccc}
\toprule
Rank & $k$ & $\Pr(\rho(B)>1)$ & Median $\rho(B)$ & Max eig. & MSPE & Edges \\
\midrule
  Full & 5  & 0.720 & 1.092 &  6.34 & 0.583 &  5.0 \\
  Full & 8  & 0.741 & 1.090 &  6.84 & 0.569 &  7.0 \\
  Full & 10 & 0.716 & 1.078 &  7.21 & 0.565 &  7.0 \\
  Full & 15 & 0.682 & 1.069 &  8.13 & 0.558 &  7.0 \\
  \midrule
  2 & 5  & 0.024 & 0.890 & 11.21 & 0.639 &  7.0 \\
  2 & 8  & 0.018 & 0.887 & 11.26 & 0.577 & 13.0 \\
  2 & 10 & 0.029 & 0.882 & 10.88 & 0.578 & 13.0 \\
  2 & 15 & 0.041 & 0.819 & 10.72 & 0.531 & 12.7 \\
  \midrule
  3 & 5  & 0.038 & 0.891 & 11.36 & 0.629 &  6.0 \\
  3 & 8  & 0.026 & 0.877 & 11.14 & 0.576 & 13.0 \\
  3 & 10 & 0.029 & 0.862 & 10.87 & 0.572 & 12.3 \\
  3 & 15 & 0.040 & 0.814 & 10.55 & 0.548 & 11.7 \\
  \midrule
  4 & 5  & 0.039 & 0.869 & 11.29 & 0.633 &  7.0 \\
  4 & 8  & 0.049 & 0.875 & 11.11 & 0.581 & 12.3 \\
  4 & 10 & 0.045 & 0.823 & 11.02 & 0.579 & 12.0 \\
  4 & 15 & 0.048 & 0.801 & 10.59 & 0.551 & 11.7 \\
\bottomrule
\end{tabular}
\end{table}

The consequence for inference is visible in the fit. An unconstrained $B$ absorbs contemporaneous co-movement that belongs in $\Omega$, and the full-rank fits reproduce a largest correlation eigenvalue of 6.3 to 8.1 against 12.89 in the data. The reduced-rank fits reach 10.6 to 11.4, closer to the data but still below it, as in the posterior predictive checks of Section~\ref{subsubsec:post_pred}. Little is given up in prediction: held-out MSPE ranges from $0.558$ to $0.583$ across the full-rank settings and from $0.531$ to $0.639$ across the reduced-rank ones, and at $k = 10$ it is $0.565$ at full rank and $0.572$ at rank 3. Apart from $k = 5$, neither the rank nor the factor count changes the fit appreciably within reduced rank, so we use $r = 3$ for consistency with the $p = 100$ analysis and $k = 10$.

\FloatBarrier

\section{Application: Supplementary Diagnostics ($p = 100$)}\label{sec:app-supplement-p100}

This appendix collects the full diagnostic output for the $p = 100$ application of Section~\ref{subsec:p100}: MCMC convergence, the complete credible-interval edge list, edge recovery by competing estimators, posterior predictive checks, paired predictive comparisons, and sensitivity analyses.

\subsection{MCMC Convergence}

\begin{figure}[h!]
\centering
\includegraphics[width=\textwidth,height=0.8\textheight,keepaspectratio,alt={Nine trace plots from four chains at p equals 100: the log-likelihood, the spectral radius of the coefficient matrix, six reported partial correlations, and the unreported real GDP to investment pair. The chains overlap in the first eight panels, apart from occasional excursions; in the spectral-radius panel one chain makes larger excursions than the others. In the last panel the chains move between a region near 0.8 and a region near 0.2 and switch rarely. Each panel heading gives its R-hat and effective sample size.}]{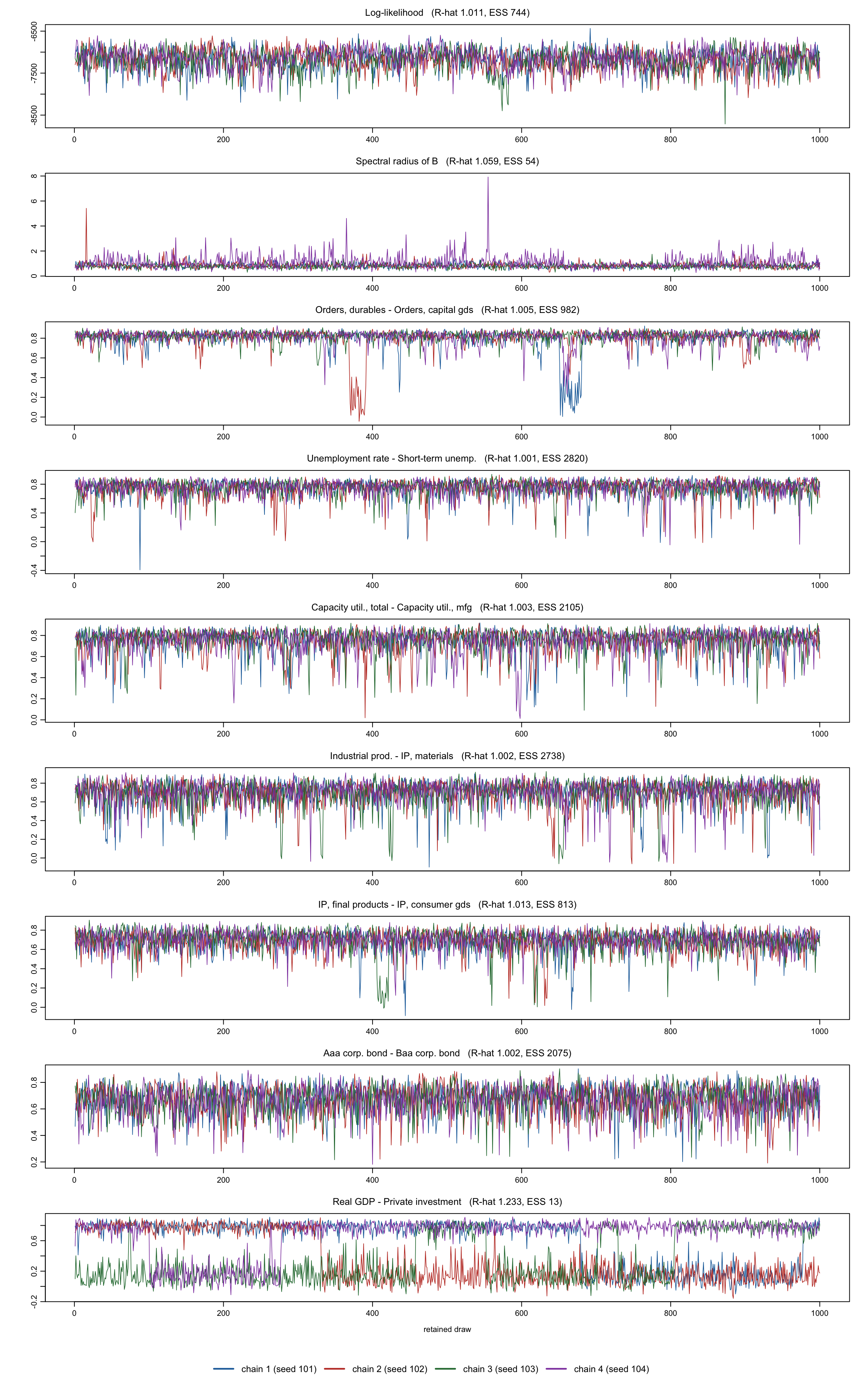}
\caption{Trace plots across four chains at $p = 100$: the log-likelihood, the spectral radius of the coefficient matrix, six reported partial correlations, and the unreported pair of real GDP and real private investment, the slowest-mixing of the 4{,}950 partial correlations in the fit. The others mix better: $4{,}912$ have $\hat{R} \le 1.05$, with median $\hat{R}$ of $1.001$. IP denotes industrial production. Each panel gives its $\hat{R}$ and effective sample size.}
\label{fig:app-trace-p100}
\end{figure}

Table~\ref{tab:app-edges-p100} lists the complete set of reported edges, those whose central 95\% posterior credible interval of the partial correlation excludes zero, ordered by magnitude, together with bulk effective sample sizes and rank-normalized $\hat{R}$ statistics from four independent chains of 100{,}000 iterations. Figure~\ref{fig:app-trace-p100} shows trace plots for the log-likelihood, the spectral radius, six reported edges, and the unreported pair of real GDP and real private investment. Mixing is good overall: across all 4{,}950 partial correlations the median $\hat{R}$ is $1.001$ and the median effective sample size is $3{,}821$, and $4{,}912$ have $\hat{R} \le 1.05$. The reported graph does not depend on the slower-mixing quantities. When the credible-interval rule is applied to each chain separately, 24 of the 26 reported edges are selected by all four chains and the other two by at least two, and no chain selects more than three edges outside the reported set. Every reported edge has $\hat{R} \le 1.06$; 21 of the 26 have $\hat{R} \le 1.013$, and the slowest is the edge between real consumption of durables and of services ($\hat{R} = 1.060$, effective sample size 44). Of the 38 with $\hat{R}$ above $1.05$, 26 exceed $1.1$. Twenty of these 26 join two NIPA series, and the remaining six join series within Industrial Production, within Inventories, Orders, and Sales, or within Employment and Unemployment. The pair of real GDP and real private investment has the largest value ($\hat{R} = 1.233$, effective sample size 13): two chains spend most of the run near $0.8$ and two near $0.2$, and each moves between the two regions only a few times. This reflects the near-identity linking GDP to its expenditure components: on the training data, regressing real GDP on real consumption, investment, government spending, exports and imports gives $R^2 = 0.997$, and the smallest eigenvalue of the correlation matrix of these six series is $0.0013$, so the precision matrix is weakly identified in that direction. The spectral radius mixes less well than at $p = 29$ ($\hat{R} = 1.059$, effective sample size 54); its posterior median is $0.818$ and $\Pr(\rho(B) > 1) = 0.223$, with most draws above one coming from one chain (seed 104, $0.466$, against $0.13$ to $0.16$ in the other three). The sampler bounds each idiosyncratic precision below at $10^{-8}$ as a numerical safeguard for the Cholesky factorization in the update of $\Lambda$; no precision reaches this bound in any chain.

\begin{table}[h!]
\centering
\caption{Reported edges at $p = 100$: all partial correlations whose central 95\% posterior interval excludes zero, ordered by magnitude, with convergence diagnostics from four independent chains of 100{,}000 iterations (burn-in 20{,}000, thinning 80). The log-likelihood gives ESS 744 and $\hat{R} = 1.011$, and the spectral radius $\rho(B)$ gives ESS 54 and $\hat{R} = 1.059$.  Series names follow Figure~\ref{fig:network_p100}.}
\label{tab:app-edges-p100}
\begin{tabular}{llccc}
\toprule
Series 1 & Series 2 & Partial cor. & ESS & $\hat R$ \\
\midrule
Orders, durables      & Orders, capital goods  &  0.807 &  982 & 1.005 \\
Unemployment rate     & Short-term unemp.      &  0.751 & 2820 & 1.001 \\
Capacity util., total & Capacity util., mfg    &  0.750 & 2105 & 1.003 \\
Industrial prod.      & IP, materials          &  0.699 & 2738 & 1.002 \\
IP, final products    & IP, consumer goods     &  0.691 &  813 & 1.013 \\
Aaa corp.\ bond       & Baa corp.\ bond        &  0.667 & 2075 & 1.002 \\
Real M1               & Real M2                &  0.648 &  194 & 1.022 \\
Fed funds             & 3-mo T-bill            &  0.639 &  850 & 1.004 \\
Emp., manufacturing   & Emp., durable mfg      &  0.632 & 2480 & 1.000 \\
Emp., goods-prod.     & Emp., construction     &  0.630 & 1276 & 1.004 \\
Unemp.\ rate, 15+ wk  & Unemp.\ 27+ weeks      &  0.616 & 1649 & 1.003 \\
IP, final products    & IP, business eq.       &  0.605 & 1813 & 1.005 \\
IP, consumer goods    & IP, business eq.       & $-0.573$ & 2641 & 1.003 \\
CPI, all items        & CPI, transport         &  0.558 & 1233 & 1.003 \\
PCE, total            & PCE, services          &  0.550 &  221 & 1.020 \\
Emp., total private   & Emp., leisure/hosp.    &  0.521 &  469 & 1.011 \\
Housing starts        & Housing starts, South     &  0.512 & 1165 & 1.002 \\
Housing starts        & Housing starts, West     &  0.504 & 1306 & 1.003 \\
PCE, total            & PCE, durables          &  0.492 &  644 & 1.009 \\
PPI, all commod.      & PPI, intermediate      &  0.486 & 2105 & 1.001 \\
3-mo T-bill           & 1-yr Treasury          &  0.467 &  356 & 1.008 \\
5-yr Treasury         & 10-yr Treasury         &  0.422 &  129 & 1.023 \\
PCE price index       & CPI, transport         &  0.419 & 1211 & 1.003 \\
PCE, durables         & PCE, services          & $-0.388$ &   44 & 1.060 \\
5-yr Treasury         & 30-yr mortgage         &  0.355 & 1071 & 1.002 \\
1-yr Treasury         & 5-yr Treasury          &  0.304 &  238 & 1.020 \\
\bottomrule
\end{tabular}
\end{table}

\FloatBarrier

\FloatBarrier

\subsection{Recovery of Reported Edges by Competing Estimators}

Table~\ref{tab:app-edge-recovery-p100} summarizes where VAPOR's 26 reported edges fall when all $4{,}950$ variable pairs are ranked by the magnitude of each method's estimated partial correlation.

\begin{table}[h!]
\centering
\caption{Recovery of VAPOR's 26 reported edges by competing estimators, $p = 100$. All $4{,}950$ variable pairs are ranked by the magnitude of each method's estimated partial correlation; entries give where VAPOR's reported edges fall in that ranking.}
\label{tab:app-edge-recovery-p100}
\begin{tabular}{lcccc}
\toprule
Method & Median rank & In top 25 & In top 100 & In top 250 \\
\midrule
    VAPOR             & 14   & 22 & 26 & 26 \\
    VAR + Glasso      & 21.5 & 14 & 24 & 24 \\
    VAR + PrecFact    & 54   &  8 & 17 & 21 \\
    Kastner (VAR+FSV) & 39.5 &  6 & 18 & 20 \\
    Sample covariance & 890.5 & 5 &  7 &  9 \\
\bottomrule
\end{tabular}
\end{table}

\FloatBarrier

\subsection{Posterior Predictive Checks}

\begin{figure}[h!]
\centering
\includegraphics[width=0.9\textwidth,alt={Nine histograms, one per summary, each showing the posterior predictive distribution with a vertical line at the observed value. Four observed values fall outside the bulk of their histogram. Kurtosis and residual lag-one autocorrelation lie far beyond every replicate; the mean and ninetieth percentile of absolute correlations miss narrowly.}]{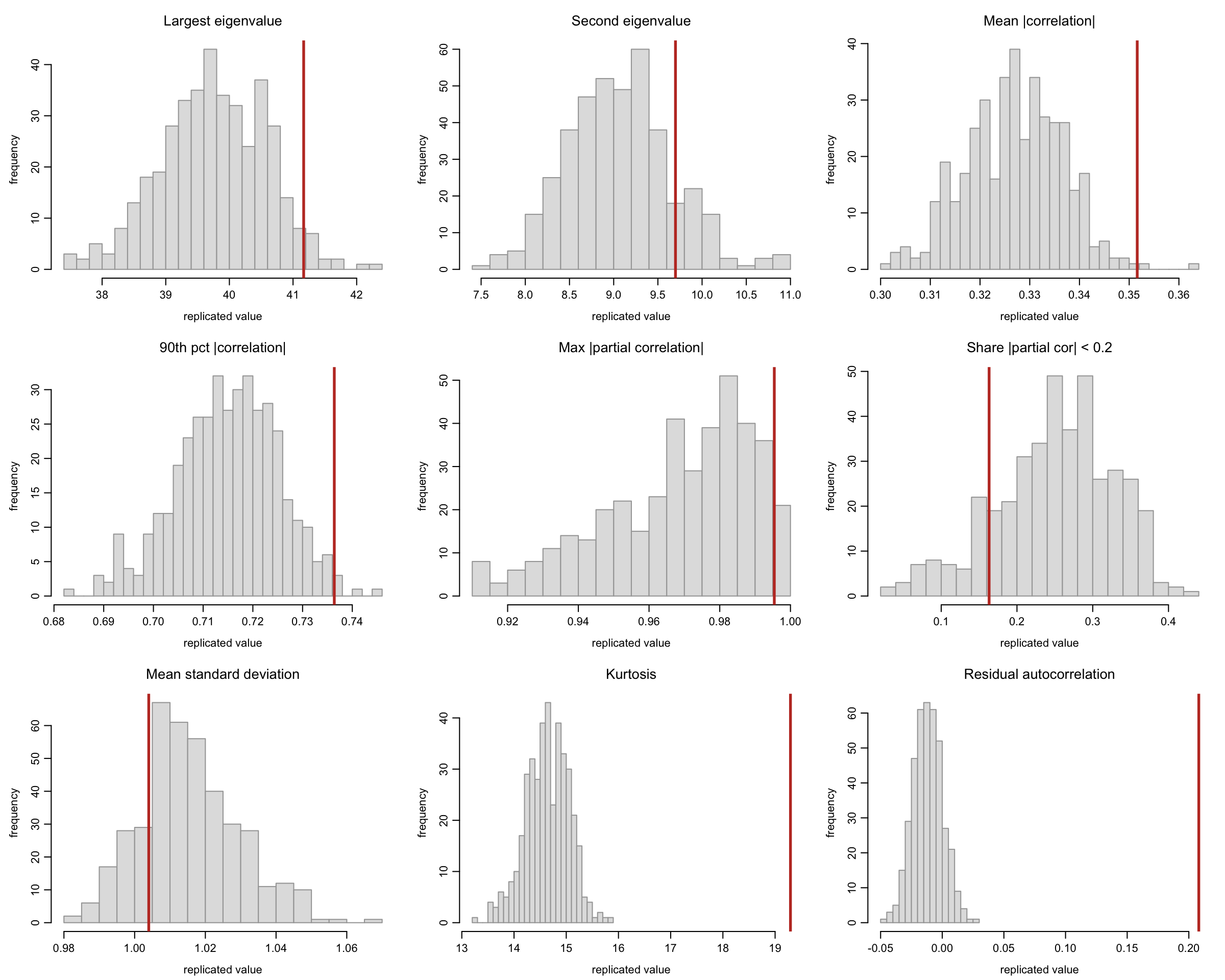}
\caption{Posterior predictive checks, $p = 100$. Histograms show the posterior predictive distribution of each of the nine summaries under the fitted model, from 400 replicated panels; vertical lines mark the observed values.}
\label{fig:app-ppc-p100}
\end{figure}

Table~\ref{tab:app-ppc-p100} and Figure~\ref{fig:app-ppc-p100} compare nine summaries of the observed panel to their posterior predictive distributions under the fitted model. Replicates are drawn one step ahead conditioning on the observed lagged values, as described in Supplement~\ref{sec:app-supplement}. At $T = p$ the sample covariance is singular, so the two partial-correlation summaries are formed from its Moore--Penrose pseudo-inverse for the observed and replicated panels alike. Five of the nine fall inside the central 95\% predictive interval, against two of nine at $p = 29$.

\begin{table}[h!]
\centering
\caption{Posterior predictive checks, $p = 100$. Observed values, posterior predictive means, and central 95\% predictive intervals for nine summaries. The two partial-correlation summaries are computed from the Moore--Penrose pseudo-inverse of the sample covariance.}
\label{tab:app-ppc-p100}
\begin{tabular}{lcccc}
\toprule
Summary & Observed & Replicate mean & \multicolumn{2}{c}{95\% interval} \\
\midrule
Largest eigenvalue        & 41.167 & 39.755 & 38.032 & 41.240 \\
Second eigenvalue         &  9.699 &  9.072 &  8.042 & 10.247 \\
Mean $|$correlation$|$    &  0.352 &  0.327 &  0.308 &  0.344 \\
90th pct.\ $|$correlation$|$ & 0.736 & 0.715 & 0.693 & 0.734 \\
Largest $|$partial cor.$|$ & 0.995 &  0.968 &  0.920 &  0.997 \\
Prop.\ $|$partial cor.$| < 0.2$ & 0.163 & 0.250 & 0.073 & 0.376 \\
Mean standard deviation   &  1.004 &  1.015 &  0.991 &  1.046 \\
Kurtosis                  & 19.294 & 14.636 & 13.751 & 15.360 \\
Residual lag-1 autocor.   &  0.208 & $-0.012$ & $-0.034$ & 0.012 \\
\bottomrule
\end{tabular}
\end{table}

Of the four that fall outside, two miss narrowly: the mean absolute correlation is $0.352$ against a predictive interval of $0.308$ to $0.344$, and the ninetieth percentile $0.736$ against $0.693$ to $0.734$. The largest eigenvalue, $41.17$ against $38.03$ to $41.24$, the second eigenvalue, $9.70$ against $8.04$ to $10.25$, and the largest partial correlation, $0.995$ against $0.920$ to $0.997$, fall inside, the first and last in the upper tail of their predictive distributions. Two miss by a wide margin. Kurtosis is $19.29$ against a predictive interval of $13.75$ to $15.36$, which reflects the Gaussian likelihood. The residuals retain lag-one autocorrelation of $0.208$ against $-0.034$ to $0.012$, larger than the $0.188$ observed at $p = 29$. At both dimensions the model understates the tails, in both the cross-section and the marginals.

\FloatBarrier

\subsection{Paired Predictive Comparison}

We repeat the paired comparison of Supplement~\ref{sec:app-supplement} at this dimension. Table~\ref{tab:dm-p100} reports the mean per-quarter difference in squared prediction error between each competitor and VAPOR, its standard error, and a Diebold--Mariano statistic with the Harvey--Leybourne--Newbold small-sample correction. A positive difference favors VAPOR. VAPOR's advantage is significant at the $5\%$ level against every competitor except VAR + Glasso, from which it is not distinguishable with either first stage ($p = 0.13$ at full rank and $0.10$ at rank 3). Against each of the others, VAPOR predicts more accurately in at least 18 of the 20 test quarters. Kastner's model is fit with the \texttt{bayesianVARs} package \citep{gruber2026bayesianvars}, which implements the factor stochastic volatility sampler of \citet{kastner2019sparse}, using the same number of factors as VAPOR, a column-wise normal-gamma prior on the factor loadings, and the package defaults otherwise, including a horseshoe prior on the VAR coefficients.

\begin{table}[h!]
\centering
\caption{Paired predictive comparison against VAPOR, $p = 100$, over 20 test quarters. Difference is the competitor's mean squared prediction error minus VAPOR's, so positive values favor VAPOR. VAPOR's mean MSPE is $0.563$. Rows marked ``rank 3'' truncate the first-stage coefficient matrix to its three leading singular directions.}
\label{tab:dm-p100}
\begin{tabular}{lccccc c}
\toprule
Method & MSPE & Difference & SE & DM & $p$ & Quarters won \\
\midrule
  VAR + Glasso            &   0.592 &   0.029 &  0.018 & 1.56 & 0.134    & 13 / 20 \\
  VAR + Glasso (rank 3)   &   0.584 &   0.021 &  0.012 & 1.73 & 0.101    & 12 / 20 \\
  Kastner (VAR+FSV)       & 218.790 & 218.227 & 36.166 & 5.88 & $<0.001$ & 20 / 20 \\
  Sample covariance       &   1.887 &   1.324 &  0.187 & 6.89 & $<0.001$ & 20 / 20 \\
  VAR + PrecFact          &   0.677 &   0.114 &  0.027 & 4.06 & 0.001    & 18 / 20 \\
  VAR + PrecFact (rank 3) &   0.701 &   0.137 &  0.031 & 4.33 & $<0.001$ & 20 / 20 \\
  PrecFact (no VAR)       &   0.666 &   0.103 &  0.024 & 4.11 & 0.001    & 20 / 20 \\
  Random walk             &   1.113 &   0.550 &  0.126 & 4.26 & $<0.001$ & 19 / 20 \\
\bottomrule
\end{tabular}
\end{table}

\FloatBarrier

\subsection{Sensitivity Analysis}

Table~\ref{tab:app-sensitivity-p100} reports held-out MSPE, the number of reported edges, and the posterior probability that the coefficient matrix is non-stationary, across factor counts, coefficient ranks, loading-prior shapes, and random seeds, each from a single chain of 100{,}000 iterations; Table~\ref{tab:app-length-p100} reports the effect of chain length. Across six random seeds at the reported configuration the held-out MSPE ranges from $0.560$ to $0.568$ and the edge count from 26 to 38. VAPOR's MSPE stays below that of every competitor in every configuration except $k = 15$ ($0.606$) and $a_L = 1.0$ ($0.607$), which are above VAR + Glasso ($0.592$). The number of reported edges rises with the factor count: 14 at $k = 15$, 29 at $k = 25$, and 34 at $k = 35$. The posterior probability that the coefficient matrix is non-stationary ranges from $0.048$ to $0.482$ across single chains and is $0.223$ for the pooled reported fit, against $0.027$ at $p = 29$; two of the six seeds (104 and 105) place close to half their draws above one. With $T = p$ the coefficient matrix and the innovation precision are less sharply separated than in the smaller panel. We fix $k = 25$, which sits in the middle of the predictive range while keeping the reported graph interpretable. Over the seven distinct configurations, seven of the 26 reported edges are selected in all seven and eight more in six; the five-year Treasury to mortgage-rate edge is selected in four, and the edges between total and manufacturing capacity utilization, between consumption of durables and of services, and between total private and leisure and hospitality employment, the least stable, in three. Across chain lengths from 7{,}500 to 100{,}000 iterations, the pooled graph has 24 to 27 edges and the MSPE ranges from $0.563$ to $0.569$.

\begin{table}[h!]
\centering
\caption{Sensitivity of the $p = 100$ results to tuning, each from a single chain of 100{,}000 iterations (seed 101 unless stated); the $k=25$, $r=3$ and $a_L=0.5$ rows are the same fit. Edges are those whose central 95\% posterior credible interval excludes zero. $\Pr(\rho(B) > 1)$ is the posterior probability that the coefficient matrix is non-stationary. The Dirichlet-Laplace shape on the VAR coefficients is fixed by dimension and not varied. The reported graph pools four chains.}
\label{tab:app-sensitivity-p100}
\begin{tabular}{lccc}
\toprule
Setting & MSPE & Edges & $\Pr(\rho(B) > 1)$ \\
\midrule
  $k=15$        & 0.606 & 14 & 0.155 \\
  $k=25$        & 0.565 & 29 & 0.132 \\
  $k=35$        & 0.565 & 34 & 0.162 \\
  $r=2$         & 0.566 & 37 & 0.048 \\
  $r=3$         & 0.565 & 29 & 0.132 \\
  $r=4$         & 0.564 & 26 & 0.245 \\
  $a_L=0.3$     & 0.571 & 36 & 0.185 \\
  $a_L=0.5$     & 0.565 & 29 & 0.132 \\
  $a_L=1.0$     & 0.607 & 35 & 0.128 \\
  seed 102      & 0.561 & 28 & 0.150 \\
  seed 103      & 0.560 & 26 & 0.112 \\
  seed 104      & 0.568 & 27 & 0.482 \\
  seed 105      & 0.561 & 28 & 0.452 \\
  seed 106      & 0.565 & 38 & 0.242 \\
\bottomrule
\end{tabular}
\end{table}

\begin{table}[h!]
\centering
\caption{Effect of chain length, $p = 100$, four chains at each length. Edges are those selected by the pooled 95\% credible-interval rule; $\hat R$ and ESS are computed across the four chains for every pair selected by at least one chain; MSPE is the mean across chains with the standard deviation in parentheses.}
\label{tab:app-length-p100}
\begin{tabular}{lcccccc}
\toprule
Iterations & Burn-in / thin & Edges & Max $\hat R$ & Median $\hat R$ & Min ESS & MSPE \\
\midrule
7{,}500   & 2{,}500 / 5   & 27 & 1.558 & 1.103 &  7 & 0.569 (0.010) \\
20{,}000  & 10{,}000 / 10 & 27 & 1.575 & 1.095 &  7 & 0.563 (0.005) \\
50{,}000  & 10{,}000 / 40 & 24 & 1.414 & 1.015 &  8 & 0.567 (0.003) \\
100{,}000 & 20{,}000 / 80 & 26 & 1.233 & 1.004 & 13 & 0.563 (0.004) \\
\bottomrule
\end{tabular}
\end{table}

\FloatBarrier

\subsection{Panel Summary}\label{subsec:app-varlist-p100}

Table~\ref{tab:data_summary} summarizes the two training panels.

\begin{table}[h!]
\centering
\caption{Summary of the two training panels, Q1 1996 to Q4 2020 ($T_{\mathrm{train}} = 100$), after transformation and standardization. Principal components are those of the correlation matrix; excess kurtosis is the fourth standardized moment minus 3.}
\label{tab:data_summary}
\begin{tabular}{lcc}
\toprule
 & $p = 29$ & $p = 100$ \\
\midrule
FRED-QD groups represented & 8 & 12 \\
Median $|$correlation$|$ & 0.35 & 0.30 \\
90th percentile $|$correlation$|$ & 0.79 & 0.74 \\
Variance explained, first PC & 44\% & 41\% \\
Variance explained, first three PCs & 66\% & 59\% \\
Condition number of sample covariance & $5.6 \times 10^{3}$ & singular (rank 99) \\
Median lag-1 autocorrelation & 0.13 & 0.16 \\
Median excess kurtosis & 7.7 & 10.2 \\
\bottomrule
\end{tabular}
\end{table}

\section{Supplementary Simulations}\label{sec:supplement_sims}

This supplement reports Simulations~A through~F in full. The competitors and evaluation metrics are those summarized in Section~\ref{sec:simulations}. The Gibbs samplers for PrecFact, VAR + PrecFact, and VAPOR run 5000 iterations with a burn-in of 2500 and thinning of 5, yielding 500 retained draws, with the priors of Section~\ref{subsec:priors} and $k = 5$ factors throughout. Kastner's vector autoregression with factor stochastic volatility is fit via \texttt{bayesianVARs} \citep{gruber2026bayesianvars} with 2000 draws after a burn-in of 1000, using five factors except in Simulation~C, where it uses the single factor of the data-generating process; the first-stage VAR(1) of the sequential methods uses the same draws and burn-in. The graphical lasso penalty is chosen to minimize Frobenius distance to the true correlation-scaled precision matrix over a logarithmically spaced grid, of 20 points from $10^{-2}$ to 1 in Simulations~A, B, and~D and 25 points from $10^{-2.5}$ to 1 in Simulations~E and~F, and an edge is selected wherever the chosen estimate is nonzero. In Simulations~A, B, and~D, graph selection for the PrecFact-based methods uses the pFDR procedure of \citet{chandra2026precfact} with $\alpha = 0.1$ and $\beta = 0.85$, and Kastner's estimate is thresholded to retain the true number of edges; in Simulations~E and~F every method's partial correlations are thresholded to retain the true number of edges. Simulation~C compares held-out prediction and selects no graph.

\subsection{Simulation A: Data Generated from the Proposed Model}

We generate data from a VAR(1) with innovation precision matrix $\Omega = \Lambda_0\Lambda_0^\top + I_p$, where $\Lambda_0 \in \mathbb{R}^{p \times k}$ is sparse with 20\% nonzero entries drawn from $\pm 1.5$. The VAR coefficient matrix $B$ has 10\% nonzero entries scaled to a target spectral radius $\rho(B)$. We fix $p = 20$, $k = 5$, and run 25 replicates per scenario. Results are reported as mean (sd) across replicates. Because every method is fit to the same simulated dataset within a replicate, we also report the mean paired per-replicate difference in Frobenius distance against VAPOR, denoted $\Delta_F$, together with its standard error.

\paragraph{Experiment A: varying autocorrelation ($T=200$).}
\begin{table}[h!]
\centering
\small
\setlength{\tabcolsep}{4pt}
\caption{Simulation A, Experiment A: varying spectral radius, $p=20$, $T=200$, 25 replicates. $\Delta_F$ is the mean paired per-replicate difference in Frobenius distance against VAPOR, with positive values favoring VAPOR. The lowest Frobenius distance in each block is shown in bold.}
\label{tab:sim-a-autocorr}
\begin{tabular}{llcccc}
\toprule
Scenario & Method & Sensitivity & Specificity & Frobenius & $\Delta_F$ (SE) \\
\midrule
$\rho = 0.10$ & PrecFact (no VAR)  & 0.555 (0.188) & 0.998 (0.004) & 0.775 (0.093) & 0.091 (0.010) \\
              & Glasso (no VAR)    & 0.825 (0.149) & 0.510 (0.138) & 0.943 (0.129) & 0.259 (0.021) \\
              & VAR + Glasso       & 0.833 (0.151) & 0.512 (0.140) & 0.929 (0.113) & 0.246 (0.019) \\
              & VAR + PrecFact     & 0.555 (0.187) & 0.998 (0.004) & 0.751 (0.077) & 0.067 (0.010) \\
              & Kastner (VAR+FSV)  & 0.547 (0.118) & 0.934 (0.021) & 1.118 (0.397) & 0.435 (0.076) \\
              & VAPOR              & 0.567 (0.154) & 0.999 (0.002) & \textbf{0.684 (0.074)} & --- \\
\midrule
$\rho = 0.50$ & PrecFact (no VAR)  & 0.398 (0.213) & 0.970 (0.024) & 1.524 (0.482) & 0.830 (0.089) \\
              & Glasso (no VAR)    & 0.675 (0.308) & 0.590 (0.216) & 1.569 (0.380) & 0.874 (0.069) \\
              & VAR + Glasso       & 0.834 (0.143) & 0.527 (0.140) & 0.932 (0.113) & 0.237 (0.018) \\
              & VAR + PrecFact     & 0.557 (0.184) & 0.999 (0.003) & 0.755 (0.076) & 0.060 (0.012) \\
              & Kastner (VAR+FSV)  & 0.536 (0.120) & 0.933 (0.022) & 1.199 (0.379) & 0.504 (0.072) \\
              & VAPOR              & 0.566 (0.157) & 1.000 (0.001) & \textbf{0.695 (0.069)} & --- \\
\midrule
$\rho = 0.75$ & PrecFact (no VAR)  & 0.304 (0.147) & 0.926 (0.036) & 2.363 (0.562) & 1.671 (0.107) \\
              & Glasso (no VAR)    & 0.407 (0.363) & 0.757 (0.242) & 2.031 (0.411) & 1.338 (0.076) \\
              & VAR + Glasso       & 0.837 (0.145) & 0.529 (0.140) & 0.935 (0.115) & 0.242 (0.017) \\
              & VAR + PrecFact     & 0.561 (0.191) & 0.998 (0.004) & 0.761 (0.082) & 0.068 (0.012) \\
              & Kastner (VAR+FSV)  & 0.536 (0.117) & 0.933 (0.020) & 1.196 (0.406) & 0.504 (0.077) \\
              & VAPOR              & 0.553 (0.157) & 1.000 (0.002) & \textbf{0.693 (0.067)} & --- \\
\midrule
$\rho = 0.89$ & PrecFact (no VAR)  & 0.236 (0.098) & 0.888 (0.050) & 2.977 (0.604) & 2.284 (0.117) \\
              & Glasso (no VAR)    & 0.146 (0.294) & 0.903 (0.225) & 2.177 (0.458) & 1.483 (0.088) \\
              & VAR + Glasso       & 0.838 (0.148) & 0.527 (0.146) & 0.935 (0.113) & 0.241 (0.018) \\
              & VAR + PrecFact     & 0.575 (0.171) & 0.998 (0.004) & 0.758 (0.077) & 0.064 (0.012) \\
              & Kastner (VAR+FSV)  & 0.529 (0.116) & 0.932 (0.021) & 1.142 (0.396) & 0.449 (0.075) \\
              & VAPOR              & 0.559 (0.160) & 1.000 (0.002) & \textbf{0.694 (0.067)} & --- \\
\midrule
$\rho = 0.95$ & PrecFact (no VAR)  & 0.238 (0.111) & 0.864 (0.061) & 3.269 (0.656) & 2.580 (0.126) \\
              & Glasso (no VAR)    & 0.053 (0.192) & 0.967 (0.139) & 2.193 (0.468) & 1.505 (0.091) \\
              & VAR + Glasso       & 0.839 (0.144) & 0.525 (0.146) & 0.934 (0.114) & 0.246 (0.019) \\
              & VAR + PrecFact     & 0.555 (0.181) & 0.998 (0.004) & 0.762 (0.082) & 0.073 (0.012) \\
              & Kastner (VAR+FSV)  & 0.543 (0.109) & 0.934 (0.020) & 1.178 (0.399) & 0.489 (0.076) \\
              & VAPOR              & 0.550 (0.169) & 1.000 (0.002) & \textbf{0.689 (0.071)} & --- \\
\midrule
$\rho = 0.99$ & PrecFact (no VAR)  & 0.239 (0.116) & 0.845 (0.068) & 3.586 (0.741) & 2.898 (0.145) \\
              & Glasso (no VAR)    & 0.009 (0.018) & 0.994 (0.014) & 2.197 (0.470) & 1.509 (0.091) \\
              & VAR + Glasso       & 0.830 (0.146) & 0.530 (0.141) & 0.931 (0.112) & 0.243 (0.019) \\
              & VAR + PrecFact     & 0.569 (0.184) & 0.999 (0.004) & 0.756 (0.080) & 0.067 (0.011) \\
              & Kastner (VAR+FSV)  & 0.547 (0.097) & 0.934 (0.019) & 1.169 (0.389) & 0.481 (0.072) \\
              & VAPOR              & 0.557 (0.158) & 1.000 (0.001) & \textbf{0.688 (0.068)} & --- \\
\bottomrule
\end{tabular}
\end{table}

VAPOR achieves the lowest Frobenius distance in all six scenarios, and every paired difference is significant at $p < 10^{-4}$. The margin over the two methods that ignore temporal structure grows with $\rho$: against PrecFact applied to the raw series it rises from 0.091 at $\rho = 0.10$ to 2.898 at $\rho = 0.99$, and against the graphical lasso applied to the raw series from 0.259 to 1.509, tracking the rise in those methods' Frobenius distances from 0.775 to 3.586 and from 0.943 to 2.197. The margins over the VAR-based competitors are flat in $\rho$: 0.237 to 0.246 against VAR + Glasso, 0.435 to 0.504 against Kastner, and 0.060 to 0.073 against VAR + PrecFact. VAR + PrecFact is the closest competitor at every level; it uses the same precision factor model but estimates it after the VAR rather than jointly, and VAPOR is closer to the truth in 18 to 22 of the 25 replicates against it, compared with 23 to 25 against the other four methods. VAPOR's specificity is 0.999 or higher throughout and its sensitivity ranges from 0.550 to 0.567, close to VAR + PrecFact. VAR + Glasso attains higher sensitivity, 0.830 to 0.839, at a specificity near 0.52. The full paired comparison, including $t$ statistics, $p$-values, and per-replicate win counts, is given in Table~\ref{tab:sim-a-paired-autocorr}.

\FloatBarrier

\paragraph{Experiment B: varying sample size ($\rho = 0.95$).}
\begin{table}[h!]
\centering
\small
\setlength{\tabcolsep}{4pt}
\caption{Simulation A, Experiment B: varying sample size, $p=20$, $\rho(B)=0.95$, 25 replicates. $\Delta_F$ is the mean paired per-replicate difference in Frobenius distance against VAPOR, with positive values favoring VAPOR. The lowest Frobenius distance in each block is shown in bold.}
\label{tab:sim-a-samplesize}
\begin{tabular}{llcccc}
\toprule
Scenario & Method & Sensitivity & Specificity & Frobenius & $\Delta_F$ (SE) \\
\midrule
$T = 50$   & PrecFact (no VAR)  & 0.095 (0.080) & 0.950 (0.030) & 3.199 (0.624) & 1.547 (0.116) \\
           & Glasso (no VAR)    & 0.022 (0.047) & 0.991 (0.019) & 2.196 (0.471) & 0.544 (0.036) \\
           & VAR + Glasso       & 0.602 (0.214) & 0.716 (0.168) & 1.711 (0.359) & 0.059 (0.024) \\
           & VAR + PrecFact     & 0.192 (0.137) & 0.994 (0.011) & 1.740 (0.305) & 0.088 (0.064) \\
           & Kastner (VAR+FSV)  & 0.396 (0.101) & 0.913 (0.021) & 1.796 (0.444) & 0.144 (0.033) \\
           & VAPOR              & 0.020 (0.039) & 1.000 (0.000) & \textbf{1.652 (0.357)} & --- \\
\midrule
$T = 100$  & PrecFact (no VAR)  & 0.170 (0.099) & 0.900 (0.043) & 3.249 (0.663) & 2.211 (0.130) \\
           & Glasso (no VAR)    & 0.035 (0.071) & 0.984 (0.040) & 2.189 (0.473) & 1.151 (0.089) \\
           & VAR + Glasso       & 0.741 (0.153) & 0.587 (0.151) & 1.321 (0.192) & 0.283 (0.030) \\
           & VAR + PrecFact     & 0.424 (0.187) & 0.996 (0.006) & 1.143 (0.116) & 0.105 (0.021) \\
           & Kastner (VAR+FSV)  & 0.484 (0.113) & 0.925 (0.021) & 1.386 (0.369) & 0.348 (0.067) \\
           & VAPOR              & 0.372 (0.145) & 1.000 (0.001) & \textbf{1.038 (0.157)} & --- \\
\midrule
$T = 200$  & PrecFact (no VAR)  & 0.238 (0.111) & 0.864 (0.061) & 3.269 (0.656) & 2.580 (0.126) \\
           & Glasso (no VAR)    & 0.053 (0.192) & 0.967 (0.139) & 2.193 (0.468) & 1.505 (0.091) \\
           & VAR + Glasso       & 0.839 (0.144) & 0.525 (0.146) & 0.934 (0.114) & 0.246 (0.019) \\
           & VAR + PrecFact     & 0.555 (0.181) & 0.998 (0.004) & 0.762 (0.082) & 0.073 (0.012) \\
           & Kastner (VAR+FSV)  & 0.543 (0.109) & 0.934 (0.020) & 1.178 (0.399) & 0.489 (0.076) \\
           & VAPOR              & 0.550 (0.169) & 1.000 (0.002) & \textbf{0.689 (0.071)} & --- \\
\midrule
$T = 500$  & PrecFact (no VAR)  & 0.339 (0.151) & 0.810 (0.076) & 3.357 (0.743) & 2.918 (0.150) \\
           & Glasso (no VAR)    & 0.106 (0.267) & 0.931 (0.190) & 2.185 (0.463) & 1.746 (0.092) \\
           & VAR + Glasso       & 0.915 (0.084) & 0.476 (0.123) & 0.606 (0.071) & 0.168 (0.010) \\
           & VAR + PrecFact     & 0.697 (0.143) & 1.000 (0.000) & 0.472 (0.056) & 0.033 (0.006) \\
           & Kastner (VAR+FSV)  & 0.613 (0.110) & 0.944 (0.019) & 1.050 (0.443) & 0.612 (0.089) \\
           & VAPOR              & 0.691 (0.131) & 1.000 (0.000) & \textbf{0.439 (0.049)} & --- \\
\midrule
$T = 1000$ & PrecFact (no VAR)  & 0.384 (0.146) & 0.782 (0.071) & 3.327 (0.706) & 3.015 (0.141) \\
           & Glasso (no VAR)    & 0.121 (0.296) & 0.923 (0.201) & 2.182 (0.461) & 1.870 (0.092) \\
           & VAR + Glasso       & 0.952 (0.044) & 0.438 (0.086) & 0.453 (0.036) & 0.141 (0.007) \\
           & VAR + PrecFact     & 0.738 (0.130) & 1.000 (0.002) & 0.335 (0.046) & 0.023 (0.004) \\
           & Kastner (VAR+FSV)  & 0.650 (0.113) & 0.949 (0.019) & 0.999 (0.454) & 0.687 (0.090) \\
           & VAPOR              & 0.736 (0.124) & 1.000 (0.001) & \textbf{0.312 (0.037)} & --- \\
\bottomrule
\end{tabular}
\end{table}

VAPOR achieves the lowest Frobenius distance at every sample size. At $T = 50$ the paired difference against VAR + PrecFact is 0.088 with standard error 0.064 ($p = 0.18$, 18 of 25 replicates) and against VAR + Glasso 0.059 with standard error 0.024 ($p = 0.022$, 17 of 25); at every sample size from $T = 100$ onward all five paired differences are significant at $p < 10^{-4}$. The margin over VAR + PrecFact is 0.088, 0.105, 0.073, 0.033, and 0.023 at $T = 50$, 100, 200, 500, and 1000. The margin over Kastner moves the other way, rising from 0.144 to 0.687 across the same range, because Kastner's Frobenius distance falls only from 1.796 to 0.999 while VAPOR's falls from 1.652 to 0.312. The margin over PrecFact applied to the raw series rises from 1.547 to 3.015, and that method's Frobenius distance stays near 3.3 at every sample size. VAPOR's specificity is 1.000 throughout. Its sensitivity is 0.020 at $T = 50$, where it selects almost no edges and still attains the lowest Frobenius distance, and rises to 0.736 at $T = 1000$, close to the 0.738 of VAR + PrecFact. VAR + Glasso reaches sensitivity 0.952 at $T = 1000$ at a specificity of 0.438. The full paired comparison is given in Table~\ref{tab:sim-a-paired-samplesize}.

\begin{figure}[t]
\centering
\includegraphics[width=\textwidth,alt={Five side-by-side 20 by 20 partial correlation matrices for one replicate: the true precision matrix, followed by VAPOR, VAR plus PrecFact, VAR plus graphical lasso, and Kastner. Variables are ordered by a hierarchical clustering of the truth, so the true matrix concentrates its nonzero entries among the first ten variables and is otherwise empty. All four methods recover that region. VAPOR reproduces its magnitudes most closely; VAR plus PrecFact is similar with more scattered background; the graphical lasso renders it faintly; and Kastner renders it more strongly than the truth.}]{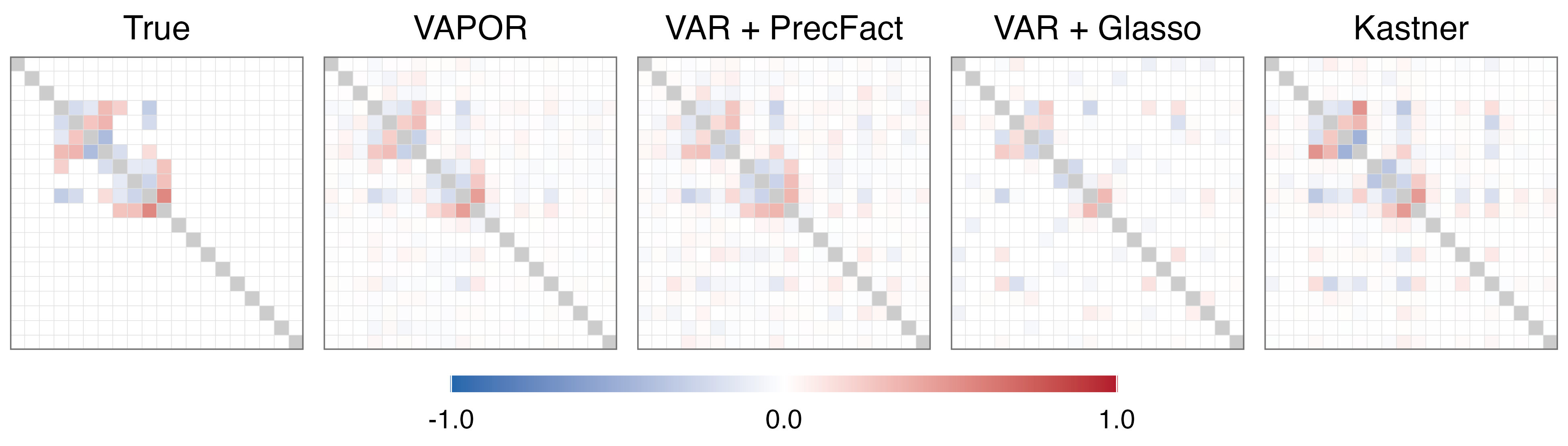}
\caption{Partial-correlation heatmaps for a representative replicate of Simulation A at $p=20$, $T=100$, $\rho(B)=0.95$. Left to right: true precision matrix, VAPOR, VAR + PrecFact, VAR + Glasso, and Kastner (the four methods with lowest Frobenius distance). Color scale as in Figure~\ref{fig:heatmaps_p29}; the diagonal is omitted. Variables are ordered by a hierarchical clustering of the true precision matrix, and the same ordering is used in every panel.}
\label{fig:sim-a-heatmap}
\end{figure}

The numerical advantage in Frobenius distance corresponds to a visible difference in recovered structure. Figure~\ref{fig:sim-a-heatmap} shows the true precision matrix alongside the four best-performing methods by Frobenius distance, displayed as partial-correlation heatmaps for a single replicate at $\rho=0.95$ and $T=100$. With variables ordered by a clustering of the true precision matrix, that matrix has two contiguous blocks of conditionally dependent variables in its upper left, with the remaining variables conditionally independent. VAPOR reproduces both blocks in the correct positions and signs, with little spurious structure elsewhere. VAR + PrecFact, which shares the same precision factor model but estimates it sequentially, recovers the same blocks but with more noise in the off-block entries. VAR + Glasso recovers little of either block, with nearly all entries close to zero. Kastner recovers both blocks but assigns comparable magnitude to off-block entries that are zero in the truth. Both attain higher Frobenius distances than VAPOR and VAR + PrecFact in Table~\ref{tab:sim-a-samplesize}.

\FloatBarrier

\paragraph{Paired comparisons.}\label{subsec:sim-a-paired}

Tables~\ref{tab:sim-a-autocorr} and~\ref{tab:sim-a-samplesize} report the mean paired difference $\Delta_F$ and its standard error. Tables~\ref{tab:sim-a-paired-autocorr} and~\ref{tab:sim-a-paired-samplesize} give the corresponding $t$ statistics, $p$-values, and per-replicate win counts. Because each replicate fits every method to the same simulated dataset, the paired standard error is smaller than the standard error of the difference between the two marginal means reported in the main tables, and the win count records how many of the 25 replicates favor VAPOR individually.

\begin{table}[h!]
\centering
\small
\caption{Simulation A, Experiment A: paired per-replicate Frobenius differences against VAPOR across 25 replicates, $p=20$, $T=200$. $\Delta_F$ is the mean difference, with positive values favoring VAPOR; $t$ and $p$ are from a paired $t$-test on 24 degrees of freedom; Wins is the number of replicates in which VAPOR is closer to the true correlation-scaled precision matrix.}
\label{tab:sim-a-paired-autocorr}
\begin{tabular}{llccccc}
\toprule
Scenario & Method & $\Delta_F$ & SE & $t$ & $p$ & Wins \\
\midrule
$\rho = 0.10$ & PrecFact (no VAR)  & 0.091 & 0.010 & 8.81 & $<10^{-4}$ & 23 / 25 \\
              & Glasso (no VAR)    & 0.259 & 0.021 & 12.50 & $<10^{-4}$ & 25 / 25 \\
              & VAR + Glasso       & 0.246 & 0.019 & 12.77 & $<10^{-4}$ & 25 / 25 \\
              & VAR + PrecFact     & 0.067 & 0.010 &  6.48 & $<10^{-4}$ & 22 / 25 \\
              & Kastner (VAR+FSV)  & 0.435 & 0.076 &  5.74 & $<10^{-4}$ & 23 / 25 \\
\midrule
$\rho = 0.50$ & PrecFact (no VAR)  & 0.830 & 0.089 &  9.36 & $<10^{-4}$ & 25 / 25 \\
              & Glasso (no VAR)    & 0.874 & 0.069 & 12.75 & $<10^{-4}$ & 25 / 25 \\
              & VAR + Glasso       & 0.237 & 0.018 & 13.52 & $<10^{-4}$ & 25 / 25 \\
              & VAR + PrecFact     & 0.060 & 0.012 &  4.92 & $<10^{-4}$ & 20 / 25 \\
              & Kastner (VAR+FSV)  & 0.504 & 0.072 &  7.02 & $<10^{-4}$ & 25 / 25 \\
\midrule
$\rho = 0.75$ & PrecFact (no VAR)  & 1.671 & 0.107 & 15.56 & $<10^{-4}$ & 25 / 25 \\
              & Glasso (no VAR)    & 1.338 & 0.076 & 17.56 & $<10^{-4}$ & 25 / 25 \\
              & VAR + Glasso       & 0.242 & 0.017 & 14.16 & $<10^{-4}$ & 25 / 25 \\
              & VAR + PrecFact     & 0.068 & 0.012 &  5.54 & $<10^{-4}$ & 18 / 25 \\
              & Kastner (VAR+FSV)  & 0.504 & 0.077 &  6.51 & $<10^{-4}$ & 25 / 25 \\
\midrule
$\rho = 0.89$ & PrecFact (no VAR)  & 2.284 & 0.117 & 19.59 & $<10^{-4}$ & 25 / 25 \\
              & Glasso (no VAR)    & 1.483 & 0.088 & 16.89 & $<10^{-4}$ & 25 / 25 \\
              & VAR + Glasso       & 0.241 & 0.018 & 13.19 & $<10^{-4}$ & 25 / 25 \\
              & VAR + PrecFact     & 0.064 & 0.012 &  5.45 & $<10^{-4}$ & 20 / 25 \\
              & Kastner (VAR+FSV)  & 0.449 & 0.075 &  5.96 & $<10^{-4}$ & 24 / 25 \\
\midrule
$\rho = 0.95$ & PrecFact (no VAR)  & 2.580 & 0.126 & 20.44 & $<10^{-4}$ & 25 / 25 \\
              & Glasso (no VAR)    & 1.505 & 0.091 & 16.58 & $<10^{-4}$ & 25 / 25 \\
              & VAR + Glasso       & 0.246 & 0.019 & 12.93 & $<10^{-4}$ & 25 / 25 \\
              & VAR + PrecFact     & 0.073 & 0.012 &  6.18 & $<10^{-4}$ & 21 / 25 \\
              & Kastner (VAR+FSV)  & 0.489 & 0.076 &  6.47 & $<10^{-4}$ & 24 / 25 \\
\midrule
$\rho = 0.99$ & PrecFact (no VAR)  & 2.898 & 0.145 & 20.01 & $<10^{-4}$ & 25 / 25 \\
              & Glasso (no VAR)    & 1.509 & 0.091 & 16.55 & $<10^{-4}$ & 25 / 25 \\
              & VAR + Glasso       & 0.243 & 0.019 & 12.54 & $<10^{-4}$ & 25 / 25 \\
              & VAR + PrecFact     & 0.067 & 0.011 &  6.01 & $<10^{-4}$ & 19 / 25 \\
              & Kastner (VAR+FSV)  & 0.481 & 0.072 &  6.65 & $<10^{-4}$ & 24 / 25 \\
\bottomrule
\end{tabular}
\end{table}

\FloatBarrier

\begin{table}[h!]
\centering
\small
\caption{Simulation A, Experiment B: paired per-replicate Frobenius differences against VAPOR across 25 replicates, $p=20$, $\rho(B)=0.95$. Columns as in Table~\ref{tab:sim-a-paired-autocorr}. The $T = 200$ block and the $\rho = 0.95$ block of Table~\ref{tab:sim-a-paired-autocorr} are the same scenario.}
\label{tab:sim-a-paired-samplesize}
\begin{tabular}{llccccc}
\toprule
Scenario & Method & $\Delta_F$ & SE & $t$ & $p$ & Wins \\
\midrule
$T = 50$   & PrecFact (no VAR)  & 1.547 & 0.116 & 13.29 & $<10^{-4}$ & 25 / 25 \\
           & Glasso (no VAR)    & 0.544 & 0.036 & 14.96 & $<10^{-4}$ & 25 / 25 \\
           & VAR + Glasso       & 0.059 & 0.024 &  2.44 & 0.022      & 17 / 25 \\
           & VAR + PrecFact     & 0.088 & 0.064 &  1.37 & 0.18      & 18 / 25 \\
           & Kastner (VAR+FSV)  & 0.144 & 0.033 &  4.33 & 0.00023     & 21 / 25 \\
\midrule
$T = 100$  & PrecFact (no VAR)  & 2.211 & 0.130 & 17.02 & $<10^{-4}$ & 25 / 25 \\
           & Glasso (no VAR)    & 1.151 & 0.089 & 13.00 & $<10^{-4}$ & 25 / 25 \\
           & VAR + Glasso       & 0.283 & 0.030 &  9.58 & $<10^{-4}$ & 23 / 25 \\
           & VAR + PrecFact     & 0.105 & 0.021 &  5.06 & $<10^{-4}$ & 22 / 25 \\
           & Kastner (VAR+FSV)  & 0.348 & 0.067 &  5.17 & $<10^{-4}$ & 22 / 25 \\
\midrule
$T = 200$  & PrecFact (no VAR)  & 2.580 & 0.126 & 20.44 & $<10^{-4}$ & 25 / 25 \\
           & Glasso (no VAR)    & 1.505 & 0.091 & 16.58 & $<10^{-4}$ & 25 / 25 \\
           & VAR + Glasso       & 0.246 & 0.019 & 12.93 & $<10^{-4}$ & 25 / 25 \\
           & VAR + PrecFact     & 0.073 & 0.012 &  6.18 & $<10^{-4}$ & 21 / 25 \\
           & Kastner (VAR+FSV)  & 0.489 & 0.076 &  6.47 & $<10^{-4}$ & 24 / 25 \\
\midrule
$T = 500$  & PrecFact (no VAR)  & 2.918 & 0.150 & 19.50 & $<10^{-4}$ & 25 / 25 \\
           & Glasso (no VAR)    & 1.746 & 0.092 & 18.89 & $<10^{-4}$ & 25 / 25 \\
           & VAR + Glasso       & 0.168 & 0.010 & 16.49 & $<10^{-4}$ & 25 / 25 \\
           & VAR + PrecFact     & 0.033 & 0.006 &  5.82 & $<10^{-4}$ & 23 / 25 \\
           & Kastner (VAR+FSV)  & 0.612 & 0.089 &  6.89 & $<10^{-4}$ & 24 / 25 \\
\midrule
$T = 1000$ & PrecFact (no VAR)  & 3.015 & 0.141 & 21.33 & $<10^{-4}$ & 25 / 25 \\
           & Glasso (no VAR)    & 1.870 & 0.092 & 20.29 & $<10^{-4}$ & 25 / 25 \\
           & VAR + Glasso       & 0.141 & 0.007 & 20.23 & $<10^{-4}$ & 25 / 25 \\
           & VAR + PrecFact     & 0.023 & 0.004 &  5.14 & $<10^{-4}$ & 22 / 25 \\
           & Kastner (VAR+FSV)  & 0.687 & 0.090 &  7.62 & $<10^{-4}$ & 25 / 25 \\
\bottomrule
\end{tabular}
\end{table}

\FloatBarrier

\subsection{Simulation B: Data Generated with Sparse Precision Matrix (No Factor Structure)}

We now consider a misspecified setting in which $\Omega$ is generated directly as a sparse positive definite matrix without any underlying factor structure. We generate a random sparse symmetric matrix via a random graph with edge probability 0.1, assign random edge weights drawn uniformly from $[-0.7, -0.3] \cup [0.3, 0.7]$, and add a scaled identity to ensure positive definiteness. The VAR coefficient matrix $B$ is generated as in Simulation A. All other settings are identical: $p=20$, $k=5$, 25 replicates. As in Simulation A we report the mean paired per-replicate difference in Frobenius distance against VAPOR, denoted $\Delta_F$, together with its standard error.

\FloatBarrier

\paragraph{Experiment A: varying autocorrelation ($T=200$).}
\begin{table}[h!]
\centering
\small
\setlength{\tabcolsep}{4pt}
\caption{Simulation B, Experiment A: varying spectral radius, $p=20$, $T=200$, 25 replicates. $\Delta_F$ is the mean paired per-replicate difference in Frobenius distance against VAPOR, with positive values favoring VAPOR and negative values favoring the competitor. The lowest Frobenius distance in each block is shown in bold.}
\label{tab:sim-b-autocorr}
\begin{tabular}{llcccc}
\toprule
Scenario & Method & Sens & Spec & Frob & $\Delta_F$ (SE) \\
\midrule
$\rho = 0.10$ & PrecFact (no VAR)  & 0.633 (0.120) & 0.985 (0.010) & 1.069 (0.159) & $0.042$ (0.012) \\
              & Glasso (no VAR)    & 0.994 (0.018) & 0.801 (0.060) & \textbf{0.633 (0.100)} & $-0.393$ (0.030) \\
              & VAR + Glasso       & 0.994 (0.018) & 0.801 (0.056) & \textbf{0.633 (0.100)} & $-0.393$ (0.030) \\
              & VAR + PrecFact     & 0.635 (0.129) & 0.985 (0.012) & 1.068 (0.179) & $0.042$ (0.009) \\
              & Kastner (VAR+FSV)  & 0.766 (0.120) & 0.977 (0.015) & 0.894 (0.217) & $-0.133$ (0.021) \\
              & VAPOR              & 0.613 (0.119) & 0.992 (0.008) & 1.027 (0.180) & --- \\
\midrule
$\rho = 0.50$ & PrecFact (no VAR)  & 0.454 (0.129) & 0.978 (0.016) & 1.337 (0.243) & $0.311$ (0.040) \\
              & Glasso (no VAR)    & 0.930 (0.060) & 0.786 (0.063) & 1.025 (0.216) & $-0.002$ (0.042) \\
              & VAR + Glasso       & 0.991 (0.022) & 0.796 (0.060) & \textbf{0.638 (0.102)} & $-0.388$ (0.029) \\
              & VAR + PrecFact     & 0.636 (0.120) & 0.984 (0.012) & 1.082 (0.183) & $0.055$ (0.009) \\
              & Kastner (VAR+FSV)  & 0.769 (0.100) & 0.977 (0.013) & 0.882 (0.213) & $-0.145$ (0.022) \\
              & VAPOR              & 0.581 (0.141) & 0.991 (0.009) & 1.027 (0.176) & --- \\
\midrule
$\rho = 0.75$ & PrecFact (no VAR)  & 0.352 (0.129) & 0.949 (0.026) & 1.853 (0.337) & $0.832$ (0.063) \\
              & Glasso (no VAR)    & 0.652 (0.328) & 0.851 (0.093) & 1.471 (0.244) & $0.450$ (0.047) \\
              & VAR + Glasso       & 0.994 (0.018) & 0.806 (0.056) & \textbf{0.639 (0.099)} & $-0.382$ (0.029) \\
              & VAR + PrecFact     & 0.649 (0.130) & 0.984 (0.012) & 1.090 (0.183) & $0.069$ (0.011) \\
              & Kastner (VAR+FSV)  & 0.767 (0.093) & 0.977 (0.013) & 0.905 (0.191) & $-0.116$ (0.016) \\
              & VAPOR              & 0.592 (0.139) & 0.992 (0.008) & 1.021 (0.168) & --- \\
\midrule
$\rho = 0.89$ & PrecFact (no VAR)  & 0.298 (0.126) & 0.925 (0.033) & 2.308 (0.443) & $1.268$ (0.086) \\
              & Glasso (no VAR)    & 0.296 (0.359) & 0.931 (0.099) & 1.649 (0.188) & $0.609$ (0.030) \\
              & VAR + Glasso       & 0.994 (0.018) & 0.805 (0.060) & \textbf{0.639 (0.099)} & $-0.402$ (0.032) \\
              & VAR + PrecFact     & 0.652 (0.132) & 0.982 (0.012) & 1.082 (0.172) & $0.041$ (0.011) \\
              & Kastner (VAR+FSV)  & 0.765 (0.114) & 0.977 (0.014) & 0.898 (0.219) & $-0.143$ (0.022) \\
              & VAPOR              & 0.585 (0.126) & 0.991 (0.009) & 1.041 (0.183) & --- \\
\midrule
$\rho = 0.95$ & PrecFact (no VAR)  & 0.263 (0.132) & 0.909 (0.033) & 2.610 (0.520) & $1.581$ (0.099) \\
              & Glasso (no VAR)    & 0.091 (0.246) & 0.982 (0.053) & 1.682 (0.168) & $0.653$ (0.024) \\
              & VAR + Glasso       & 0.994 (0.018) & 0.803 (0.059) & \textbf{0.638 (0.099)} & $-0.392$ (0.029) \\
              & VAR + PrecFact     & 0.630 (0.157) & 0.982 (0.015) & 1.085 (0.169) & $0.055$ (0.008) \\
              & Kastner (VAR+FSV)  & 0.774 (0.110) & 0.977 (0.014) & 0.899 (0.229) & $-0.131$ (0.022) \\
              & VAPOR              & 0.588 (0.111) & 0.992 (0.008) & 1.030 (0.172) & --- \\
\midrule
$\rho = 0.99$ & PrecFact (no VAR)  & 0.305 (0.120) & 0.877 (0.062) & 2.909 (0.679) & $1.872$ (0.137) \\
              & Glasso (no VAR)    & 0.087 (0.246) & 0.983 (0.050) & 1.693 (0.153) & $0.656$ (0.023) \\
              & VAR + Glasso       & 0.994 (0.018) & 0.802 (0.058) & \textbf{0.636 (0.100)} & $-0.400$ (0.027) \\
              & VAR + PrecFact     & 0.629 (0.103) & 0.983 (0.012) & 1.088 (0.180) & $0.051$ (0.013) \\
              & Kastner (VAR+FSV)  & 0.765 (0.103) & 0.977 (0.013) & 0.888 (0.210) & $-0.149$ (0.018) \\
              & VAPOR              & 0.575 (0.150) & 0.992 (0.007) & 1.037 (0.163) & --- \\
\bottomrule
\end{tabular}
\end{table}

VAR + Glasso attains the lowest Frobenius distance at every autocorrelation level, between 0.633 and 0.639, under the oracle penalty tuning described in Section~\ref{sec:simulations} and at a specificity near 0.80. Kastner is second at every level, between 0.882 and 0.905, ahead of VAPOR by 0.116 to 0.149 with $p < 10^{-4}$ throughout and with VAPOR closer to the truth in at most 4 of the 25 replicates. VAPOR is ahead of VAR + PrecFact at every level, by 0.041 to 0.069 ($p \le 0.002$). All four VAR-based methods have Frobenius distances that are flat in $\rho$, while the two methods that ignore temporal structure degrade as $\rho$ rises: VAPOR's margin over PrecFact applied to the raw series grows from 0.042 at $\rho = 0.10$ ($p = 0.002$) to 1.872 at $\rho = 0.99$, and its margin over the graphical lasso applied to the raw series from $-0.393$ to 0.656. VAPOR retains the highest specificity, 0.991 to 0.992, at the lowest sensitivity among the VAR-based methods, 0.575 to 0.613, against 0.991 to 0.994 for VAR + Glasso at a specificity near 0.80. When $\Omega$ carries no factor structure the precision factor parameterization does not recover it more accurately than Kastner's covariance factor model or than oracle-tuned graphical lasso applied to vector-autoregressive residuals, and the advantage VAPOR holds in Simulation A does not extend to this setting.

\FloatBarrier

\paragraph{Experiment B: varying sample size ($\rho = 0.95$).}
\begin{table}[h!]
\centering
\small
\setlength{\tabcolsep}{4pt}
\caption{Simulation B, Experiment B: varying sample size, $p=20$, $\rho(B)=0.95$, 25 replicates. $\Delta_F$ is the mean paired per-replicate difference in Frobenius distance against VAPOR, with positive values favoring VAPOR and negative values favoring the competitor. The lowest Frobenius distance in each block is shown in bold, under oracle penalty tuning for the graphical lasso.}
\label{tab:sim-b-samplesize}
\begin{tabular}{llcccc}
\toprule
Scenario & Method & Sens & Spec & Frob & $\Delta_F$ (SE) \\
\midrule
$T = 50$   & PrecFact (no VAR)  & 0.082 (0.086) & 0.965 (0.030) & 2.534 (0.546) & $0.997$ (0.103) \\
           & Glasso (no VAR)    & 0.094 (0.145) & 0.976 (0.040) & 1.685 (0.147) & $0.148$ (0.021) \\
           & VAR + Glasso       & 0.767 (0.135) & 0.838 (0.045) & \textbf{1.284 (0.115)} & $-0.253$ (0.020) \\
           & VAR + PrecFact     & 0.124 (0.101) & 0.994 (0.008) & 1.770 (0.216) & $0.232$ (0.039) \\
           & Kastner (VAR+FSV)  & 0.419 (0.092) & 0.946 (0.014) & 1.572 (0.182) & $0.035$ (0.034) \\
           & VAPOR              & 0.000 (0.000) & 1.000 (0.000) & 1.537 (0.107) & --- \\
\midrule
$T = 100$  & PrecFact (no VAR)  & 0.166 (0.085) & 0.929 (0.044) & 2.571 (0.525) & $1.281$ (0.105) \\
           & Glasso (no VAR)    & 0.110 (0.247) & 0.971 (0.077) & 1.688 (0.146) & $0.398$ (0.025) \\
           & VAR + Glasso       & 0.948 (0.052) & 0.804 (0.055) & \textbf{0.925 (0.135)} & $-0.366$ (0.028) \\
           & VAR + PrecFact     & 0.313 (0.134) & 0.992 (0.014) & 1.362 (0.176) & $0.072$ (0.019) \\
           & Kastner (VAR+FSV)  & 0.620 (0.089) & 0.964 (0.015) & 1.152 (0.183) & $-0.138$ (0.028) \\
           & VAPOR              & 0.162 (0.106) & 0.998 (0.006) & 1.290 (0.156) & --- \\
\midrule
$T = 200$  & PrecFact (no VAR)  & 0.263 (0.132) & 0.909 (0.033) & 2.610 (0.520) & $1.581$ (0.099) \\
           & Glasso (no VAR)    & 0.091 (0.246) & 0.982 (0.053) & 1.682 (0.168) & $0.653$ (0.024) \\
           & VAR + Glasso       & 0.994 (0.018) & 0.803 (0.059) & \textbf{0.638 (0.099)} & $-0.392$ (0.029) \\
           & VAR + PrecFact     & 0.630 (0.157) & 0.982 (0.015) & 1.085 (0.169) & $0.055$ (0.008) \\
           & Kastner (VAR+FSV)  & 0.774 (0.110) & 0.977 (0.014) & 0.899 (0.229) & $-0.131$ (0.022) \\
           & VAPOR              & 0.588 (0.111) & 0.992 (0.008) & 1.030 (0.172) & --- \\
\midrule
$T = 500$  & PrecFact (no VAR)  & 0.383 (0.146) & 0.857 (0.073) & 2.593 (0.587) & $1.681$ (0.123) \\
           & Glasso (no VAR)    & 0.142 (0.310) & 0.975 (0.060) & 1.675 (0.170) & $0.762$ (0.031) \\
           & VAR + Glasso       & 1.000 (0.000) & 0.801 (0.055) & \textbf{0.405 (0.039)} & $-0.508$ (0.031) \\
           & VAR + PrecFact     & 0.746 (0.114) & 0.977 (0.011) & 0.940 (0.170) & $0.028$ (0.014) \\
           & Kastner (VAR+FSV)  & 0.838 (0.096) & 0.983 (0.012) & 0.731 (0.227) & $-0.181$ (0.022) \\
           & VAPOR              & 0.752 (0.106) & 0.981 (0.013) & 0.913 (0.177) & --- \\
\midrule
$T = 1000$ & PrecFact (no VAR)  & 0.450 (0.137) & 0.833 (0.072) & 2.623 (0.548) & $1.766$ (0.116) \\
           & Glasso (no VAR)    & 0.149 (0.338) & 0.970 (0.069) & 1.682 (0.162) & $0.825$ (0.031) \\
           & VAR + Glasso       & 1.000 (0.000) & 0.788 (0.055) & \textbf{0.291 (0.039)} & $-0.566$ (0.040) \\
           & VAR + PrecFact     & 0.777 (0.097) & 0.972 (0.014) & 0.898 (0.207) & $0.041$ (0.019) \\
           & Kastner (VAR+FSV)  & 0.839 (0.090) & 0.984 (0.012) & 0.687 (0.237) & $-0.170$ (0.023) \\
           & VAPOR              & 0.789 (0.111) & 0.975 (0.015) & 0.857 (0.200) & --- \\
\bottomrule
\end{tabular}
\end{table}

VAR + Glasso attains the lowest Frobenius distance at every sample size under oracle penalty tuning, falling from 1.284 at $T = 50$ to 0.291 at $T = 1000$ at a specificity between 0.788 and 0.838. Kastner is second from $T = 100$ onward, ahead of VAPOR by 0.131 to 0.181 with $p < 10^{-4}$ at each of those sample sizes. At $T = 50$ the ordering differs: VAPOR selects no edges at all, with sensitivity 0.000 and specificity 1.000, and the resulting complete shrinkage gives a Frobenius distance of 1.537, second to VAR + Glasso, ahead of VAR + PrecFact by 0.232 ($p < 10^{-4}$) and indistinguishable from Kastner (0.035, $p = 0.31$, 14 of 25 replicates). VAPOR is ahead of VAR + PrecFact at every sample size from $T = 100$, by 0.028 to 0.072, significantly at $T = 100$, 200, and 1000 ($p \le 0.04$) but not at $T = 500$ ($p = 0.056$). VAPOR's margin over the two methods that ignore temporal structure grows with $T$, from 0.997 to 1.766 against PrecFact on the raw series and from 0.148 to 0.825 against the graphical lasso on the raw series, at $p < 10^{-4}$ throughout. Its sensitivity rises from 0.000 to 0.789 and its specificity falls from 1.000 to 0.975 across the range. As in Experiment A, the precision factor parameterization does not recover a precision matrix without factor structure more accurately than the covariance factor model or than oracle-tuned graphical lasso applied to vector-autoregressive residuals.

\FloatBarrier

\subsection{Simulation C: Data Generated from a Factor Stochastic Volatility Model}

We now generate data directly from the model of \citet{kastner2020sparse}, in which the VAR errors follow a single-factor stochastic volatility structure with time-varying covariance $\Sigma_t = \lambda\lambda^\top \exp(h_t) + \mathrm{diag}(\exp(g_t))$. Factor loadings are drawn from $\mathcal{N}(0.001, 0.001^2)$, idiosyncratic log-variances have mean $-12$ and persistence between 0.85 and 0.98, and the factor log-variance has persistence 0.99. VAPOR is misspecified here: it assumes a static $\Omega$ while the true innovation covariance is time-varying.

We fix $p=20$, $T=200$, and vary the sparsity of $B$ across three scenarios following \citet{kastner2020sparse}: sparse (1\% off-diagonal density), intermediate (10\%), and dense (80\%). Since no single true $\Omega$ exists, we evaluate all methods on held-out prediction MSPE: at each of the final 20\% of time points, 30\% of variables are held out and predicted from the remaining observed variables using the estimated precision matrix. VAPOR and the PrecFact-based methods use $k=5$ factors; Kastner's method is fit at a single factor, matching the data-generating process, so it is the correctly specified estimator in this setting. Results are reported as mean (sd) over 25 replicates, together with the mean paired per-replicate difference in MSPE against VAPOR, denoted $\Delta_{\mathrm{MSPE}}$, and its standard error.

\begin{table}[h!]
\centering
\small
\caption{Simulation C: data generated from the model of \citet{kastner2020sparse}, held-out prediction MSPE (lower is better), $p=20$, $T=200$, 25 replicates. $\Delta_{\mathrm{MSPE}}$ is the mean paired per-replicate difference against VAPOR, with positive values favoring VAPOR and negative values favoring the competitor. The lowest MSPE in each block is shown in bold.}
\label{tab:sim-c}
\begin{tabular}{llcc}
\toprule
Scenario & Method & MSPE & $\Delta_{\mathrm{MSPE}}$ (SE) \\
\midrule
Sparse (1\%)        & PrecFact (no VAR)   & 0.9093 (0.1250) & $-0.0004$ (0.0037) \\
                    & VAR + PrecFact      & 0.9041 (0.1295) & $-0.0055$ (0.0025) \\
                    & VAPOR               & 0.9097 (0.1264) & --- \\
                    & Kastner (VAR+FSV)   & \textbf{0.8520 (0.1492)} & $-0.0577$ (0.0104) \\
                    & Sample covariance   & 0.9190 (0.1543) & $0.0093$ (0.0141) \\
\midrule
Intermediate (10\%) & PrecFact (no VAR)   & 0.9007 (0.1437) & $-0.0078$ (0.0031) \\
                    & VAR + PrecFact      & 0.8992 (0.1325) & $-0.0094$ (0.0062) \\
                    & VAPOR               & 0.9085 (0.1413) & --- \\
                    & Kastner (VAR+FSV)   & \textbf{0.8403 (0.1283)} & $-0.0683$ (0.0171) \\
                    & Sample covariance   & 0.8985 (0.1474) & $-0.0100$ (0.0212) \\
\midrule
Dense (80\%)        & PrecFact (no VAR)   & 0.7066 (0.2062) & $-0.1462$ (0.0330) \\
                    & VAR + PrecFact      & 0.8531 (0.3504) & $0.0003$ (0.0071) \\
                    & VAPOR               & 0.8527 (0.3393) & --- \\
                    & Kastner (VAR+FSV)   & 0.7427 (0.3266) & $-0.1101$ (0.0176) \\
                    & Sample covariance   & \textbf{0.4668 (0.1551)} & $-0.3859$ (0.0690) \\
\bottomrule
\end{tabular}
\end{table}

Kastner's method attains the lowest MSPE in the sparse and intermediate scenarios, ahead of VAPOR by 0.058 and 0.068 respectively ($p < 10^{-4}$ and $p = 0.0005$), with VAPOR predicting more accurately in 2 and 1 of 25 replicates. In those two scenarios the remaining four methods are close: every paired difference between VAPOR and PrecFact, VAR + PrecFact, or the sample covariance is at most 0.010 in magnitude, and two reach significance, both favoring the competitor (VAR + PrecFact in the sparse scenario, 0.006, $p = 0.035$; PrecFact in the intermediate scenario, 0.008, $p = 0.019$). The dense scenario reverses the ordering. The sample covariance attains the lowest MSPE, 0.467 against 0.853 for VAPOR ($p < 10^{-4}$, 0 of 25 replicates), and PrecFact applied to the raw series also predicts more accurately than VAPOR, by 0.146 ($p = 0.0002$); VAR + PrecFact and VAPOR remain indistinguishable (0.0003, $p = 0.96$). Standard deviations for VAPOR, VAR + PrecFact, and Kastner in this scenario are two to three times those in the other two, so the ranking among the structured methods there is less firmly established than the gap to the sample covariance. Together, Simulations A, B, and C indicate that the precision factor parameterization helps when $\Omega$ carries factor structure and does not when the innovation covariance is time-varying or the coefficient matrix is dense.

\FloatBarrier

\subsection{Simulation D: Varying Signal Strength}

We fix $p=20$, $k=5$, $T=200$, $\rho(B)=0.95$ and vary the magnitude of nonzero entries in $\Lambda_0$, denoted \texttt{val}, across $\{0.5, 1.0, 1.5, 2.5, 4.0\}$. Higher \texttt{val} produces stronger partial correlations in $\Omega$ and a higher signal-to-noise ratio for graph recovery. The data-generating process is otherwise identical to Simulation A. Results are reported as mean (sd) over 25 replicates, together with the mean paired per-replicate difference in Frobenius distance against VAPOR, denoted $\Delta_F$, and its standard error.

\begin{table}[h!]
\centering
\small
\setlength{\tabcolsep}{3.5pt}
\caption{Simulation D: varying signal strength, $p=20$, $k=5$, $T=200$, $\rho(B)=0.95$, 25 replicates. The mean absolute nonzero partial correlation in $\Omega$ is 0.058, 0.166, 0.262, 0.394, and 0.498 across the five values of \texttt{val} in increasing order. $\Delta_F$ is the mean paired per-replicate difference in Frobenius distance against VAPOR, with positive values favoring VAPOR. The lowest Frobenius distance in each block is shown in bold.}
\label{tab:sim-d}
\begin{tabular}{llcccc}
\toprule
Scenario & Method & Sens & Spec & Frob & $\Delta_F$ (SE) \\
\midrule
\texttt{val} $= 0.5$ & PrecFact (no VAR)  & 0.123 (0.062) & 0.888 (0.048) & 2.651 (0.708) & $2.146$ (0.145) \\
                     & Glasso (no VAR)    & 0.003 (0.011) & 0.999 (0.003) & 0.508 (0.127) & $0.002$ (0.015) \\
                     & VAR + Glasso       & 0.154 (0.129) & 0.956 (0.038) & \textbf{0.471 (0.122)} & $-0.035$ (0.013) \\
                     & VAR + PrecFact     & 0.010 (0.027) & 1.000 (0.002) & 0.512 (0.124) & $0.006$ (0.009) \\
                     & Kastner (VAR+FSV)  & 0.270 (0.091) & 0.896 (0.016) & 0.569 (0.112) & $0.063$ (0.015) \\
                     & VAPOR              & 0.015 (0.034) & 1.000 (0.000) & 0.506 (0.109) & --- \\
\midrule
\texttt{val} $= 1.0$ & PrecFact (no VAR)  & 0.166 (0.073) & 0.884 (0.053) & 2.873 (0.689) & $2.192$ (0.134) \\
                     & Glasso (no VAR)    & 0.006 (0.015) & 0.997 (0.007) & 1.417 (0.327) & $0.736$ (0.061) \\
                     & VAR + Glasso       & 0.653 (0.193) & 0.706 (0.119) & 0.818 (0.135) & $0.136$ (0.022) \\
                     & VAR + PrecFact     & 0.329 (0.185) & 0.998 (0.004) & 0.763 (0.102) & $0.082$ (0.011) \\
                     & Kastner (VAR+FSV)  & 0.461 (0.112) & 0.923 (0.020) & 0.888 (0.188) & $0.207$ (0.032) \\
                     & VAPOR              & 0.351 (0.177) & 0.998 (0.003) & \textbf{0.681 (0.077)} & --- \\
\midrule
\texttt{val} $= 1.5$ & PrecFact (no VAR)  & 0.238 (0.111) & 0.864 (0.061) & 3.269 (0.656) & $2.580$ (0.126) \\
                     & Glasso (no VAR)    & 0.053 (0.192) & 0.967 (0.139) & 2.193 (0.468) & $1.505$ (0.091) \\
                     & VAR + Glasso       & 0.839 (0.144) & 0.525 (0.146) & 0.934 (0.114) & $0.246$ (0.019) \\
                     & VAR + PrecFact     & 0.555 (0.181) & 0.998 (0.004) & 0.762 (0.082) & $0.073$ (0.012) \\
                     & Kastner (VAR+FSV)  & 0.543 (0.109) & 0.934 (0.020) & 1.178 (0.399) & $0.489$ (0.076) \\
                     & VAPOR              & 0.550 (0.169) & 1.000 (0.002) & \textbf{0.689 (0.071)} & --- \\
\midrule
\texttt{val} $= 2.5$ & PrecFact (no VAR)  & 0.333 (0.147) & 0.849 (0.049) & 3.882 (0.777) & $3.158$ (0.146) \\
                     & Glasso (no VAR)    & 0.307 (0.404) & 0.779 (0.298) & 3.115 (0.569) & $2.391$ (0.104) \\
                     & VAR + Glasso       & 0.942 (0.067) & 0.360 (0.105) & 0.968 (0.110) & $0.244$ (0.013) \\
                     & VAR + PrecFact     & 0.751 (0.101) & 0.999 (0.002) & \textbf{0.702 (0.066)} & $-0.022$ (0.019) \\
                     & Kastner (VAR+FSV)  & 0.599 (0.120) & 0.942 (0.021) & 1.830 (0.691) & $1.105$ (0.130) \\
                     & VAPOR              & 0.718 (0.114) & 1.000 (0.001) & 0.724 (0.113) & --- \\
\midrule
\texttt{val} $= 4.0$ & PrecFact (no VAR)  & 0.400 (0.146) & 0.835 (0.060) & 4.387 (0.859) & $3.500$ (0.167) \\
                     & Glasso (no VAR)    & 0.435 (0.439) & 0.673 (0.343) & 3.769 (0.621) & $2.882$ (0.117) \\
                     & VAR + Glasso       & 0.965 (0.033) & 0.297 (0.045) & 0.970 (0.105) & $0.083$ (0.023) \\
                     & VAR + PrecFact     & 0.829 (0.090) & 0.997 (0.006) & \textbf{0.680 (0.080)} & $-0.207$ (0.032) \\
                     & Kastner (VAR+FSV)  & 0.620 (0.117) & 0.945 (0.020) & 2.453 (0.774) & $1.566$ (0.148) \\
                     & VAPOR              & 0.778 (0.102) & 0.999 (0.002) & 0.887 (0.137) & --- \\
\bottomrule
\end{tabular}
\end{table}

VAPOR attains the lowest Frobenius distance at \texttt{val} $= 1.0$ and $1.5$, where the mean absolute partial correlation is 0.17 and 0.26, and its margin over every competitor there is significant at $p < 10^{-4}$. The advantage does not extend to either end of the range. At \texttt{val} $= 0.5$, where the mean absolute partial correlation is 0.06, no method recovers more than 27\% of the true edges and the five methods other than PrecFact applied to the raw series are separated by less than 0.10 in Frobenius distance; VAR + Glasso is ahead of VAPOR by 0.035 ($p = 0.012$). At \texttt{val} $= 2.5$ and $4.0$, VAR + PrecFact overtakes VAPOR, by 0.022 ($p = 0.26$) and 0.207 ($p < 10^{-4}$, with VAPOR closer to the truth in 1 of 25 replicates). Kastner's Frobenius distance rises monotonically with signal strength, from 0.569 to 2.453, and VAR + Glasso's specificity falls from 0.956 to 0.297 as its sensitivity rises from 0.154 to 0.965. VAPOR's specificity is 0.998 or above across the full range. The gain from joint estimation is largest where edges are detectable but individually uncertain, and the sequential approach is preferable once the partial correlations are strong enough to be estimated reliably in two steps.

\FloatBarrier

\subsection{Simulation E: Approximately Low-Rank VAR Coefficients Under Limited Sample Size}

We consider a setting in which the VAR coefficient matrix is approximately low rank while the innovation precision matrix follows the proposed precision factor model. Specifically, the innovations satisfy
\[
\Omega=\Lambda_0\Lambda_0^\top+I_p,
\]
where $\Lambda_0\in\mathbb{R}^{p\times k}$ is sparse with nonzero entries equal to $\pm1.5$, exactly as in Simulation A.

Unlike the previous simulations, the VAR coefficient matrix is generated as
\[
B = UV^\top + E,
\]
where $U,V\in\mathbb{R}^{p\times r_0}$ with $r_0=3$, yielding an effective rank-three coefficient matrix. A Gaussian perturbation equal to 5\% of the Frobenius norm of the low-rank component is added so that $B$ is approximately, rather than exactly, low rank. The resulting matrix is rescaled to have spectral radius $\rho(B)=0.7$.

This structure is designed to stress the low-sample regime for which the reduced-rank parameterization is intended: because full-rank estimation of $B$ requires $T$ comfortably larger than $p$, the full-rank sampler is not applicable here and is omitted. We fit VAPOR with the reduced-rank parameterization at rank $r=3$ under both the ridge and Dirichlet-Laplace priors on $V$, and compare against Kastner's factor stochastic volatility VAR and the graphical lasso.\footnote{Sample covariance and the sequential VAR-plus-graphical-lasso estimators degrade sharply at $T\le p$ and are omitted for clarity.}

As in Section~\ref{subsec:reduced-rank}, we fix the rank rather than estimate it; here $r=3$ coincides with the effective rank of the data-generating $B$. We evaluate two axes. First, holding $p=50$ fixed, we vary the sample size across $T/p\in\{1.0,1.5,2.0\}$ to trace how recovery improves as data accumulate. Second, holding $T/p=1.0$ fixed, we vary the dimension across $p\in\{50,100,150\}$ to check whether any advantage depends on $p$ being small. Each setting is averaged over 25 replicates.

Performance is measured by the Frobenius distance between the estimated and true correlation-scaled precision matrices, $\|\bar R-R_0\|_F$; the relative Frobenius error of the estimated coefficient matrix, $\|\hat B-B_0\|_F/\|B_0\|_F$ (reported for VAPOR only, since it isolates the mechanism); and edge-recovery sensitivity and specificity computed at an oracle threshold that retains the true number of edges. We also report the mean paired per-replicate difference in $\|\bar R-R_0\|_F$ against VAPOR with the Dirichlet-Laplace prior. Representative estimated precision matrices are shown as partial-correlation heatmaps.

\begin{table}[t]
\centering
\caption{Simulation E results, averaged over 25 replicates (standard deviations in parentheses). $\|\bar R-R_0\|_F$ is the Frobenius error of the correlation-scaled precision matrix; $\|\hat B-B_0\|_F/\|B_0\|_F$ is the relative coefficient error (VAPOR only); Sens and Spec are edge-recovery sensitivity and specificity at an oracle threshold retaining the true number of edges. The lowest Frobenius error in each block is shown in bold. Against VAPOR (DL), every paired difference to Kastner and the graphical lasso is positive and significant at $p<10^{-4}$, and the two VAPOR variants are statistically indistinguishable in every block but $(50,100)$. Full-rank VAPOR is omitted because full-rank estimation of $B$ requires $T$ well above $p$. The graphical lasso's Sens/Spec are omitted as degenerate (see text).}
\label{tab:sim-e}
\small
\begin{tabular}{llcccc}
\toprule
$(p,T)$ & Method & $\|\bar R-R_0\|_F$ & $\|\hat B-B_0\|_F/\|B_0\|_F$ & Sens & Spec \\
\midrule
\multirow{4}{*}{$(50,50)$}
  & VAPOR (DL)    & \textbf{3.95 (0.41)} & 0.70 (0.10) & 0.30 (0.08) & 0.95 (0.01)\\
  & VAPOR (ridge) & 3.96 (0.40) & 1.09 (0.54) & 0.29 (0.07) & 0.95 (0.01)\\
  & Kastner       & 4.60 (0.46) & --          & 0.13 (0.04) & 0.93 (0.01)\\
  & Glasso        & 4.35 (0.46) & --          & --          & --\\
\midrule
\multirow{4}{*}{$(50,75)$}
  & VAPOR (DL)    & 3.25 (0.50) & 0.57 (0.15) & 0.43 (0.09) & 0.96 (0.01)\\
  & VAPOR (ridge) & \textbf{3.24 (0.49)} & 0.58 (0.30) & 0.44 (0.10) & 0.96 (0.01)\\
  & Kastner       & 4.49 (0.48) & --          & 0.14 (0.04) & 0.93 (0.01)\\
  & Glasso        & 4.32 (0.47) & --          & --          & --\\
\midrule
\multirow{4}{*}{$(50,100)$}
  & VAPOR (DL)    & 2.69 (0.43) & 0.48 (0.15) & 0.55 (0.08) & 0.97 (0.01)\\
  & VAPOR (ridge) & \textbf{2.63 (0.44)} & 0.44 (0.21) & 0.57 (0.08) & 0.97 (0.01)\\
  & Kastner       & 4.48 (0.48) & --          & 0.15 (0.04) & 0.93 (0.01)\\
  & Glasso        & 4.26 (0.46) & --          & --          & --\\
\midrule
\multirow{4}{*}{$(100,100)$}
  & VAPOR (DL)    & \textbf{6.85 (1.30)} & 0.64 (0.08) & 0.47 (0.09) & 0.95 (0.01)\\
  & VAPOR (ridge) & 6.93 (1.22) & 0.94 (0.33) & 0.47 (0.08) & 0.95 (0.01)\\
  & Kastner       & 9.60 (1.01) & --          & 0.11 (0.02) & 0.92 (0.01)\\
  & Glasso        & 9.47 (1.00) & --          & --          & --\\
\midrule
\multirow{4}{*}{$(150,150)$}
  & VAPOR (DL)    & \textbf{8.10 (1.30)} & 0.57 (0.07) & 0.58 (0.06) & 0.96 (0.01)\\
  & VAPOR (ridge) & 8.12 (1.37) & 0.78 (0.25) & 0.58 (0.06) & 0.96 (0.01)\\
  & Kastner       & 14.57 (1.42)& --          & 0.09 (0.02) & 0.92 (0.01)\\
  & Glasso        & 14.46 (1.41)& --          & --          & --\\
\bottomrule
\end{tabular}
\end{table}

Across every setting VAPOR recovers the precision matrix more accurately than both competitors, and the margin widens with dimension. At $p=50$ the paired improvement over Kastner in $\|\bar R-R_0\|_F$ ranges from 0.65 to 1.79 across the three sample sizes; by $p=150$ it reaches 6.47, where VAPOR attains roughly half the error of both Kastner and the graphical lasso ($8.10$ versus $14.57$ and $14.46$). Every paired difference against Kastner and the graphical lasso is significant at $p<10^{-4}$, with VAPOR closer to the truth in all 25 replicates of every cell. The same ordering holds for edge recovery: at matched specificity near $0.95$, VAPOR's sensitivity is two to six times that of Kastner throughout. The coefficient error $\|\hat B-B_0\|_F/\|B_0\|_F$ shows that reduced-rank estimation recovers the dynamics well even at $T=p$, which explains the precision-matrix advantage: accurate estimation of $B$ yields clean residuals from which the factor-structured precision matrix can be identified.

Holding $p=50$ fixed and increasing $T/p$ from $1.0$ to $2.0$, VAPOR's precision error falls steadily from $3.95$ to $2.69$, while Kastner's stays near $4.5$, so only VAPOR improves as data accumulate. Holding $T/p=1.0$ fixed and increasing $p$, VAPOR's relative advantage grows rather than shrinks, so its advantage at $p=50$ does not depend on the dimension being small. The ridge and Dirichlet-Laplace priors on $V$ perform near-identically on precision recovery, their paired difference never exceeding $0.09$ in Frobenius distance and reaching significance only at $(50,100)$; the Dirichlet-Laplace prior yields somewhat more stable coefficient estimates, and we adopt it as the default. The graphical lasso's oracle-threshold sensitivity and specificity are omitted as degenerate. Under Frobenius-optimal tuning its selected estimate is exactly diagonal in 11 to 17 of the 25 replicates at each setting, and its mean Frobenius error is within $0.10$ of that of the identity matrix ($\|R_0 - I\|_F = 4.36$, $9.48$, and $14.48$ at $p = 50$, $100$, and $150$), so thresholding it to retain the true number of edges is an arbitrary tie-break.

\begin{figure}[t]
\centering
\includegraphics[width=\textwidth,alt={Five side-by-side 150 by 150 partial correlation matrices for one replicate at T equal to p: the true precision matrix, followed by VAPOR under the Dirichlet-Laplace prior, VAPOR under the ridge prior, Kastner, and the graphical lasso. The true matrix holds two dense clusters of nonzero entries, one near the top left and one near the bottom right, against an otherwise empty background. Both VAPOR panels reproduce both clusters at the same positions and comparable magnitudes. The Kastner and graphical lasso panels show no visible structure anywhere.}]{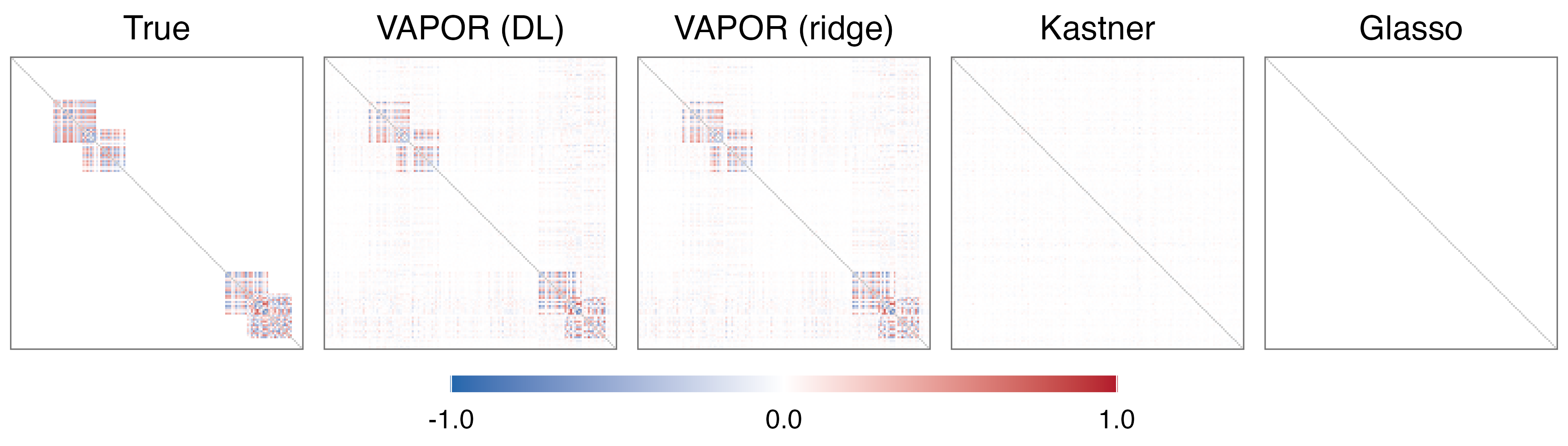}
\caption{Partial-correlation heatmaps for a representative replicate of Simulation E at $p=150$, $T=150$ ($T/p=1.0$). Left to right: true precision matrix, VAPOR (Dirichlet-Laplace prior, rank $3$), VAPOR (ridge prior, rank $3$), Kastner, and the graphical lasso. Color encodes the partial correlation from $-1$ to $+1$ as shown in the scale; the diagonal is omitted.}
\label{fig:sim-e-heatmaps-p150}
\end{figure}

The Frobenius and edge-recovery metrics are summarized numerically in Table~\ref{tab:sim-e}, but the qualitative difference between the methods is most apparent visually. Figure~\ref{fig:sim-e-heatmaps-p150} shows the true precision matrix alongside the estimates from VAPOR, Kastner, and the graphical lasso, displayed as partial-correlation heatmaps for a single representative replicate at $p=150$, $T=150$. The true matrix has clear block structure, with groups of variables that load on shared factors appearing as dense diagonal blocks. VAPOR reproduces these blocks in the correct positions with the correct signs under both the ridge and Dirichlet-Laplace priors, while Kastner returns a diffuse estimate in which the block structure is absent and the graphical lasso returns a diagonal estimate with no off-diagonal entries. VAPOR's lower Frobenius error therefore reflects recovery of the block structure: it identifies which groups of variables are conditionally dependent, and the competing methods do not.

\FloatBarrier

\subsection{Simulation F: The $T<p$ Regime}

Simulation E established that VAPOR recovers the precision matrix more accurately than competing methods once $T \geq p$, with the advantage widening as the dimension grows. We now test the limits of the model by moving into the regime $T<p$. This is the setting the reduced-rank parameterization is ultimately designed for: full-rank estimation of $B$ is infeasible, and even the innovation precision matrix is difficult to recover for any method, since the strongest partial correlations are hard to distinguish from sampling variability. We fix the ratio at $T/p = 0.7$ and vary the dimension across $p\in\{50, 100, 150\}$, keeping the data-generating process, fitted rank, and evaluation metrics identical to Simulation E. 

\begin{table}[h!]
\centering
\caption{Simulation F ($T<p$): fixed ratio $T/p=0.7$, varying dimension, 25 replicates (standard deviations in parentheses). $\|\bar R-R_0\|_F$ is the Frobenius error of the correlation-scaled precision matrix; $\|\hat B-B_0\|_F/\|B_0\|_F$ is the relative coefficient error (VAPOR only); Sens and Spec are edge-recovery sensitivity and specificity at an oracle threshold retaining the true number of edges. Full-rank VAPOR is omitted as it is not applicable at $T\le p$. The graphical lasso's Sens/Spec are omitted as degenerate (see text).}
\label{tab:sim-f}
\centering
\small
\begin{tabular}{llcccccc}
\toprule
$(p,T)$ & Method & $\|\bar R-R_0\|_F$ & $\|\hat B-B_0\|_F/\|B_0\|_F$ & Sens & Spec \\
\midrule
\multirow{4}{*}{$(50,35)$}
  & VAPOR (DL)    & 4.36 (0.36) & 0.84 (0.08) & 0.20 (0.06) & 0.94 (0.01)\\
  & VAPOR (ridge) & \textbf{4.32 (0.38)} & 0.90 (0.27) & 0.20 (0.06) & 0.94 (0.01)\\
  & Kastner       & 4.75 (0.48) & --          & 0.12 (0.04) & 0.93 (0.01)\\
  & Glasso        & 4.35 (0.46) & --          & --          & --\\
\midrule
\multirow{4}{*}{$(100,70)$}
  & VAPOR (DL)    & \textbf{8.95 (1.05)} & 0.79 (0.05) & 0.23 (0.04) & 0.93 (0.02)\\
  & VAPOR (ridge) & 8.98 (1.06) & 0.79 (0.15) & 0.23 (0.05) & 0.93 (0.02)\\
  & Kastner       & 9.63 (1.00) & --          & 0.09 (0.02) & 0.92 (0.01)\\
  & Glasso        & 9.48 (1.01) & --          & --          & --\\
\midrule
\multirow{4}{*}{$(150,105)$}
  & VAPOR (DL)    & \textbf{13.47 (1.55)} & 0.77 (0.05) & 0.24 (0.05) & 0.93 (0.01)\\
  & VAPOR (ridge) & 13.56 (1.54) & 0.72 (0.11) & 0.23 (0.05) & 0.93 (0.01)\\
  & Kastner       & 14.58 (1.44) & --          & 0.09 (0.01) & 0.92 (0.01)\\
  & Glasso        & 14.48 (1.43) & --          & --          & --\\
\bottomrule
\end{tabular}
\end{table}

The $T<p$ regime is the hardest setting we consider, and recovery is difficult for every method, with absolute Frobenius errors far larger than in Simulation E. At $p = 50$ VAPOR and the graphical lasso are not separable: under the Dirichlet-Laplace prior they differ by $-0.010$ in paired Frobenius distance ($p = 0.75$, 13 of 25 replicates), the ridge prior is nominally lowest at $4.32$, and Kastner trails VAPOR by $0.389$ ($p < 10^{-4}$). From $p = 100$ onward VAPOR attains the lowest precision error and the paired differences against both competitors are significant at $p < 10^{-4}$, with the margin over Kastner rising from $0.681$ to $1.109$ and that over the graphical lasso from $0.526$ to $1.009$ between $p = 100$ and $150$. The advantage therefore appears only once the dimension is large enough for the factor structure to be worth exploiting, even though the sample-to-dimension ratio is held fixed. Edge recovery separates the methods more cleanly: VAPOR's sensitivity is between $1.7$ and $2.7$ times that of Kastner at every dimension, at comparable specificity. The two priors on $V$ differ significantly only at $p = 50$, where ridge is ahead by $0.037$ ($p = 0.002$); elsewhere they are indistinguishable, and as in Simulation E the Dirichlet-Laplace prior gives the more stable coefficient estimates, with standard deviations of $0.05$ to $0.08$ against $0.11$ to $0.27$ for ridge. As in Simulation E, the selected graphical lasso estimate is exactly diagonal in 16 to 21 of the 25 replicates at each dimension, and its mean error is within $0.02$ of that of the identity matrix.

\FloatBarrier

\end{document}